\documentclass[a4paper,11pt,preprintnumbers]{article}
\pdfoutput=1
\usepackage{jheppub}

\usepackage{mathrsfs}
\usepackage{amsfonts}
\usepackage{amsmath,amssymb,bm}
\usepackage{epsfig}
\usepackage{wrapfig}
\usepackage{graphicx,xcolor}
\usepackage[figurename=figure]{caption}
\usepackage{subcaption}
\usepackage[colorlinks=true,citecolor=blue,linkcolor=blue,urlcolor=blue]{hyperref}
\usepackage[normalem]{ulem}
\usepackage{xcolor}
\usepackage{ulem}
\usepackage{multirow}
\usepackage{physics}
\usepackage{ulem,xpatch}
\usepackage{titlesec}
\usepackage{mathtools}
\usepackage{slashed}
\usepackage{adjustbox}
\usepackage{lipsum}

\renewcommand{\slash}[1]{#1 \hspace{-0.45em} / }

\newcommand{\sgn}[1]{{\rm sgn}\left(#1\right)}

\newcommand{\beq}{\begin{equation}}
\newcommand{\eeq}{\end{equation}}
\newcommand{\bse}{\begin{subequations}}
\newcommand{\ese}{\end{subequations}}
\newcommand{\LQCD}{ \Lambda_{\rm QCD} }
\newcommand{\nn}{\nonumber}

\def\B{{\rm B}}

\begin{document}

\title{Exclusive production of a pair of high transverse momentum photons in pion-nucleon collisions for extracting generalized parton distributions}

\author[a,b]{Jian-Wei Qiu}
\author[c]{Zhite Yu}

\affiliation[a]{Theory Center, Jefferson Lab,
Newport News, Virginia 23606, USA}
\affiliation[b]{Department of Physics, William \& Mary,
Williamsburg, Virginia 23187, USA}
\affiliation[c]{Department of Physics and Astronomy, 
Michigan State University, East Lansing, MI 48824, USA}

\emailAdd{jqiu@jlab.org}
\emailAdd{yuzhite@msu.edu} 

\date{\today}
\preprint{JLAB-THY-22-3617, MSUHEP-22-018}

\abstract{
We show that exclusive production of a pair of high transverse momentum photons in pion-nucleon collisions can be systematically studied in QCD factorization approach if the photon's transverse momentum $q_T$ with respect to the colliding pion is much greater than $\Lambda_{\rm QCD}$. We demonstrate that the leading power non-perturbative contributions to the scattering amplitudes of this exclusive process are process-independent and can be systematically factorized into universal pion's distribution amplitudes (DAs) and nucleon's generalized parton distributions (GPDs), which are convoluted with corresponding infrared safe and perturbatively calculable short-distance hard parts.  The correction to this factorized expression is suppressed by powers of $1/q_T$.  We demonstrate quantitatively that this new type of exclusive processes is not only complementary to existing processes for extracting GPDs, but also capable of providing an enhanced sensitivity to the dependence of both DAs and GPDs on the active parton's momentum fraction $x$. We also introduce additional, but the same type of exclusive observables to enhance our capability to explore GPDs, in particular, their $x$-dependence.
}


\arxivnumber{2205.07846}

\maketitle
\flushbottom

\section{Introduction}
\label{s.intro}

The proton and neutron, known as nucleons, are the fundamental building blocks of all atomic nuclei, and 
themselves are emerged as strongly interacting and relativistic bound states of quarks and gluons of Quantum Chromodynamics (QCD). Understanding the internal structure of nucleons in terms of their constituents, quarks and gluons, and their interactions has been one of the central goals of modern particle and nuclear physics.  However, owing to the color confinement of QCD, it has been an unprecedented intellectual challenge to explore and quantify the structure of nucleons without being able to see quarks and gluons directly.  QCD color interaction is so strong at a typical hadronic scale ${\cal O}(\Lambda_{\rm QCD})\sim 1/R$ with a typical hadron radius $R\sim 1$~fm that 
any cross section with identified hadron(s) cannot be calculated fully in QCD perturbation theory.

Fortunately, with the help of asymptotic freedom of QCD by which the color interaction becomes weaker and calculable perturbatively at short distances, the QCD factorization theorem~\cite{Collins:1989gx} has been developed to factorize the dynamics at different momentum scales to identify good cross sections (or good physical observables) whose leading non-perturbative dynamics can be organized into universal distribution functions, while other non-perturbative contributions are shown to be suppressed by inverse power of the large momentum transfer of the collision. Predictions follow when cross sections with different hard scatterings but the same nonperturbative distributions are compared.  It is the QCD factorization for physical scattering processes with a large momentum transfer $Q\gg 1/R$ that has enabled us to probe the particle (or partonic) nature of quarks and gluons at the short-distance, and to connect them to observed hadron(s) in terms of universal distribution functions.  With a set of well determined universal distribution functions to find a quark ($q$), antiquark ($\bar{q}$),  or gluon ($g$) with a momentum fraction $x$ inside a colliding hadron of momentum $p$ with $xp\sim Q$, known as the parton distribution functions (PDFs) $f_{i/h}(x,\mu^2)$ for finding a parton of type $i=q,\bar{q},g$ inside a colliding hadron $h$ probed at a hard scale $\mu\sim Q$, QCD factorization formalism has been extremely successful in interpreting high energy experimental data from all facilities around the world, covering many orders in kinematic reach in both $x$ and $Q$ and as large as 15 orders of magnitude in difference in the size of observed scattering cross sections, which is a great success story of QCD and the Standard Model at high energy and has given us the confidence and the tools to discover the Higgs particle in proton-proton collisions \cite{CMS:2012qbp,ATLAS:2012yve}, and to search for the new physics \cite{CidVidal:2018eel}.

However, the probe with a large momentum transfer $Q\, (\gg 1/R)$ is so localized in space that it is not very sensitive to the details of confined three-dimensional (3D) internal structure of the colliding hadron, in which a confined parton should have a characteristic transverse momentum scale $\langle k_T\rangle \sim 1/R \ll Q$ and an uncertainty in transverse position $\langle b_T\rangle  \sim R \gg 1/Q$.  Recently, new and more precise data are becoming available for {\it two-scale} observables with a hard scale $Q$ to localize the collision to probe the partonic nature of quarks and gluons along with a soft scale to be sensitive to the dynamics taking place at ${\cal O}(1/R)$.  In addition, theoretical advances over the past decades have resulted in the development of QCD factorization formalism for two types of two-scale observables, distinguished by their inclusive or exclusive nature, which enables quantitative matching between the measurements of such two-scale observables and the 3D internal partonic structure of a colliding hadron.  For inclusive two-scale observables, one well-studied example is the production of a massive boson that decays into a pair of measured leptons in hadron-hadron collisions (known as the Drell-Yan process), as a function of the pair's invariant mass $Q$ and transverse momentum $Q_T$ in the Lab frame \cite{Collins:1984kg}.  When $Q \gg 1/R$, the production is dominated by the annihilation of one active parton from one colliding hadron with another active parton from the other colliding hadron, including quark-antiquark annihilation to a vector boson ($\gamma$, $W/Z$) or gluon-gluon fusion to a Higgs particle.  When $Q\gg Q_T \gtrsim 1/R$, the measured transverse momentum of the pair is sensitive to the transverse momenta of the two colliding partons before they annihilate into the massive boson, providing the opportunity to extract the information on the active parton's transverse motion inside the colliding hadron, which is encoded in transverse momentum dependent (TMD) PDFs (or simply, TMDs), $f_{i/h}(x,k_T,\mu^2)$ \cite{Collins:2011zzd}.  Like PDFs, TMDs are universal distribution functions to find a quark (or gluon) with a momentum fraction $x$ and transverse momentum $k_T$ from a colliding hadron of momentum $p$ with $xp\sim \mu\sim Q \gg k_T$, and describe the 3D motion of this active parton, its flavor dependence and its correlation with the property of the colliding hadron, such as its spin \cite{Bacchetta:2006tn,Diehl:2015uka,Sivers:1989cc,Collins:1992kk,Qiu:1991pp}.  However, the probed transverse momentum $k_T$ of the active parton in the hard collision is {\it not} the same as the intrinsic or confined transverse momentum of the same parton inside a bound hadron. When the colliding hadron is broken by the large momentum transfer of the collision, a parton shower (the collision induced partonic radiation) is developed during the collision, generating additional transverse momentum to the probed active parton, which is encoded in the QCD evolution of the TMDs and could be non-perturbative, depending on the hard scale $Q$ and the phase space available for the shower~\cite{Collins:1984kg,Qiu:2000hf}. With more data from current and future experiments, including lepton-hadron semi-inclusive deep inelastic scatterings, better understanding of the scale dependence of TMDs could provide us with valuable information on the confined motion of quarks and gluons inside a bound hadron \cite{Accardi:2012qut, AbdulKhalek:2021gbh, Liu:2021jfp}.

Without breaking the colliding hadron, the exclusive observables could provide different aspects of the hadron's internal structure.  Since any cross section with identified hadron(s) cannot be calculated fully in QCD perturbation theory, it is necessary to have a hard scale $Q\gg 1/R$ for good exclusive observables for studying hadron's partonic structure.  One classic example of exclusive hadronic observables is the high energy elastic $\pi$-scattering from atomic electrons \cite{Akerlof:1967zza}, from which the electromagnetic form factor $F_{\pi}(Q^2)$ of the pion could be extracted as a function of the invariant mass of the exchanged virtual photon momentum $q$ in the collision with $Q^2\equiv -q^2 \geq 0$.  But, with the size and limited range of $Q^2$, the extracted form factor $F_{\pi}(Q^2)$ did not reveal much information on the partonic nature of the pion. On the other hand, when $Q^2\gg 1/R^2$, $F_{\pi}(Q^2)$ could be factorized in terms of a convolution of two pion distribution amplitudes (DAs), $\phi_{\pi}(x,\mu)$ with momentum fraction $x$ for an active quark, $1-x$ for the corresponding antiquark and factorization scale $\mu$, along with a perturbatively calculable short-distance coefficient function, as seen in eq.~\eqref{eq:pionFF}\footnote{where instead of $x$, variables $z_1$ and $z_2$ are used for parton momentum fractions of DAs.}.
The contributions from the pion's partonic states beyond a pair of active quark and antiquark are expected to be suppressed by powers of $1/(QR)$ \cite{Brodsky:1989pv}. Various experimental efforts have been devoted to measure the pion form factors at larger momentum transfers, from which the pion DAs could be extracted \cite{JeffersonLabFpi:2000nlc,JeffersonLabFpi-2:2006ysh,JeffersonLabFpi:2007vir}. However, with the \textit{localized} single hard interaction from the exchanged virtual photon, the factorized pion form factor $F_{\pi}(Q^2)$ is not very sensitive to the detailed shape of $\phi_{\pi}(x,\mu)$ as a function of $x$, other than the integral of $\phi_{\pi}(x,\mu)$ over $x$;
see the discussion following eq.~\eqref{eq:H4pionFF}.

\begin{figure}[htb]
\begin{center}
\begin{minipage}[c]{0.25\textwidth}
\includegraphics[width=0.99\textwidth]{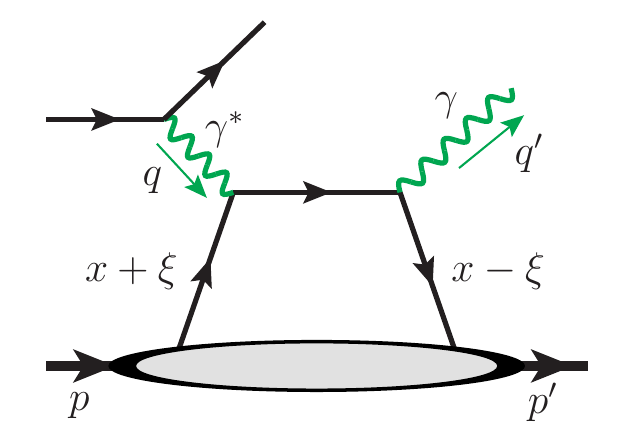}
\\
\end{minipage}
$+ ...$\hskip 0.03\textwidth
\begin{minipage}[c]{0.25\textwidth}
\includegraphics[width=0.99\textwidth]{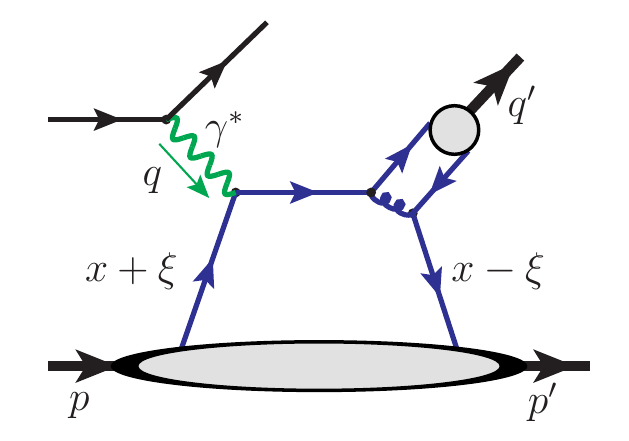}
\\
\end{minipage}
$+ ...$\hskip 0.03\textwidth
\begin{minipage}[c]{0.25\textwidth}
\includegraphics[width=0.99\textwidth]{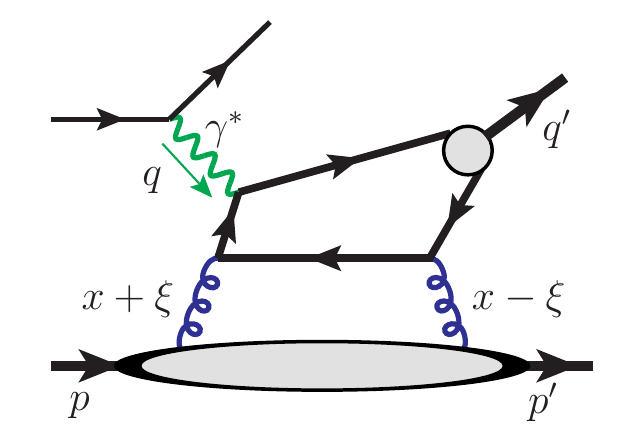}
\\
\end{minipage}
$+ ...$ \\
\hskip -0.055\textwidth (a) \hskip 0.29\textwidth (b) \hskip 0.29\textwidth (c) 
\caption{Sample two-scale observables from exclusive deeply virtual lepton-hadron scattering:
(a) deeply virtual Compton scattering (DVCS), 
(b) deeply virtual meson production (DVMP), and
(c) deeply virtual heavy quarkonium production (DVQP).
}
\label{figuredvlhs}
\end{center}
\end{figure} 

Nucleon's internal structure could be much more rich and complex than the pion structure.  QCD factorization of hard exclusive processes involving nucleons, such as large angle exclusive hadronic scattering, could be worked out, but, the corresponding calculations are much more difficult \cite{Brodsky:1989pv}.  On the other hand, exclusive lepton-nucleon scattering with a virtual photon of invariant mass $Q^2\gg 1/R^2$ could provide various two-scale observables, such as those in figure~\ref{figuredvlhs}, where the hard scale is $Q^2\equiv -q^2$ and the soft scale is $t\equiv (p-p')^2$.  When $Q^2\gg |t|$, which is 
equivalent to requiring the time scale of the partonic hard collision $\sim 1/Q$ to be much shorter than the lifetime of the exchanged partonic states $\sim 1/\sqrt{|t|}$, these two-scale exclusive processes are dominated by the exchange of an active $q\bar{q}$ or $gg$ pair, as shown in figure~\ref{figuredvlhs}, and can be systematically treated in QCD factorization approach~{\cite{Collins:1997hv,Collins:1996fb,Collins:1998be}.  The hadronic properties of the diffracted nucleon, the bottom part of the diagrams in figure~\ref{figuredvlhs}, could be represented by generalized parton distribution functions (or simply, GPDs), $f_{i/h}(x,\xi,t,\mu)$, where $\xi\equiv (p-p')^+/(p+p')^+$. The $(p-p')^+ = 2\xi [(p+p')^+/2]$ represents a total light-cone momentum transfer between the diffracted nucleon $h$ and the hard partonic collision, where the light-cone components are defined as $v^{\pm}=(v^0\pm v^z)/\sqrt{2}$ for any four vector $v^\mu$.  The GPDs were introduced by D.~M\"{u}ller {\it et al.} in 1994~\cite{Muller:1994ses}, and their important roles in charactering hadron's partonic structure were further established by pioneering work in \cite{Ji:1996ek,Radyushkin:1997ki} and many years' theoretical development since then, which could be summarized in the reviews~\cite{Goeke:2001tz,Diehl:2003ny,Belitsky:2005qn,Boffi:2007yc} and references therein.

By Fourier transforming the transverse component of the momentum transfer $(p-p')_T$ to position space $b_T$ in the forward limit, $p'^+\to p^+$ (or $\xi\to 0$), the transformed GPD as a function of $b_T$ provides a transverse spatial distribution of quarks or gluons inside a colliding hadron at different values of momentum fraction $x$ \cite{Burkardt:2000za}, That is, measuring GPDs could provide an opportunity to study QCD tomography to obtain images of the transverse spatial densities of quarks and gluons slicing at different momentum fraction $x$ inside a colliding hadron.  Their spatial $b_T$ dependence could allow us to define an effective hadron radius in terms of its quark (or gluon) spatial distributions, $r_q(x)$ (or $r_g(x)$), as a function of $x$, in contrast to its electric charge radius, allowing us to ask some interesting questions, such as should $r_q(x) > r_g(x)$ or vice versa, and could $r_g(x)$ saturate if $x\to 0$, which could reveal valuable information on how quarks and gluons are bounded inside a hadron.  
Although we could expect that $r_q(x)$ (or $r_g(x)$) is small at large $x$ and increases when $x$ decreases, as demonstrated in explicit model calculations \cite{Burkardt:2002hr}, it is the precise knowledge of GPDs as functions of the parton flavor and kinematic variables, $(x,\xi, t)$, that is needed for us to address these kinds of interesting and fundamental questions about the hadron, in particular, the proton and neutron, the fundamental building blocks of our visible world.

However, as clearly evident from the leading order diagrams in figure~\ref{figuredvlhs}, the scattering with the exchange of a single virtual photon in figure~\ref{figuredvlhs} is effectively an exclusive $2\to 2$ process: $\gamma^*(q)+h(p)\to X(q')+h'(p')$ with a final-state particle $X=\gamma, \pi, {\rm J}/\psi, ...$, whose momentum is uniquely fixed by the virtual photon momentum $q$ and total momentum transfer $(p-p')$ from the diffracted hadron (or $\xi$- and $t$-dependence of GPDs).  Any sensitivity to the dependence of GPDs on the momentum fraction $x$,
 which is proportional to the relative momentum of the active quark and antiquark in figure~\ref{figuredvlhs}(a) and (b), or the two gluons in figure~\ref{figuredvlhs}(c), has to come from high order contribution and scale dependence of the process.  More specifically, let's consider the deeply virtual Compton scattering (DVCS), first introduced in~\cite{Ji:1996nm}, as sketched in figure~\ref{figuredvlhs}(a). 
The DVCS cross section can be naturally expressed in terms of Compton form factors (CFFs), which are then factorized as convolutions of GPDs with perturbatively calculable coefficients according to QCD factorization~\cite{Radyushkin:1997ki,Ji:1998xh,Collins:1998be}.  Extracting full details of GPDs from CFFs is a challenging inverse or deconvolution problem~\cite{Kumericki:2016ehc}.  Due to the lack of sensitivity on the $x$-dependence for CFFs, it was shown \cite{Bertone:2021yyz} that based on a next-to-leading order analysis and a careful study of evolution effects, the reconstruction of GPDs from DVCS measurements does not possess a unique solution.  Actually, two sample GPDs with different $x$-dependence can both fit the same CFFs \cite{Bertone:2021yyz}.

\begin{figure}[htb]
\begin{center}
\begin{tabular}{ccc}
\includegraphics[scale=0.6]{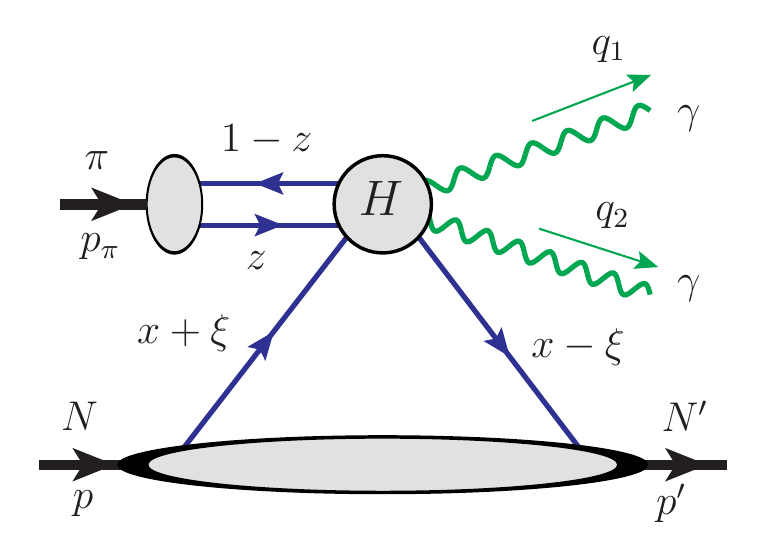}
& \includegraphics[scale=0.6]{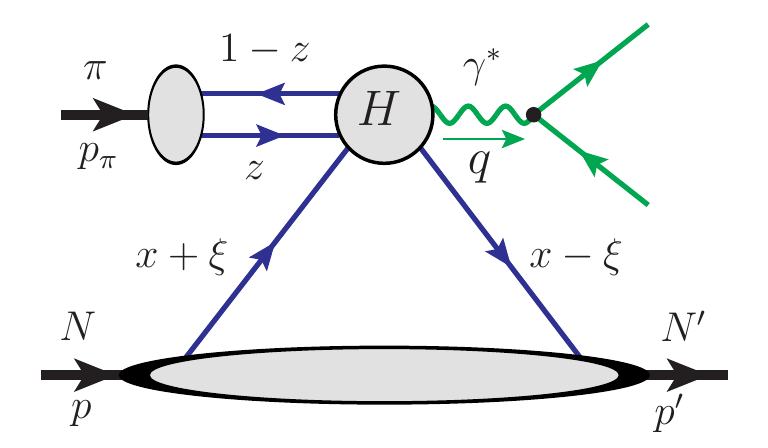}
& \includegraphics[scale=0.6]{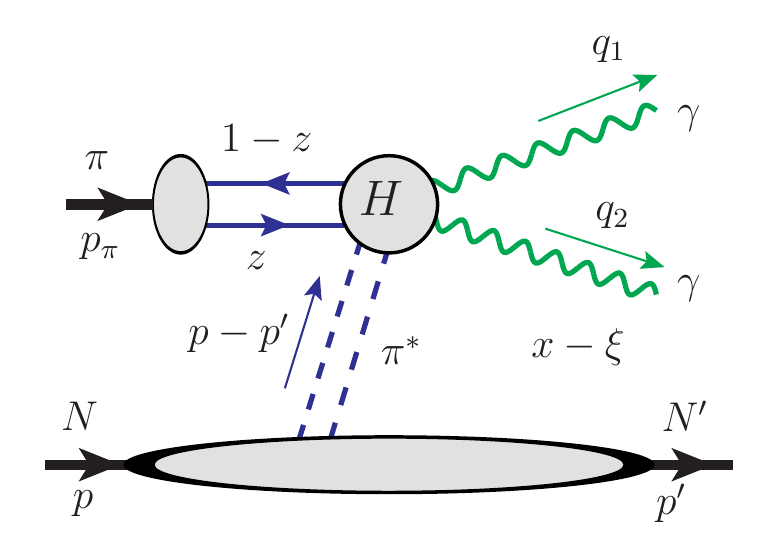}
\\
(a) & (b) & (c)
\end{tabular}
\caption{
Exclusive massive photon-pair (a) and lepton-pair (b) production in pion-nucleon collision, and (c) the photon-pair productions when $|t| \equiv |(p-p')^2| \to 0$.
}
\label{figureempp}
\end{center}
\end{figure} 

Meanwhile, new exclusive diffractive processes have been introduced to enhance our capability to extract various GPDs from experimental measurements.  Instead of the lepton-nucleon scattering in figure~\ref{figuredvlhs}, it was proposed to study the diffractive photo-production of a massive photon pair: $\gamma(q)+N(p_1)\to \gamma(k_1)+\gamma(k_2)+N'(p_2)$ with the pair's invariant mass 
$M_{\gamma\gamma}\gg\Lambda_{\rm QCD}$~\cite{Pedrak:2017cpp,Pedrak:2020mfm,Grocholski:2021man,Grocholski:2022rqj}.  Similarly, the diffractive photo-production of a massive photon and meson pair: $\gamma(q)+N(p_1)\to \gamma(k_1)+\rho(k_2)+N'(p_2)$~\cite{Boussarie:2016qop} and 
$\gamma(q)+N(p_1)\to \gamma(k_1)+\pi^\pm(k_2)+N'(p_2)$~\cite{Duplancic:2018bum}, as well as the diffractive production of two jets with a large invariant mass~\cite{Golec-Biernat:1998exl,Braun:2005rg,Ji:2016jgn} were also proposed.  Unlike the lepton-scattering processes in figure~\ref{figuredvlhs}, whose factorization was proved by Collins {\it et al.}~\cite{Collins:1997hv,Collins:1996fb,Collins:1998be}, the challenge for these new processes has been the lack of the same level of justification for the QCD factorization.

In this paper, we study exclusive pion-nucleon diffractive production of a pair of high transverse momentum photons: $\pi(p_\pi)+N(p)\to \gamma(q_1) +\gamma(q_2) +N'(p')$, as sketched in figure~\ref{figureempp}(a), with the photon's transverse momentum with respect to the collision axis between the colliding pion and the quark-antiquark pair from the diffracted nucleon being $q_T=|q_{1T}| = |q_{2T}| \gg \Lambda_{\rm QCD}$. 
Similar to the exclusive Drell-Yan process in pion-nucleon collision in figure~\ref{figureempp}(b)~\cite{Berger:2001zn}, or the exclusive deeply virtual lepton-hadron scattering processes in figure~\ref{figuredvlhs}, the $\pi N$ scattering process of our consideration in figure~\ref{figureempp}(a) is also a $2\to 3$ exclusive process with a diffractive nucleon.  Instead of measuring the lepton pair from the decay of a massive virtual photon in figure~\ref{figureempp}(b), or the scattered lepton to have the deeply virtual photon in figure~\ref{figuredvlhs}, the hard scale of this new type of exclusive two-scale processes is provided by the large transverse momentum $q_T$, which flows between the two back-to-back photons. The soft scale of this new type of two-scale processes is provided by $t=(p-p')^2$, the invariant mass squared of momentum transfer from the diffractive nucleon, which is the same as the soft scale of those exclusive processes in figure~\ref{figuredvlhs} and \ref{figureempp}(b).
With $q_T\gg \sqrt{-t}$, we demonstrate that this new observable can be systematically studied in terms of QCD factorization approach with the same level of justification as those in figure~\ref{figuredvlhs}, and our factorization arguments can be generalized to the similar type of exclusive processes, including some mentioned above.  We also show that this observable can be not only a good probe of the factorized GPDs, complementary to those known exclusive processes, but also capable of providing more sensitivity to the much needed $x$-dependence of GPDs.

When the hard scale $q_T$ is sufficiently large, the diffractive scattering on the nucleon $N(p)$ is likely dominated by an exchange of a quark-antiquark pair, as indicated in figure~\ref{figureempp}(a), pulling more physically polarized partons into the hard collision would be suppressed by powers of $q_T^{-1}/R$.  Depending on the momentum flow of the active quark and antiquark, there are two distinctive kinematic regions for this exclusive process: (1) both active quark and antiquark have their momenta flowing into the hard part, as indicated in figure~\ref{figureempp}(c), and (2) only one of the active partons (quark or antiquark) has its momentum entering into the hard collision while the other has its momentum flowing out the hard collision to recombine with the spectators to form the diffracted hadron $N'(p')$, as sketched in figure~\ref{figureempp}(a).  As explained in Sec.~\ref{s.pih2gg}, the factorization proof for these two regions requires different consideration due to the characteristic difference of soft gluons in the Glauber region.  Once factorized, the region (1) gets contribution from the ERBL region of GPDs, while the other is relevant to the GPDs' DGLAP region~\cite{Diehl:2003ny}. 
When $|t|\to 0$, while $q_T^2 \gg \Lambda_{\rm QCD}^2$, the diffractive scattering with the nucleon in figure~\ref{figureempp}(c) is kinematically similar to the Sullivan process in lepton-nucleon scattering \cite{Sullivan:1971kd} and becomes sensitive to the nucleon's pion cloud. The production of the massive photon-pair in this kinematic regime ($|t|\to 0$) could be viewed approximately as an annihilation of a real pion and a virtual (or almost real) pion of the colliding nucleon.  To help present our justification of QCD factorization for exclusive massive photon-pair production in pion-nucleon collision
in figure~\ref{figureempp}(a), we first demonstrate how the exclusive scattering amplitude of a simpler exclusive process, $\pi^+(p_1)+\pi^-(p_2)\to \gamma(q_1)+\gamma(q_2)$ with $q_T\gg \Lambda_{\rm QCD}$ in figure~\ref{fig:pipi_LO}, can be systematically factorized into a convolution of two pion DAs along with an infrared safe and perturbatively calculable short-distance coefficient in Sec.~\ref{s.toy}.  With the large transverse momentum flow from one photon to the other through the hard scattering, interfering with the relative momentum flow between the active quark and antiquark of the colliding pion(s), the $q_T$ distribution of one of the two produced photons (or the equivalent $\cos\theta$ distribution of the photon with respect to the collision axis in the pair's rest frame) can be sensitive to the momentum difference between the quark and antiquark of colliding pion(s), providing the sensitivity to the shape of factorized pion DAs.  

In Sec.~\ref{s.pih2gg}, we extend our collinear factorization arguments for the single-scale exclusive process: $\pi^+(p_1)+\pi^-(p_2)\to \gamma(q_1)+\gamma(q_2)$ to the two-scale exclusive observable: $\pi(p_\pi) + N(p) \to \gamma(q_1)+\gamma(q_2)+N'(p')$ with $q_T\gg \sqrt{|t|}$.  With the nonlocal color coherence between the incoming and the outgoing (or diffracted) nucleon, the $N$ and $N'$, we need additional discussions and reasoning for justifying the factorization of soft gluon interactions for this two-scale observable.  We argue that when $q_T \gg \sqrt{|t|}$, the leading contribution to exclusive scattering amplitude of $\pi(p_\pi) + N(p) \to \gamma(q_1)+\gamma(q_2)+N'(p')$ can be factorized into the universal GPDs convoluted with a pion DA along with infrared safe and perturbatively calculable coefficients.  The corrections to this factorized expression is suppressed by powers of $|t|/q_T^2$. 
We show that by extending the $\pi^+\pi^-$ process to $\pi N$ process, the scattering amplitude develops both real and imaginary parts, both of which contribute to the cross section, and contains contributions from both unpolarized and polarized GPDs.
Consequently, this new type of two-scale exclusive processes can be sensitive to both unpolarized and polarized GPDs.

In Sec.~\ref{s.numerical}, we demonstrate numerically the sensitivity of this new type of exclusive high transverse momentum observables to the functional forms of pion DAs and nucleon GPDs in terms of their $x$-dependence.  
We introduce a flexible parametrization for DAs and a simplified version of the GK model for nucleon GPDs \cite{Goloskokov:2005sd,Goloskokov:2007nt,Goloskokov:2009ia} with parameters to adjust their dependence on the parton momentum fraction.  With our perturbatively calculated short-distance coefficients and our models for nucleon GPDs and pion DAs, we show explicitly how sensitive this exclusive production of a pair of high-$q_T$ photons can be to the shape of nucleon GPDs and pion DAs as functions of $x$.  
We also point out that such sensitivity could be enhanced with improved 
high-order calculation of the short-distance coefficients so that they are more perturbatively reliable  
at the end points where the momentum fraction of active parton from DAs and GPDs vanishes.  
Finally, in Sec.~\ref{s.outlook}, we present our summary and outlook on opportunities to measure this new type of exclusive process at J-PARC and other facilities.  We also discuss possibilities of additional two-scale observables of this type, which have the hard scale provided by the large transverse momentum $q_T$ of two exclusively produced ``back-to-back" final-state particles (or jets) with $q_T\gg\sqrt{|t|}\gtrsim\Lambda_{\rm QCD}$. The results of the hard coefficients are presented in the Appendix.

\section{Exclusive production of a pair of high transverse momentum photons in a $\pi^+\pi^-$ annihilation}
\label{s.toy}

Exclusive production of a pair of high transverse momentum photons in $\pi^+\pi^-$ annihilation, as sketched in figure~\ref{fig:pipi_LO}, 
has a single observed hard scale, $q_T$, the transverse momentum of one of the two produced photons with respect to the $\pi^+\pi^-$ collision axis. 
The large scale $q_T$ leads to a point-like interaction that is sensitive to the partonic structure of the pions.
It is then natural to consider QCD collinear factorization approach for studying this exclusive process.  
We show in this section that when $q_T \gg \Lambda_{\rm QCD}$, the scattering amplitude of this exclusive process can be factorized in terms of two pion DAs and a perturbatively calculable hard part, 
with corrections suppressed by powers of $1/q_T$.
One of the main steps in deriving the factorization is to deform the soft gluon momenta out of the Glauber region. This is straightforward for the $\pi^+\pi^-$ annihilation process because there is no pinch in the Glauber region, as we will show below. When we generalize the factorization formalism to the diffractive $\pi N$ process in Sec.~\ref{s.pih2gg}, an additional kinematic region, referred to as DGLAP region for GPD, appears, for which the soft gluon momentum is partly pinched in the Glauber region, and some modification is needed to prove the factorization.

\begin{figure}[htb]
\begin{center}
\hskip0.05in
\begin{minipage}[c]{0.35\textwidth}
\includegraphics[width=0.99\textwidth]{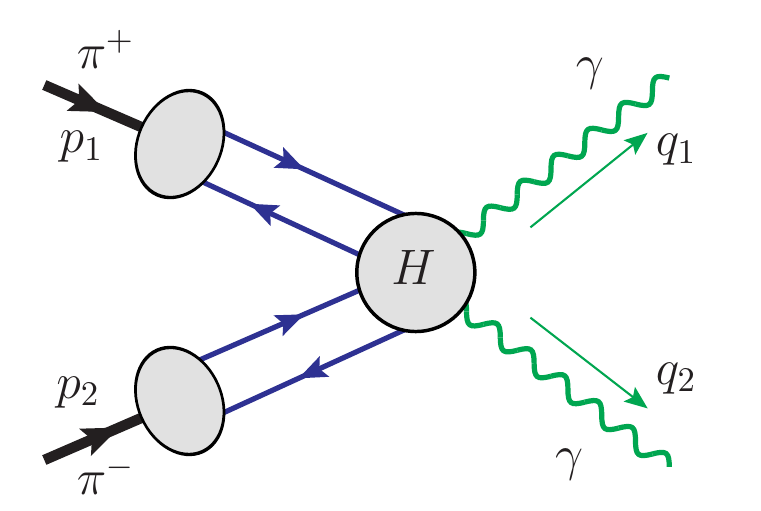}
\end{minipage}
\hskip 0.02\textwidth
\begin{minipage}[c]{0.61\textwidth}
\includegraphics[width=0.99\textwidth]{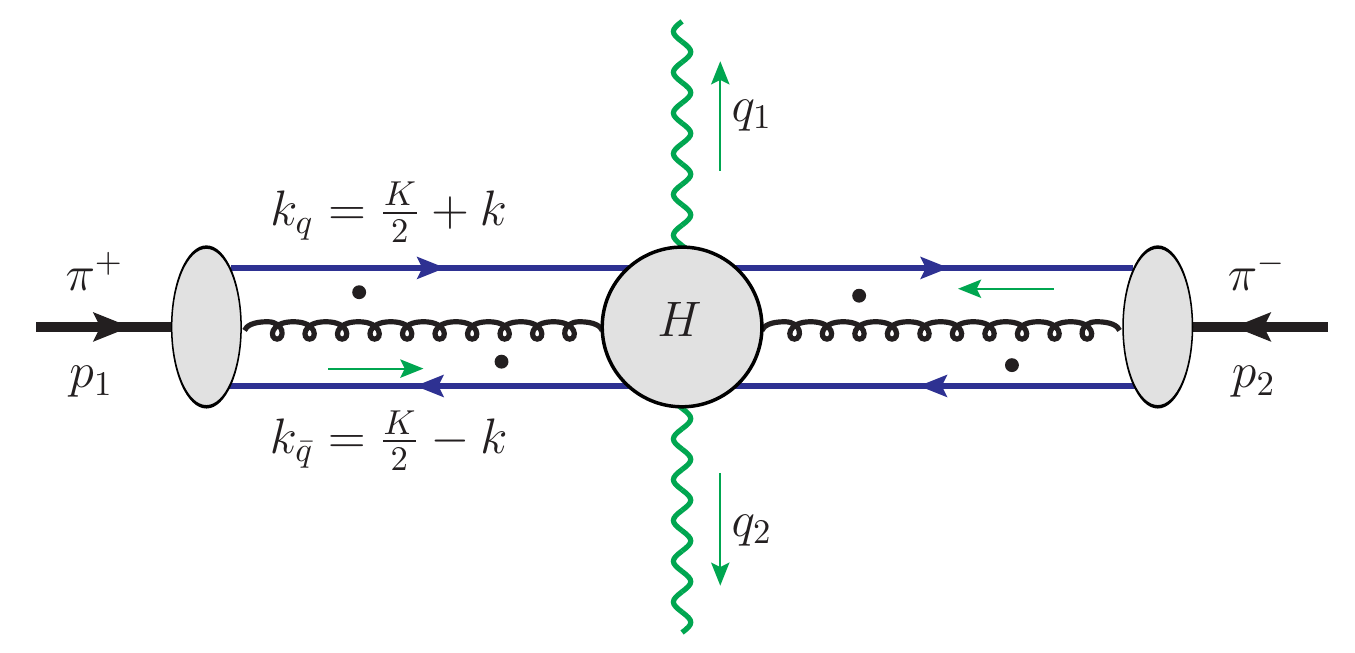}
\end{minipage}
\\
\vspace{0.1in}
\hskip -0.12\textwidth (a) 
\hskip 0.46\textwidth (b) 
\caption{
(a) Exclusive massive photon-pair production in $\pi^+\pi^-$ annihilation, and (b) 
sample diagram to show the existence of perturbative pinch needed to separate the dynamics at different scales.
}
\label{fig:pipi_LO}
\end{center}
\end{figure}

\subsection{The process and corresponding kinematics}
\label{ss.definion}

We study the exclusive production of a pair of high transverse momentum back-to-back photons in $\pi^+\pi^-$ annihilation 
in the center-of-mass (CM) frame of the collision,
\beq
\pi^+(p_1) + \pi^-(p_2) \longrightarrow \gamma(q_1) + \gamma(q_2)\, ,
\eeq
as sketched in figure~\ref{fig:pipi_LO}(a), where $\pi^+$ moves along $+\hat{z}$ direction and $\pi^-$ along $-\hat{z}$ direction.
The scattering amplitude of this exclusive process is defined as
\beq
{\cal M}_{\pi^+\pi^- \to \gamma(\lambda)\gamma(\lambda')} 
\equiv \varepsilon^{\lambda *}_{\mu}(q_1)\, \varepsilon^{\lambda' *}_{\nu}(q_2) \, {\cal M}^{\mu\nu}_{\pi^+\pi^- \to \gamma \gamma} \, ,
\eeq
where $\varepsilon^{\lambda}_{\mu}(q)$ is the polarization vector for a photon of momentum $q$ and polarization $\lambda$.

In the CM frame of this $2\to 2$ process, the energy of the colliding pion is the same as the energy of the observed photon and equal to $\sqrt{s}/2$ with 
$s=(p_1+p_2)^2 = (q_1+q_2)^2 = \hat{s}_{\gamma\gamma} \gg \Lambda_{\rm QCD}^2$.  
By requiring $q_T\gg\Lambda_{\rm QCD}$, we can safely neglect the pion mass $m_{\pi}$ in the following discussion of the leading power QCD factorization of this process in the power expansion of $1/q_T$.

Using the light-cone coordinates defined with respect to the $\hat{z}$ axis, we define all the relevant momenta as follows,
\bse \label{eq:kin}
\begin{align}
p_1 &= 
\left( p_1^+, \; \frac{m_{\pi}^2}{2 p_1^+} , \; \bm{0}_T \right)
\simeq 
\left( p_1^+, \; 0^- , \; \bm{0}_T \right), \\
p_2 &= 
\left( \frac{m_{\pi}^2}{2 p_2^-} ,\; p_2^-, \; \bm{0}_T \right)
\simeq 
\left( 0^+,\; p_2^-, \; \bm{0}_T \right), \\
q_1 &= 
\left( q_1^+,\; \frac{q_T^2}{2 q_1^+},\; -\bm{q}_T \right), \\
q_2 &= 
\left( \frac{q_T^2}{2 q_2^-},\; q_2^-,\; \bm{q}_T \right), 
\end{align}\ese
where $q_T\equiv |\bm{q}_T|$.
Introducing the light-cone unit vectors, 
\beq
\bar{n}^{\mu}=(1,0,\bm{0}_T),\quad
n^{\mu}=(0,1,\bm{0}_T),\quad
n^{\mu}_T=(0,0,\bm{1}_T),
\label{eq:n nb}
\eeq 
with 
$\bar{n}^2=n^2=0$, 
$\bar{n}\cdot n =1$, 
$\bar{n}\cdot n_T = n\cdot n_T=0$, 
and $n_T^2=-1$,
we can express the momenta of colliding pions as
$p_1^\mu = \sqrt{s/2}~\bar{n}^\mu$ and 
$p_2^\mu = \sqrt{s/2}~{n}^\mu$
in the CM frame.
Similarly, the observed photon momenta $q_1$ and $q_2$ are fully determined once $\bm{q}_T$ is specified, 
\bse\label{eq:kin q12}
\begin{align}
q_1 &= \left( \frac{p_1^+}{2} \left( 1 \pm \sqrt{1-\kappa} \right),\,
				\frac{p_2^-}{2} \left( 1 \mp \sqrt{1-\kappa} \right),\,
				-\bm{q}_T 
		\right)	\, ,	\\
q_2 &= \left( \frac{p_1^+}{2} \left( 1 \mp \sqrt{1-\kappa} \right),\,
				\frac{p_2^-}{2} \left( 1 \pm \sqrt{1-\kappa} \right),\,
				\bm{q}_T 
		\right)	\, ,
\end{align}
\ese
where $\kappa=4q_T^2/s  \leq 1$, $p_1^+ = p_2^- = \sqrt{s/2}$ in the  CM frame, and the $\pm$ solution refers to $q_1$ goes to the forward ($+\hat{z}$) or backward ($-\hat{z}$) direction.

\subsection{All-order factorization of exclusive scattering amplitude}
\label{sec:factorization1}

The development of factorized cross sections starts with an examination of scattering amplitudes in terms of general properties of Feynman diagrams in QCD perturbation theory.  When $q_T \sim \sqrt{s}$ becomes large, the exclusive $\pi^+\pi^-$ annihilation process, as sketched in figure~\ref{fig:pipi_LO}(a), is associated with two distinctive scales: 
(1) the hard scale $q_T$ characterizing the short-distance (perturbative) hard collision to produce the massive photon pair, as shown by the middle blob of the diagram in figure~\ref{fig:pipi_LO}(b), and 
(2) the soft scale ${\cal O}(m_{\pi})\sim \Lambda_{\rm QCD}$ characterizing the long-distance (non-perturbative) hadronic dynamics associated with the colliding pions.  A consistent separation of QCD dynamics 
taking place at these two distinctive scales can lead to a factorization formalism, which is an approximation up to corrections suppressed in power of $m_{\pi}/q_T$.  
The validity of perturbative QCD factorization formalism requires the suppression of quantum interference between the dynamics 
taking place at these two different momentum scales.  That is, the dominant contributions to the factorized formalism should necessarily come from the phase space where the active parton(s) linking the dynamics at two different scales are forced onto their mass shells, and are consequently long-lived compared to the time scale of the hard collision.  
For example, for the exclusive scattering amplitude in figure~\ref{fig:pipi_LO}(b), the suppression of quantum interference between the dynamics taking place in the middle blob at ${\cal O}(q_T)$ and the blobs on its left and right associated with the colliding pions requires us to demonstrate that the active quark, antiquark and gluon(s) from the colliding pions are effectively forced to be near their mass shells.  

However, all internal loop momentum integrals to any scattering amplitude are defined by contours in complex momentum space, and it is only at momentum configurations where some subset of loop momenta are pinched that the contours are forced to or near mass-shell poles that correspond to long-distance behavior. These ‘‘pinch surfaces’’ in multidimensional momentum space can be classified according to their {\it reduced diagrams}, found by contracting off-shell lines to points, from which we then derive the factorization formalism.  

\subsubsection{Reduced diagrams and leading pinch surfaces}
\label{sec:rd}

Reduced diagrams specify the regions in the multidimensional loop momentum space that give dominant contributions to the loop integrals.  Such leading regions are more conveniently realized in cut diagram notation of inclusive cross sections, in which graphical contributions to the cross sections are represented by the scattering amplitude to the left of the final state cut and the complex conjugate amplitude to the right. In the complex conjugate graphs all roles of momentum integrals are reversed with an opposite sign of $i\epsilon$, which are responsible for the pinched poles associated with initial- and final-state interactions.  However, for the factorization of exclusive scattering amplitudes, like the one in figure~\ref{fig:pipi_LO}(b), all partons are internal and virtual.  Their pinched poles, if there is any, do not come from the pair of the same propagators in the amplitude and its complex conjugate amplitude, since the momentum flows through them do not have to be the same in the amplitude and its complex conjugate amplitude.  For the exclusive scattering amplitudes, like the one in figure~\ref{fig:pipi_LO}(b), it is the integration of the relative momentum of any two active partons that pinches their momenta to be approximately on mass-shell if the invariant mass of these two active partons from the colliding pion is much smaller than their total energy.  

We illustrate this pinch of loop momenta by using the sample diagram in figure~\ref{fig:pipi_LO}(b) and labeling the active quark and antiquark momenta from $\pi^+$ on the left as $k_q=K/2 + k$ and $k_{\bar{q}}=K/2-k$, respectively.  The scattering amplitude in figure~\ref{fig:pipi_LO}(b) then takes the form,
\begin{eqnarray}
\label{eq:amplitude}
{\cal M}(q_T,s)
&\propto &
\int \frac{d^4K}{(2\pi)^4} \int \frac{d^4k}{(2\pi)^4}\, {\rm Tr}{\Big [}
\hat{R}_{\pi^-}(p_2,l_j)\otimes_{l_j}
\hat{H}(K,k,k_i; l_j; q_T,s) 
\nonumber\\
&& {\hskip 0.6in}\otimes_{k_i} 
\frac{\gamma\cdot(K/2+k)}{(K/2+k)^2+i\epsilon}\,
\hat{D}_{\pi^+}(p_1,K,k,k_i) \,
\frac{-\gamma\cdot(K/2-k)}{(K/2-k)^2+i\epsilon}
{\Big ]}	\, ,
\end{eqnarray}
where $\hat{H}$ and $\hat{R}_{\pi^-}$ represent the middle blob and right-hand-side of the diagram, respectively, the $\otimes_{l_j}$ and $\otimes_{k_i}$ indicate the convolution of parton momenta $l_j$ and $k_i$, respectively, with $i,j=1,2,...$, $\hat{D}_{\pi^+}$ represents the DA of the $\pi^+$ of momentum $p_1$, and $K$ and $k$ are the total and relative momentum of the active quark-antiquark pair on the left.  If the total momentum of the pair is dominated by $K^+$, we can identify the relevant perturbative contribution from the integration of $k$ in eq.~\eqref{eq:amplitude} by examining the pole structure of its $k^-$ integration.  From the denominators of eq.~\eqref{eq:amplitude}, we have the two poles for $k^-$,
\bse\label{eq:pinch}
\begin{align}
k^- &= \frac{-1}{K^+} \left[ \frac{K^2}{4} 
    - \frac{k_T^2 }{ 1  + 2k^+/K^+} \right] 
    - i\epsilon \, \sgn{K^+ + 2k^+}
{\hskip 0.2in}
 \rightarrow\ \ 0 - i\epsilon,
                 \\
k^- &= \frac{1}{K^+} \left[ \frac{K^2}{4}
     - \frac{k_T^2 }{ 1 - 2k^+/K^+} \right] 
     + i\epsilon \, \sgn{K^+ - 2k^+}
{\hskip 0.2in}
\rightarrow -0 + i\epsilon \, ,
\end{align}
\ese
where we neglected the quark mass and overall transverse momentum of the pair $K_T$.  These two denominators pinch the $k^-$ integral, when the total energy of the pair (or its light-cone momentum $K^+$) is much larger than the virtuality of the pair, so long as we are away from the region $k^+ \to \pm K^+/2$, where the quark (or antiquark) of the pair carries all the momentum
while the other carries none.
We should assume that this region is strongly suppressed by the $\pi$'s DA when $p_1^+\gg m_\pi$.  It is then clear from eq.~\eqref{eq:pinch} that the contributions from the diagram in figure~\ref{fig:pipi_LO}(b) are forced into the region of phase space where the active quark and antiquark are both close to their mass shells.  The same consideration can be applied to any pair of almost parallel active partons from the nonperturbative blob either on the left or the right in figure~\ref{fig:pipi_LO}(b).  That is, at the amplitude level, pinches happen among each pair of collinear partons from either the $\pi^+$ or $\pi^-$ side, as long as their total energy (or light-cone plus or minus momentum) is much greater than their invariant mass, which means that those partons evolve well before they enter the short-time hard interaction.
Therefore, it is possible to factorize the two non-perturbative blobs associated with $\pi^+$ and $\pi^-$, respectively, from the short-distance hard scattering process.

\begin{figure}[htbp]
    \centering
    \includegraphics[scale=0.75]{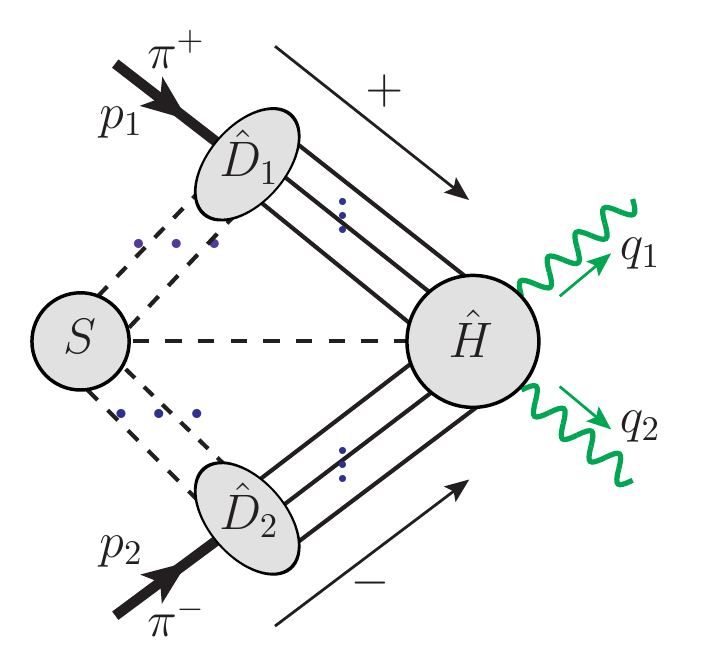}
    \caption{General reduced diagram for the scattering amplitude of exclusive annihilation process for $\pi^+\pi^-\to\gamma\gamma$. Both dashed and solid lines can represent quarks and/or gluons. }
    \label{fig:pi-pi-reduced-graph}
\end{figure}

The generalization of the above pinch analysis leads to the so-called Libby-Sterman analysis~\cite{Libby:1978qf, Libby:1978bx}, by which
all loop momenta can be categorized into three groups: hard, collinear and soft, which we do not repeat here.  Each external particle is associated with a group of collinear lines. 
With the assumption that the two observed high-$q_T$ photons are produced in the same hard scattering, the relevant reduced diagrams for the exclusive 
$\pi^+\pi^-\to\gamma\gamma$ scattering amplitude are illustrated in figure~\ref{fig:pi-pi-reduced-graph}.   At the pinch surfaces, there are two groups of collinear lines associated with the directions of colliding $\pi^+$ and $\pi^-$, respectively, shown as solid lines, and a hard part for the exclusive production of two back-to-back high-$q_T$ photons.  We can have an arbitrary number of collinear lines attaching the collinear subgraph $\hat{D}_1$ or $\hat{D}_2$ to the hard subgraph $\hat{H}$.  In addition, we can have an arbitrary number of soft lines attaching to $\hat{D}_1, \hat{D}_2$ and/or $\hat{H}$, represented by the dashed lines. 

Important contributions to the exclusive scattering amplitude come from the neighborhood of the pinch surfaces characterized by the reduced diagram in figure~\ref{fig:pi-pi-reduced-graph}, but, not all of them contribute to the leading power term in $1/q_T$ expansion.  The leading pinch surfaces, contributing to the leading power, can be identified and determined by performing a power-counting analysis for the neighborhood of the reduced diagram in figure~\ref{fig:pi-pi-reduced-graph}.  

We characterize these regions of momentum space by introducing dimensionless scaling variables, denoted as $\lambda$, which control the relative rates at which components of loop momenta vanish near the pinch surfaces.  
A leading region is the one for which a vanishing region of loop momentum space near a pinch surface produces leading power behavior in $1/q_T$ expansion.  For the loop momenta $k_i$ and $l_j$, which attach $\hat{D}_1$ and $\hat{D}_2$ to the $\hat{H}$, respectively, we choose 
\begin{equation}
    k_i \sim \left( 1,\, \lambda^2,\, \lambda \right) Q
    \quad {\rm and} \quad
    l_j \sim \left( \lambda^2,\,
    1,\,  \lambda \right)Q\, ,
\label{eq:co-scaling}
\end{equation}
with $Q\sim {\cal O}(q_T)  \gg \Lambda_{\rm QCD}$ as a characteristic hard scale and $\lambda \sim {\cal O}(\Lambda_{\rm QCD}/Q)$.  We have $k_i^2\sim {\cal O}(\lambda^2Q^2)\to 0$ and $l_j^2\sim {\cal O}(\lambda^2Q^2)\to 0$ to quantify how the loop momenta approach to the pinch surface as $\lambda\to 0$. 
We choose the momentum of the soft loops to have the following scaling behavior,
\begin{equation}
    k_s \sim \left( \lambda_s,\, \lambda_s,\, \lambda_s \right) Q\, ,
\label{eq:soft-scaling}
\end{equation}
with all components vanishing at the same rate, maintaining $k_s^2\sim {\cal O}(\lambda_s^2 
Q^2)\to 0$.
In principle, the two scaling variables, $\lambda$ and $\lambda_s$, need not be the same or related.   In our discussion of power-counting, we choose $\lambda_s \sim \lambda^2$. Considering the sample diagram in figure~\ref{fig:soft-momentum}, we have
\begin{equation}
(k_i+k_s)^2 \sim 2k_i^+ k_s^-[\sim \lambda^2Q^2]+ {\cal O}(\lambda^3)\, , 
\quad
(l_j-k_s)^2 \sim - 2l_j^- k_s^+[\sim \lambda^2Q^2] + {\cal O}(\lambda^3)\, . 
\label{eq:soft-approx}
\end{equation}
That is, for the leading power contribution, we only need to keep the components $k_s^-$ and $k_s^+$ for soft gluon momentum $k_s$ to enter the $\hat{D}_1$ and $\hat{D}_2$, respectively.  
A more comprehensive discussion including power-counting for subdivergences, as some loop lines approach the mass-shell faster than others, can be found in~\cite{Sterman:1978bi}.

\begin{figure}[htbp]
    \centering
    \includegraphics[scale=0.6]{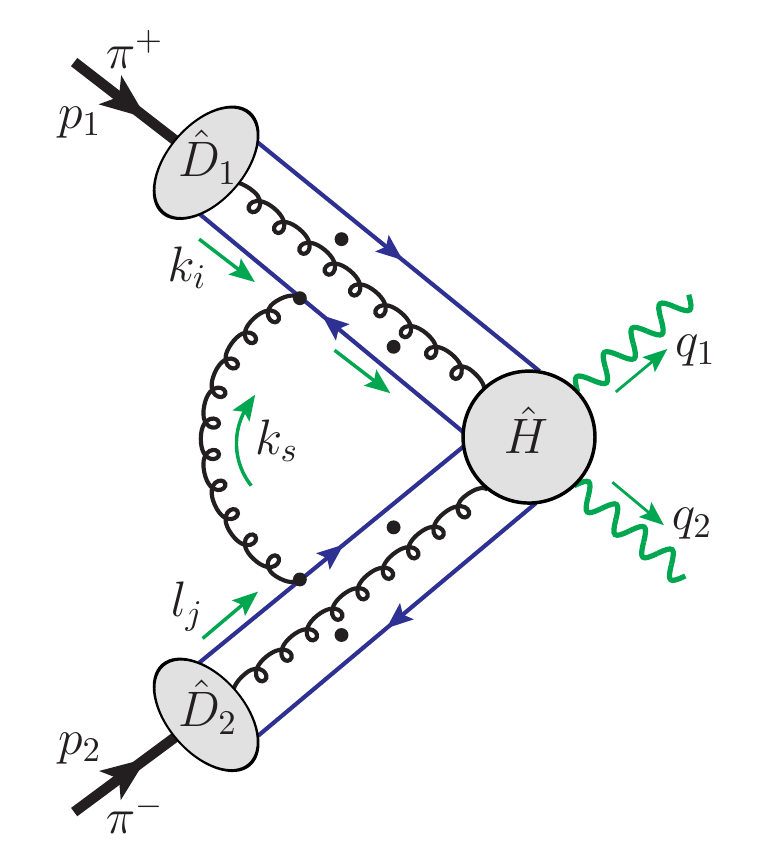}
    \caption{Sample diagram for identifying the leading soft contribution.
    }
    \label{fig:soft-momentum}
\end{figure}

Following the power-counting analysis arguments in~\cite{Collins:2011zzd, Collins:1996fb}, we obtain the scaling behavior for the reduced diagram in figure~\ref{fig:pi-pi-reduced-graph} as
\begin{align}
    {\cal M}_{\pi^+\pi^-\to\gamma\gamma}\sim
    \hat{H}\otimes \hat{D}_1\otimes \hat{D}_2\otimes  S \
    \propto\
    \lambda^{\alpha}	\, ,
\label{eq:scale1}
\end{align}
where $\otimes$ indicates both the contraction of Lorentz indices, spinor indices and convolution of loop momenta, and the power $\alpha$ is given by
\begin{align}
    \alpha =\, & 
     N_{q_{(D_1\to H)}}  + N_{q_{(D_2\to H)}} -2
     + N_{{\bf g}_{(D_1\to H)}}  + N_{{\bf g}_{(D_2\to H)}} 
     \nn\\
   &
    + N_{q_{(S\to D_1)}} + N_{q_{(S\to D_2)}}
    + 3 N_{q_{(S\to H)} }
    + N_{{\bf g}_{(S\to D_1)}} + N_{{\bf g}_{(S\to D_2)}} 
    + 2 N_{g_{(S\to H)}}	\, ,
\label{eq:pc1}
\end{align}
where $N_{q_{(A\to B)}}$, $N_{g_{(A\to B)}}$ and $N_{{\bf g}_{(A\to B)}}$ represent, respectively, the number of quarks, gluons and physically polarized gluons connecting from subgraph $A$ to $B$.  
It is clear from eqs.~\eqref{eq:scale1} and \eqref{eq:pc1} that the leading pinch surfaces (or leading regions) to the scattering amplitude are those in figure~\ref{fig:pipi LR} with the minimal power $\alpha=2$. 
Given the fact that the meson $\pi^{\pm}$ has one valence quark and antiquark, we must have a pair of quark and antiquark lines out of both $\hat{D}_1$ and $\hat{D}_2$ for this exclusive scattering process in figure~\ref{fig:pipi LR}. At $\alpha=2$, all the gluons linking $\hat{D}_1$ (and $\hat{D}_2$) to $\hat{H}$ (and $S$) are longitudinally polarized. 

\begin{figure}[htbp]
\centering
\begin{subfigure}{.35\textwidth}
  \centering
  \includegraphics[scale=0.7]{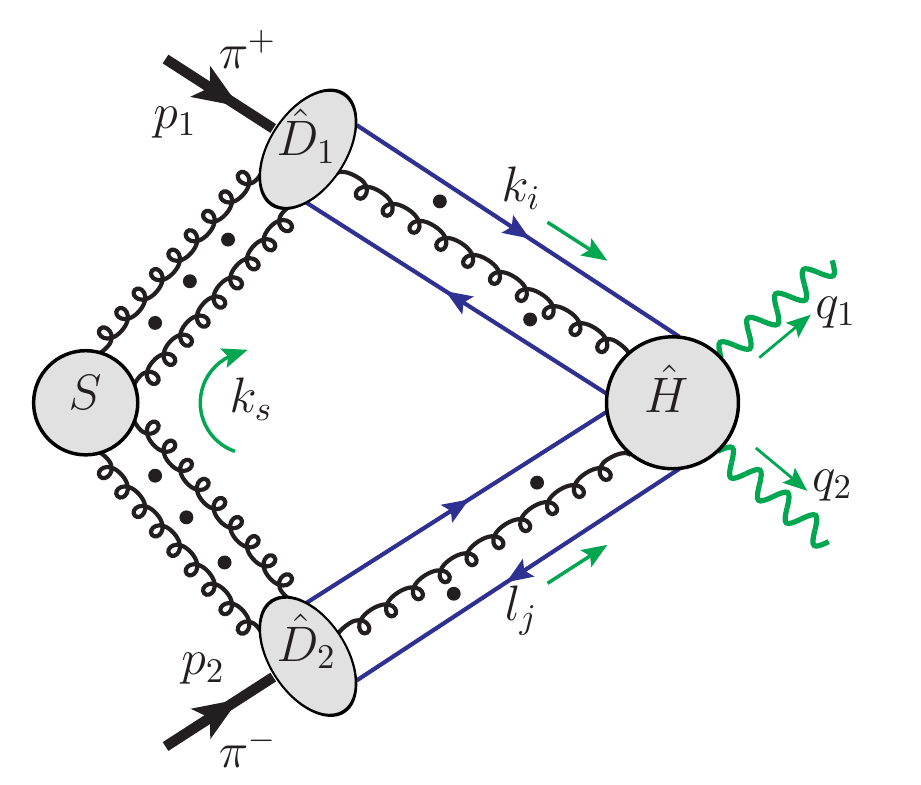} 
  \caption{}
  \label{fig:LR1}
\end{subfigure}
{\hskip 0.6in}
\begin{subfigure}{.35\textwidth}
  \centering
   \includegraphics[scale=0.7]{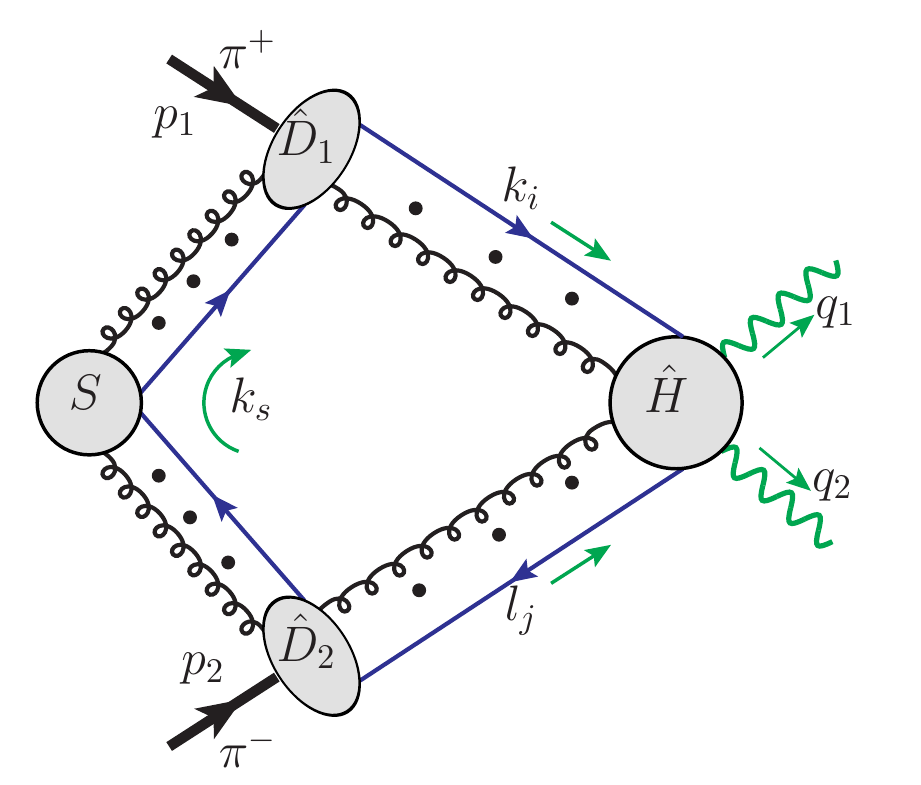} 
  \caption{}
  \label{fig:LR2}
\end{subfigure}
\caption{Two possible leading regions for the exclusive $\pi^+\pi^-$ annihilation to two photons. An arbitrary number of {\it longitudinally} polarized gluons can connect the collinear subgraph $\hat{D}_1$ or $\hat{D}_2$ to $\hat{H}$ or $S$.
    }
\label{fig:pipi LR}
\end{figure}

Although the pinch surfaces in figure~\ref{fig:LR2} are expected to provide the leading contributions from the perturbative power-counting analysis, reasonable arguments would make these contributions power suppressed.  One simple argument is to recognize  that with the quark lines from the soft factor $S$, these contributions are likely proportional to the end point of the pion wave function, where one of the two valence quarks carries almost no momentum.  Since the pion wave function is expected to vanish at this point, we could conclude that figure~\ref{fig:LR2} does not contribute at the leading power, but, might impact factorization at higher powers.
Another possible argument for the contribution from figure~\ref{fig:LR2} to be power suppressed could be achieved by studying the situation in which the soft loop momentum is scaled with $\lambda_s \sim \lambda$~\cite{Collins:1996fb}.

\subsubsection{Approximations}
\label{sec:approx}

With the leading region identified as figure~\ref{fig:LR1}, we introduce some controllable approximations to pick up the leading power contributions from the Feynman diagrams to prepare ourselves for performing the factorization of the collinear and soft gluons in next two subsections, respectively.

We first introduce two sets of auxiliary vectors to help extract leading contributions from the collinear and soft regions, respectively, 
\begin{subequations}\label{eq:aux vec}
        \begin{align}
            & 
            w_1=\left(1,0,\bm{0}_T \right),\\
            &
            w_2=\left(0,1,\bm{0}_T \right),\\
            &
            n_1=w_1-e^{-2y_1}w_2
                =\left(1,-e^{-2y_1},\bm{0}_T \right),\\
            &
            n_2=w_2-e^{2y_2}w_1
                =\left(-e^{2y_2},1,\bm{0}_T \right) ,
         \end{align}
\end{subequations}
where the non-light-like vectors $n_1$ and $n_2$ are introduced to regulate rapidity divergence with 
finite parameters $y_1$ and $y_2$ to keep them slightly off light cone.  To avoid confusion of notations, in this subsection, we do not use the $n$ and $\bar{n}$ as in eq.~\eqref{eq:n nb}.

The leading reduced diagram in figure~\ref{fig:LR1} can be formally expressed and approximated at the leading power as,
\begin{align}
    {\cal M}_{\pi^+\pi^-\to\gamma\gamma}
    & = 
    \hat{H}^{\{a\},\{b\},\{\mu\},\{\nu\}}_{i,j;\,m,n}(k_q,k_{\bar{q}},\{k\};l_q,l_{\bar{q}},\{l\};q_1,q_2)
    \nn \\
    &{\hskip 0.3in} \times 
    \hat{D}_{1\, {j,i,\{\mu\},\{\rho\}}}^{\,\{a\},\{c\}}(k_q, k_{\bar q}, \{k\}; \{k_{s}\})\, 
    \hat{D}_{2\, {n,m,\{\nu\},\{\sigma\}}}^{\,\{b\},\{d\}}(l_q, l_{\bar q}, \{l\}; \{l_{s}\}) 
    \nn \\
    &{\hskip 0.3in} \times 
    S^{\{c\}, \{d\}, \{ \rho \}, \{ \sigma \}}(\{ k_{s} \}, \{ l_{s} \})  
    \label{eq:rc-full}  \\
    & \approx
    \left[ \hat{H}^{\{a\},\{b\},\{\mu\},\{\nu\}}_{i,j; \, m,n}(\hat{k}_q, \hat{k}_{\bar{q}}, \{ \hat{k} \}; \hat{l}_q, \hat{l}_{\bar{q}}, \{ \hat{l} \}; q_1,q_2) \,
    \cdot \{ \hat{k}_{\mu} \} \cdot \{ \hat{l}_{\nu} \} \right]
    \nn \\
    & {\hskip 0.3in}  \times
    \bigg\{ \frac{w_{2}^{\bar{\mu}}}{k\cdot w_2 + i\epsilon} \bigg\} \cdot 
    \left[ \hat{D}^{\,\{a\},\{c\}}_{1\, {j,i,\{\bar{\mu}\},\{\bar{\rho}\}}}(k_q, k_{\bar q}, \{ k \}; \{ \hat{k}_{s} \}) \cdot \{\hat{k}_{s}^{\,\bar{\rho}}\}\right]
    \nn \\ 
    & {\hskip 0.3in} \times
   \bigg\{ \frac{w_{1}^{\bar{\nu}}}{l \cdot w_1 + i\epsilon}\bigg\} \cdot
    \left[\hat{D}^{\,\{b\},\{d\}}_{2\, {n,m,\{\bar{\nu}\},\{\bar{\sigma}\}} }(l_q, l_{\bar q}, \{ l \}; \{\hat{l}_{s}\}) \cdot \{\hat{l}_{s}^{\,\bar{\sigma}}\} \right]
    \nn \\
    & {\hskip 0.3in} \times 
    \left[\left\{\frac{n_{1\rho}}{k_{s}\cdot n_1 + i\epsilon} \right\} \cdot
           \left\{\frac{n_{2\sigma}}{l_{s}\cdot n_2 + i\epsilon} \right\} \cdot
    S^{\{c\},\{d\},\{\rho\},\{\sigma\}}(\{k_{s}\},\{l_{s}\}) 
    \right]	\, ,
    \label{eq:rc-app}  
\end{align}
where $k_q$ and $k_{\bar{q}}$ ($l_q$ and $l_{\bar{q}}$) are the active quark and antiquark momenta from the collinear part $\hat{D}_1$ ($\hat{D}_2$) to the hard part $\hat{H}$, respectively, $\{k\}=k_1,k_2,...$ ($\{l\}=l_1,l_2,...$) are momenta of longitudinally polarized gluons flowing from $\hat{D}_1$ ($\hat{D}_2$) into $\hat{H}$, $\{k_{s}\}=k_{s_1},k_{s_2},... $ ($\{l_{s}\}= l_{s_1},l_{s_2},... $) are soft gluon momenta flowing from $S$ into $\hat{D}_1$ ($\hat{D}_2$);
$\{\mu\}=\mu_1,\mu_2,...$ ($\{\nu\}=\nu_1,\nu_2,...$) are Lorentz indices for gluons attached from $\hat{D}_1$ ($\hat{D}_2$) to $\hat{H}$, and $\{\rho\}=\rho_1,\rho_2,...$ ($\{\sigma\}=\sigma_1,\sigma_2,...$)  are Lorentz indices for gluons attached from $S$ to $\hat{D}_1$ ($\hat{D}_2$); 
and $i,j$ ($m,n$) are color indices for active quark and antiquark, $\{a\}=a_1,a_2,...$ ($\{b\}=b_1,b_2,...$) are color indices for gluons linking $\hat{D}_1$ ($\hat{D}_2$) and $\hat{H}$, and $\{c\}=c_1,c_2,...$ ($\{d\}=d_1,d_2,...$) are color indices for gluons linking $S$ to $\hat{D}_1$ ($\hat{D}_2$).  
In eq.~\eqref{eq:rc-app}, we used some simplified notations,
\begin{align}
\{\hat{k}_\mu\} 
& \equiv \prod_{i=1,2,...} \hat{k}_{i\mu_i}\, ,
\quad
\{\hat{l}_\nu\} 
\equiv \prod_{i=1,2,...} \hat{l}_{i\nu_i}\, ,
\nn \\
\{\hat{k}_{s}^{\bar{\rho}}\} 
& \equiv \prod_{i=1,2,...} \hat{k}_{s_i}^{\bar{\rho}_i}\, ,
\quad
\{\hat{l}_{s}^{\bar{\sigma}}\} 
\equiv \prod_{i=1,2,...} \hat{l}_{s_i}^{\bar{\sigma}_i} \, ;
\label{eq:prod_mom}
\end{align}
and similarly,
\begin{align}
 \left\{\frac{w_{2}^{\bar{\mu}}}{k\cdot w_2 + i\epsilon}\right\}
 & = \prod_{i=1,2,...} \frac{w_{2}^{\bar{\mu}_i}}{k_i\cdot w_2 + i\epsilon}\, ,
\quad
 \left\{\frac{w_{1}^{\bar{\nu}}}{l\cdot w_1+ i\epsilon }\right\}
  = \prod_{i=1,2,...} \frac{w_{1}^{\bar{\nu}_i}}{l_i\cdot w_1 + i\epsilon}\, ,
\nn \\
 \left\{\frac{n_{1\rho}}{k_s\cdot n_1 + i\varepsilon}\right\}
 & = \prod_{i=1,2,...} \frac{n_{1\rho_i}}{k_{s_i}\cdot n_1 + i\varepsilon}\, ,
 \quad
  \left\{\frac{n_{2\sigma}}{l_s\cdot n_2 + i\varepsilon}\right\}
  = \prod_{i=1,2,...} \frac{n_{2\sigma_i}}{l_{s_i}\cdot n_2 + i\varepsilon}\, .
    \label{eq:eikonal}  
\end{align}

In deriving eq.~\eqref{eq:rc-app}, we made the following approximations for all parton momenta to pick up leading power contribution,
\begin{subequations}\label{eq:coll-m}
        \begin{align}
            k_{i}\mapsto 
            \hat{k}_{i} 
            &= w_1 \frac{k_{i}\cdot w_2}{w_1\cdot w_2} 
             = \left(k_{i}^+, 0, \bm{0}_T \right),
             &&i = q, \bar{q}, 1, 2, \cdots,
        \label{eq:k1H} \\
            l_{j}\mapsto 
            \hat{l}_{j} 
            &= w_2 \frac{l_{j}\cdot w_1}{w_1\cdot w_2}   
             = \left(0, l_{j}^-, \bm{0}_T \right),
             &&j = q, \bar{q}, 1, 2, \cdots,
        \label{eq:k2H} \\
            k_{s_i}\mapsto 
            \hat{k}_{s_i} 
            &= w_2 \frac{k_{s_i}\cdot n_1}{w_2\cdot n_1}   
             = \left(0, k_{s_i}^- - e^{-2y_1} k_{s_i}^+, \bm{0}_T \right),
             &&i = 1, 2, \cdots,
        \label{eq:ks}\\
            l_{s_j}\mapsto 
            \hat{l}_{s_j} 
            &= w_1 \frac{l_{s_j}\cdot n_2}{w_1\cdot n_2}  
             = \left(l_{s_j}^+ - e^{2y_2} l_{s_j}^-, 0, \bm{0}_T \right),
             &&j = 1, 2, \cdots.
        \label{eq:ks2}
        \end{align}
\end{subequations}
In eqs.~\eqref{eq:rc-full} and \eqref{eq:rc-app}, corresponding convolution of loop momenta (or momentum components) are suppressed.  In deriving the first three lines of eq.~\eqref{eq:rc-app}, we used
\begin{subequations}\label{eq:HC}
      \begin{align}
          &  \hat{H}^{\{\mu\},\{\nu\}}(k_q,k_{\bar{q}},\{k\};l_q,l_{\bar{q}},\{l\};q_1,q_2)\, 
          \hat{D}_{1{\{\mu\},\{\rho\}}}(k_q, k_{\bar q}, \{k\}; \{k_{s}\})
           \label{eq:HC1} \\
          & \quad\quad 
            \mapsto
            \hat{H}^{\{\mu\},\{\nu\}}(\hat{k}_q, \hat{k}_{\bar{q}}, \{\hat{k}\}; \hat{l}_q, \hat{l}_{\bar{q}}, \{ \hat{l} \}; q_1,q_2)
            \cdot \bigg\{ \frac{ \hat{k}_{\mu} w_{2}^{\bar{\mu}} }{ k \cdot w_2 + i\epsilon} \bigg\} \cdot
            \hat{D}_{1 { \{ \bar{\mu} \}, \{ \rho \} } }(k_q, k_{\bar q}, \{ k \}; \{ k_{s} \}) \, ,
         	\nn\\
          &  \hat{H}^{\{\mu\},\{\nu\}}(k_q, k_{\bar{q}}, \{ k \}; l_q, l_{\bar{q}}, \{l\}; q_1,q_2) \,
          \hat{D}_{2 { \{ \nu \}, \{ \sigma \} } }(l_q, l_{\bar q}, \{ l \}; \{ l_{s} \})
           \label{eq:HC2} \\
           & \quad\quad \mapsto
            \hat{H}^{\{ \mu \}, \{ \nu \}}(\hat{k}_q, \hat{k}_{\bar{q}}, \{ \hat{k} \}; \hat{l}_q, \hat{l}_{\bar{q}}, \{ \hat{l} \}; q_1, q_2)
            \cdot \bigg\{ \frac{ \hat{l}_{\nu} w_{1}^{ \bar{ \nu } } }{l \cdot w_1 + i \epsilon} \bigg\} \cdot
            \hat{D}_{2 {\{ \bar{\nu} \}, \{ \sigma \} }}(l_q, l_{\bar q}, \{ l \}; \{ l_{s} \})
            \nn
     \end{align}
\end{subequations}
to pick up gluons' Lorentz components that give the leading power contribution in the Feynman gauge, where we suppressed the color indices.  In eq.~\eqref{eq:HC}, the $i\varepsilon$ prescription is chosen such that the poles of $k_i\cdot w_2=k_i^+$ and $l_j\cdot w_1=l_j^-$ introduced by the inserted factors do not affect the deformation of soft Glauber gluon momentum discussed later. Even though we are considering the collinear gluons now, the same momenta can also reach soft region and especially the Glauber region, which are treated coherently, and the approximators must be applied on the whole momentum integration regions in order for the use of subtraction formalism for the overlapping regions~\cite{Collins:2011zzd}.
In deriving the last line of eq.~\eqref{eq:rc-app}, we used
\begin{subequations}\label{eq:SC}
\begin{align}
          &  \hat{D}_{1{\{\mu\},\{\rho\}}}(k_q, k_{\bar q}, \{ k \}; \{ k_{s} \}) \,  S^{\{\rho\},\{\sigma\}} (\{k_{s}\},\{l_{s}\}) 
          \nn \\
          & \quad\quad \mapsto
            \hat{D}_{1{\{\mu\}, \{\rho}\}}(k_q, k_{\bar q}, \{ k \}; \{\hat{k}_s\})  \cdot
            \bigg\{\frac{\hat{k}_{s}^{\rho}\, n_{1\bar{\rho}}}{k_{s}\cdot n_1 + i\varepsilon}\bigg\} \cdot
            S^{\{\bar{\rho}\},\{\sigma\}}(\{k_{s}\},\{l_{s}\})\, ,
        \label{eq:SC1}\\
          &  \hat{D}_{2{\{\nu\},\{\sigma\}}}(l_q, l_{\bar q}, \{ l \}; \{l_{s}\}) \, S^{\{\rho\},\{\sigma\}} (\{k_{s}\}, \{l_{s}\})
          \nn \\
            &  \quad\quad \mapsto
            \hat{D}_{2{\{\nu\},\{\sigma\}}}(l_q, l_{\bar q}, \{ l \}; \{\hat{l}_{s}\}) \cdot
            \bigg\{\frac{\hat{l}_{s}^{\sigma}\, n_{2\bar{\sigma}}}{l_{s}\cdot n_2 + i\varepsilon}\bigg\} \cdot
            S^{\{\rho\},\{\bar{\sigma}\}}(\{k_{s}\},\{l_{s}\}) \, ,
        \label{eq:SC2}
\end{align}  
\end{subequations}      
where the color indices are again suppressed and the sign of $i\varepsilon$ as well as the space-like $n_{1,2}$ vectors in eq.~\eqref{eq:SC}
are chosen to be compatible with the contour deformation of ``$+$'' and ``$-$'' components of soft momenta when they are in the Glauber region as discussed later. With the large $\{k^+\}$ in $\hat{D}_1$ and $\{l^-\}$ in $\hat{D}_2$, we only need to keep $\{k_s^-\}$ in $\hat{D}_1$ and $\{l_s^+\}$ in $\hat{D}_2$, respectively, in eq.~\eqref{eq:rc-app} for ensuring the leading power contributions, as indicated in eq.~\eqref{eq:soft-approx}\footnote{In the actual treatment, we also use the space-like vectors $n_1$ and $n_2$ to introduce some small components $\{k_s^+\}$ ($\{l_s^-\}$) in $\hat{D}_1$($\hat{D}_2$) to regulate rapidity divergences, which does not affect the leading-power accuracy.}.  
This is justified for the canonical scaling in eq.~\eqref{eq:soft-scaling}, but may not be valid for the soft momenta $k_s$ in the Glauber region, where we have soft gluons with
\begin{equation}
    k_s^{\rm Glauber}\sim 
    \left(\lambda^2,\,\lambda^2,\,\lambda \right) Q
\label{eq:Glauber-scaling}
\end{equation}
connecting $\hat{D}_1$ and $\hat{D}_2$. See figure~\ref{fig:soft-momentum} as an example, where the propagator $(k_i+k_s^{\rm Glauber})^2$ [or $(l_j-k_s^{\rm Glauber})^2$] can get additional ${\cal O}(\lambda^2)$ leading contribution from ${\bm k}_{iT}\cdot {\bm k}_{sT}^{\rm Glauber}$ [or ${\bm l}_{jT}\cdot {\bm k}_{sT}^{\rm Glauber}$] and $({\bm k}_{sT}^{\rm Glauber})^2$ terms. 
As part of the soft region that also gives leading-power contribution, the Glauber region endangers factorization since it forbids the approximations made in eq.~\eqref{eq:rc-app} or \eqref{eq:SC} which are key to the use of Ward identities (to be discussed in the next two subsections) to factorize soft gluons out of the collinear subgraphs $\hat{D}_1$ and $\hat{D}_2$.

Fortunately, in the Glauber region, we can approximate the propagator of the soft gluon of momentum $k_s$ as $1/(k_s^2+i\varepsilon) \mapsto 1/(- {\bm k}_{sT}^2+i\varepsilon)$ to be independent of $k_s^{+}$ and $k_s^{-}$. 
Then with neglect of $k_s^+$ in $\hat{D}_1$ and $k_s^-$ in $\hat{D}_2$, the only poles of $k_s^+$ ($k_s^-$) come from the propagators in $\hat{D}_2$ ($\hat{D}_1$), which all lie on one side of integration contour in the complex plane because all the collinear parton lines in $\hat{D}_2$ ($\hat{D}_1$) move {\it into} the hard part with positive minus (plus) momenta, as a special feature of $\pi^+\pi^-\to \gamma\gamma$ annihilation process.
The integrations of $k_s^{+}$ and $k_s^{-}$ are thus not pinched in the Glauber region, so that we can get out of it by deforming the contours of integration over $k_s^{+}$, $k_s^{-}$.
Specifically, for a soft gluon of momentum $k_s$ in the Glauber region to enter $\hat{D}_1$, we deform the $k_{s}^-$ contour as $k_{s}^- \mapsto k_{s}^- + i \mathcal{O}(\lambda Q)$, and for a soft gluon of momentum $l_s$ in the Glauber region to enter $\hat{D}_2$, we deform the $l_{s}^+$ contour as $l_{s}^+ \mapsto l_{s}^+ + i \mathcal{O}(\lambda Q)$.
This deformation keeps all the components, $k_s^{+}$, $k_s^{-}$ and ${\bm k}_{sT}$ (or $l_s^{+}$, $l_s^{-}$ and ${\bm l}_{sT}$), effectively in the same order $\sim {\cal O}(\lambda Q)$, allowing us to keep only $\{ k_s^- \}$ in $\hat{D}_1$, and $\{ l_s^+ \}$ in $\hat{D}_2$.
Unfortunately, for the meson-baryon case, $\pi N\to \gamma\gamma N'$ to be discussed in the next section, the soft gluon momentum component $k_s^-$ can be trapped in the Glauber region if $N$ is moving in the ``$+$" direction, and additional discussion is needed for treating the Glauber region.

For extracting the leading power contribution from the spinor of active quark entering $\hat{H}$ from $\hat{D}_1$ or leaving $\hat{H}$ into $\hat{D}_2$, we insert the following spinor projector~\cite{Collins:2011zzd},
\begin{equation}
            \mathcal{P}_A = \frac{1}{2}\gamma^-\gamma^+ \, .
\label{eq:spinorPA}
\end{equation}
For a quark line entering $\hat{H}$ from $\hat{D}_2$ or leaving $\hat{H}$ into $\hat{D}_1$, we have corresponding spinor projector,
\begin{equation}
            \mathcal{P}_B = \frac{1}{2}\gamma^+\gamma^- \, .
\label{eq:spinorPB}
\end{equation}
These projectors effectively project out the largest components of the spinor indices of active quarks and antiquarks, which give the leading power contributions to the exclusive scattering amplitude in the $1/q_T$ expansion.

Among all approximations listed above, the biggest error comes from neglecting the transverse components of active parton momenta entering into $H$, which leads to an error of order $\LQCD/q_T$.

The approximate expression in eq.~\eqref{eq:rc-app}, with the spin projections in eqs.~\eqref{eq:spinorPA} and \eqref{eq:spinorPB} applied to active quark and antiquark lines, represents the leading power contribution to the exclusive $\pi\pi\to\gamma\gamma$ scattering amplitude in $1/q_T$ expansion.  In next two subsections, we demonstrate that this leading power contribution can be factorized into a convolution of two universal pion distribution amplitudes with a perturbatively calculable short-distance hard part that produces the two observed high transverse momentum photons.

\begin{figure}[htbp]
\centering
\begin{align*}
\adjincludegraphics[valign=c, scale=0.45]{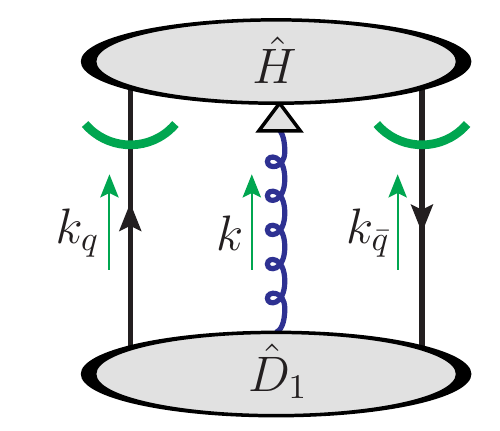}
=
- \adjincludegraphics[valign=c, scale=0.45]{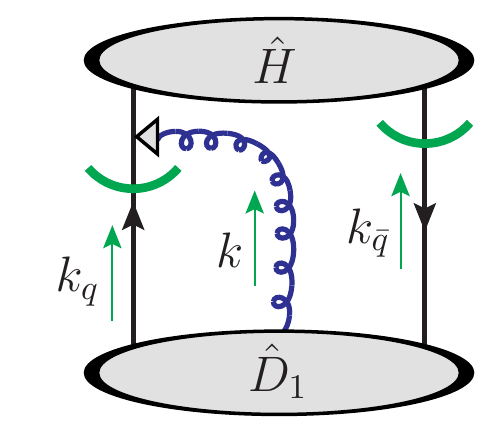}
- \adjincludegraphics[valign=c, scale=0.45]{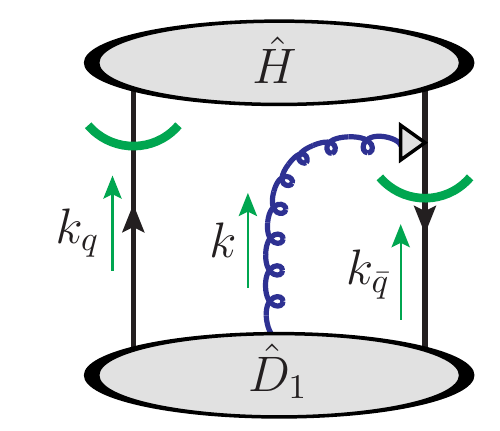}
= \adjincludegraphics[valign=c, scale=0.45]{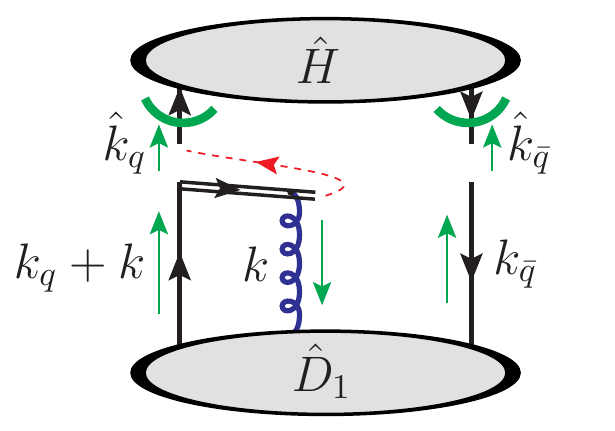} +
\adjincludegraphics[valign=c, scale=0.45]{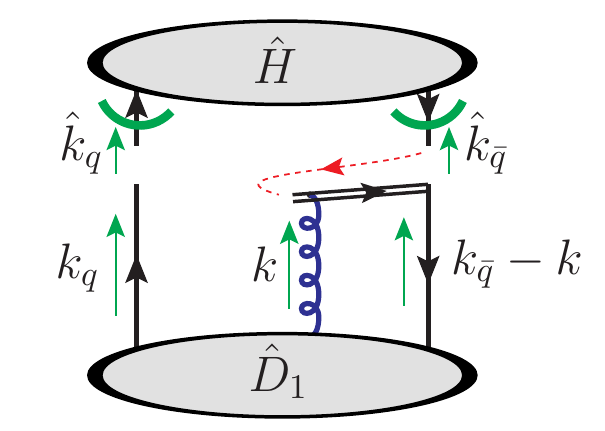} 
\end{align*}
\caption{Graphic representation of the two steps to detach a longitudinally polarized collinear gluon from the $\hat{D}_1$ to the $\hat{H}$, and reconnect it to corresponding gauge links of the $\hat{D}_1$. The triangles at the end of gluon lines mean the use of Ward identity, and the red thin dashed lines represent the color flows.
}
\label{fig:ward_identity}
\end{figure}

\begin{figure}[htbp]
\centering
\includegraphics[scale=0.65]{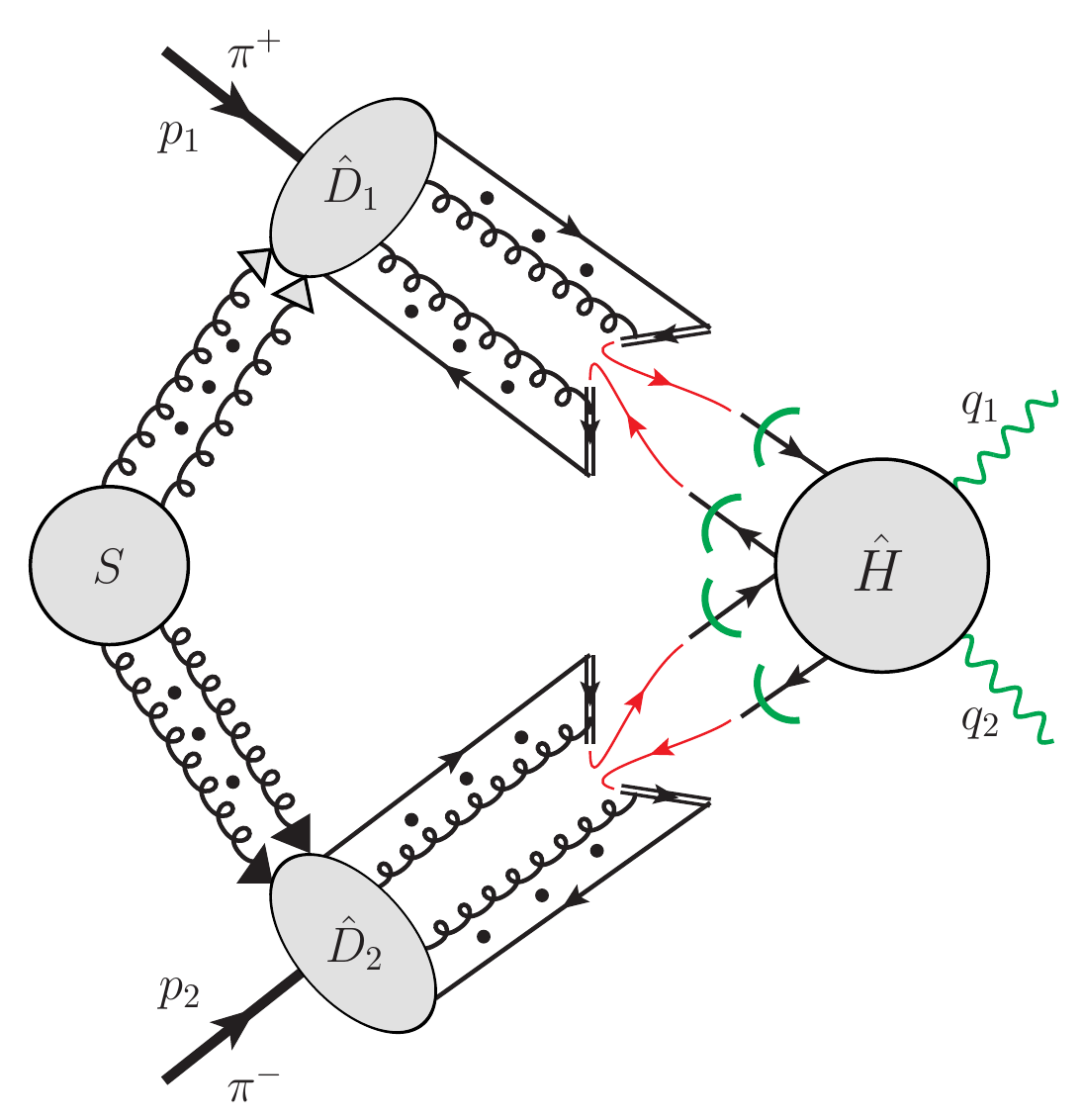}
\caption{Factorization of longitudinally polarized collinear gluons to the $H$ into the Wilson lines in the fundamental representation of the SU(3) color. The arrows on the Wilson line indicate the color flow.
See text for the meaning of other symbols.
}
\label{fig:ward-col}
\end{figure}

\subsubsection{Ward identity and factorization of collinear gluons}

With the leading power contribution to the scattering amplitude given in eq.~\eqref{eq:rc-app}, we can use Ward identity to factorize all collinear and longitudinally polarized gluons from the hard part $\hat{H}$.  

From the first line in eq.~\eqref{eq:rc-app}, all Lorentz indices of attached gluon lines are effectively contracted by corresponding gluon momenta, $\hat{H}^{\{\mu\},\{\nu\}}(\hat{k}_q,\hat{k}_{\bar{q}},\{\hat{k}\};\hat{l}_q, \hat{l}_{\bar{q}}, \{\hat{l}\};q_1,q_2) \{\hat{k}_{\mu}\} \{\hat{l}_{\nu}\}$, which enables the use of Ward identity.  We will first consider the situation with one longitudinally polarized collinear gluon of momentum $k$ from $\hat{D}_1$ to $\hat{H}$, as shown in figure~\ref{fig:ward_identity}.  As an identity, summing over all the possible attachments of a longitudinally polarized gluon to the $\hat{H}$ is equivalent to attaching it to the active quark and antiquark lines of the $\hat{H}$ with a minus sign, as illustrated by the first equal sign in the graphic representation in figure~\ref{fig:ward_identity}.  With the spinor projectors in eqs.~\eqref{eq:spinorPA} and \eqref{eq:spinorPB} for the active quark and antiquark lines linking the $\hat{H}$ and $\hat{D}_1$ and the use of the graphic Ward identity, we can move the attachment of a longitudinally polarized gluon to an active quark (or antiquark) line of the $\hat{H}$ to a gauge link of the same active quark (or antiquark) of the $\hat{D}_1$ along the direction $w_2$, as illustrated by the second equal sign in the graphic representation in figure~\ref{fig:ward_identity}, multiplied by the same $\hat{H}$ without the gluon attachment while its active quark (or antiquark) momentum is adjusted, 
\begin{subequations}\label{eq:Hcollinear}
\begin{align}
& \hat{H}^{a,\{b\},\mu,\{\nu\}}_{i,j;\,m,n}(\hat{k}_q,\hat{k}_{\bar{q}},\hat{k};\hat{l}_q,\hat{l}_{\bar{q}},\{\hat{l}\};q_1,q_2)\, \hat{k}_{\mu}
\left[\frac{w_{2}^{\bar{\mu}}}{k\cdot w_2 + i\epsilon }\right]
\left[\mathcal{P}_A \hat{D}_{1\, {j,i,\bar{\mu},\{\rho\}}}^{\,a,\{c\}}(k_q,k_{\bar{q}},k;\{k_{s}\}) \mathcal{P}_B\right]
\nn \\
& {\hskip 0.4in} = 
\hat{H}^{\{b\},\{\nu\}}_{i,j';\,m,n}(\hat{k}_q+\hat{k},\hat{k}_{\bar{q}};\hat{l}_q,\hat{l}_{\bar{q}},\{\hat{l}\};q_1,q_2)
\nn \\
&  {\hskip 0.7in} \times
\left(\frac{-i}{k\cdot w_2 + i\epsilon} \right)
\left(-ig_s(t^a)_{j'j}w_2^{\bar{\mu}}\right)
\left[\mathcal{P}_A \hat{D}_{1\, {j,i,\bar{\mu},\{\rho\}}}^{\,a,\{c\}}(k_q,k_{\bar{q}},k;\{k_{s}\}) \mathcal{P}_B\right]
\nn\\
&  {\hskip 0.6in} +
\hat{H}^{\{b\},\{\nu\}}_{i',j;\,m,n}(\hat{k}_q,\hat{k}_{\bar{q}}+\hat{k};\hat{l}_q,\hat{l}_{\bar{q}},\{\hat{l}\};q_1,q_2)
\nn \\
&  {\hskip 0.7in} \times
\left(\frac{i}{k\cdot w_2 + i\epsilon} \right)
\left(-ig_s(t^a)_{ii'}w_2^{\bar{\mu}}\right)
\left[\mathcal{P}_A \hat{D}_{1\, {j,i,\bar{\mu},\{\rho\}}}^{\,a,\{c\}}(k_q,k_{\bar{q}},k;\{k_{s}\}) \mathcal{P}_B\right]
\label{eq:HcollinearD1}\\
& {\hskip 0.4in} = 
\hat{H}^{\{b\},\{\nu\}}_{i,j;\,m,n}(\hat{k}_q,\hat{k}_{\bar{q}};\hat{l}_q,\hat{l}_{\bar{q}},\{\hat{l}\};q_1,q_2)
\nn \\
&  {\hskip 0.5in} \times \bigg(
\frac{-i}{-k\cdot w_2 + i\epsilon}
\left(-ig_s(t^a)_{jj'}w_2^{\bar{\mu}}\right)
\left[\mathcal{P}_A \hat{D}_{1\, {j',i,\bar{\mu},\{\rho\}}}^{\,a,\{c\}}(k_q + k, k_{\bar{q}},k;\{k_{s}\}) \mathcal{P}_B\right]
\nn\\
&  {\hskip 0.7in} +
\frac{i}{k\cdot w_2 + i\epsilon}
\left(-ig_s(t^a)_{i'i}w_2^{\bar{\mu}}\right)
\left[\mathcal{P}_A \hat{D}_{1\, {j,i',\bar{\mu},\{\rho\}}}^{\,a,\{c\}}(k_q,k_{\bar{q}}-k,k;\{k_{s}\}) \mathcal{P}_B\right] 
\bigg)\, ,
\label{eq:HcollinearD2}
\end{align}
\end{subequations}
where $g_s$ is the strong coupling constant and $t^a$ is the generator for the fundamental representation of SU(3) color. In order to formally factor out the $\hat{H}$ without attachment from $\hat{D}_1$ in eq.~\eqref{eq:HcollinearD2}, we shifted the active quark and antiquark momenta in $\hat{D}_1$ accordingly. 
And in the second line of eq.~\eqref{eq:HcollinearD2}, we also reversed the gluon momentum $k$ such that it flows along the same direction of the gauge link. Eq.~\eqref{eq:Hcollinear} and its graphic representation in figure~\ref{fig:ward_identity} clearly indicate that the attachment of a longitudinally polarized gluon of momentum $\hat{k}$ from $\hat{D}_1$ to $\hat{H}$ can be effectively detached from $\hat{H}$ and connected to the gauge links of active quark of momentum $k_q$ and antiquark of momentum $k_{\bar{q}}$ along the direction of $w_2$ with its momentum effectively flowing through the active quark (or antiquark) line into the $\hat{H}$. Similarly, by applying the Ward identity to the attachment of a longitudinally polarized gluon of momentum $\hat{l}$ from $\hat{D}_2$ to $\hat{H}$, we can effectively detach this gluon from the $\hat{H}$ with its momentum effectively flowing through the active quark (or antiquark) line from $\hat{D}_2$, as sketched in figure~\ref{fig:ward-col}, where the hooks on the external quark lines of $\hat{H}$ indicate the amputation with the spinor projectors in eqs.~\eqref{eq:spinorPA} and \eqref{eq:spinorPB}.  

The attachment of multiple collinear and longitudinally polarized gluons of momenta $\hat{k}_i^{\mu_i}$ with $i=1,2, ...,I$ from $\hat{D}_1$ ($\hat{l}_j^{\nu_j}$ with $j=1,2, ...,J$ from $\hat{D}_2$) to $\hat{H}$ can be detached in the same way, by summing over all possible attachments and using the Ward identity multiple times. 
The corresponding factor from detaching such gluons from $\hat{D}_1$ ($\hat{D}_2$) to $\hat{H}$ can be combined with the factor in front of $\hat{D}_1$ ($\hat{D}_2$) in the second (third) line of eq.~\eqref{eq:rc-app} to make up the eikonal vertices and eikonal propagators that match the expansion of a product of two gauge links in the fundamental representation along the direction of $w_2$ ($w_1$), one from the active quark of momentum $k_q$ ($l_q$) and the other to the active antiquark of momentum $k_{\bar{q}}$ ($l_{\bar{q}}$), to the order with a total of $I$ ($J$) gluons~\cite{Collins:1985ue}.  
And then by summing over all possible numbers of attachments of collinear and longitudinally polarized gluons with $I, J=1,2,...,\infty$, we are able to detach all of them from the $\hat{H}$ and attach them to two gauge links along the direction of $w_2$ ($w_1$), or the Wilson lines in momentum space, one from the active quark of momentum $k_q$ ($l_q$) and the other to the active antiquark of momentum $k_{\bar{q}}$ ($l_{\bar{q}}$).
That is, we are able to factorize all attachments of collinear and longitudinally polarized gluons from $\hat{D}_1$ (or $\hat{D}_2$) to the $\hat{H}$ into the well-defined gauge links that become a part of the $\hat{D}_1$ (or $\hat{D}_2$), as shown in figure~\ref{fig:ward-col}, where the red thin lines indicate the color flow between collinear subgraphs, $\hat{D}_1$ and $\hat{D}_2$, and the hard subgraph, $\hat{H}$. 

As pointed out above, the Wilson lines in figure~\ref{fig:ward-col} are in the fundamental $\bm{3}$ or $\bar{ \bm{3} }$ color representation, indicated by the arrows on the Wilson line which denote the color flow. 
With the signs of $i\epsilon$ necessitated by the deformation out of the Glauber region, we have the Wilson line as 
\begin{align}
\Phi_{ij}(\infty, x;-n) 
= {\cal P}\exp\left\{ig \int_0^{\infty} \dd{\lambda} n^{\mu}
	A^a_{\mu} \left(x - \lambda n\right) (t^a_{ij}) \right\},
\label{eq:gaugelink}
\end{align}
where $t^a$ is again the generator for the fundamental representation of SU(3) color and will be suppressed in the rest of this paper, and the indices $i,j$ are color indices in fundamental representation. This Wilson line in eq.~\eqref{eq:gaugelink} points from $x$ to $-\infty$ along the direction $-n^{\mu} (n^0> 0)$, and is a past-pointing Wilson line, like those in parton distribution functions from factorized Drell-Yan process, which is consistent with the picture that all the collinear parton lines from colliding pions come from the past to the hard collision, $\hat{H}$, to produce a pair of high transverse momentum photons.

\begin{figure}[htbp]
    \centering
     \includegraphics[scale=0.65]{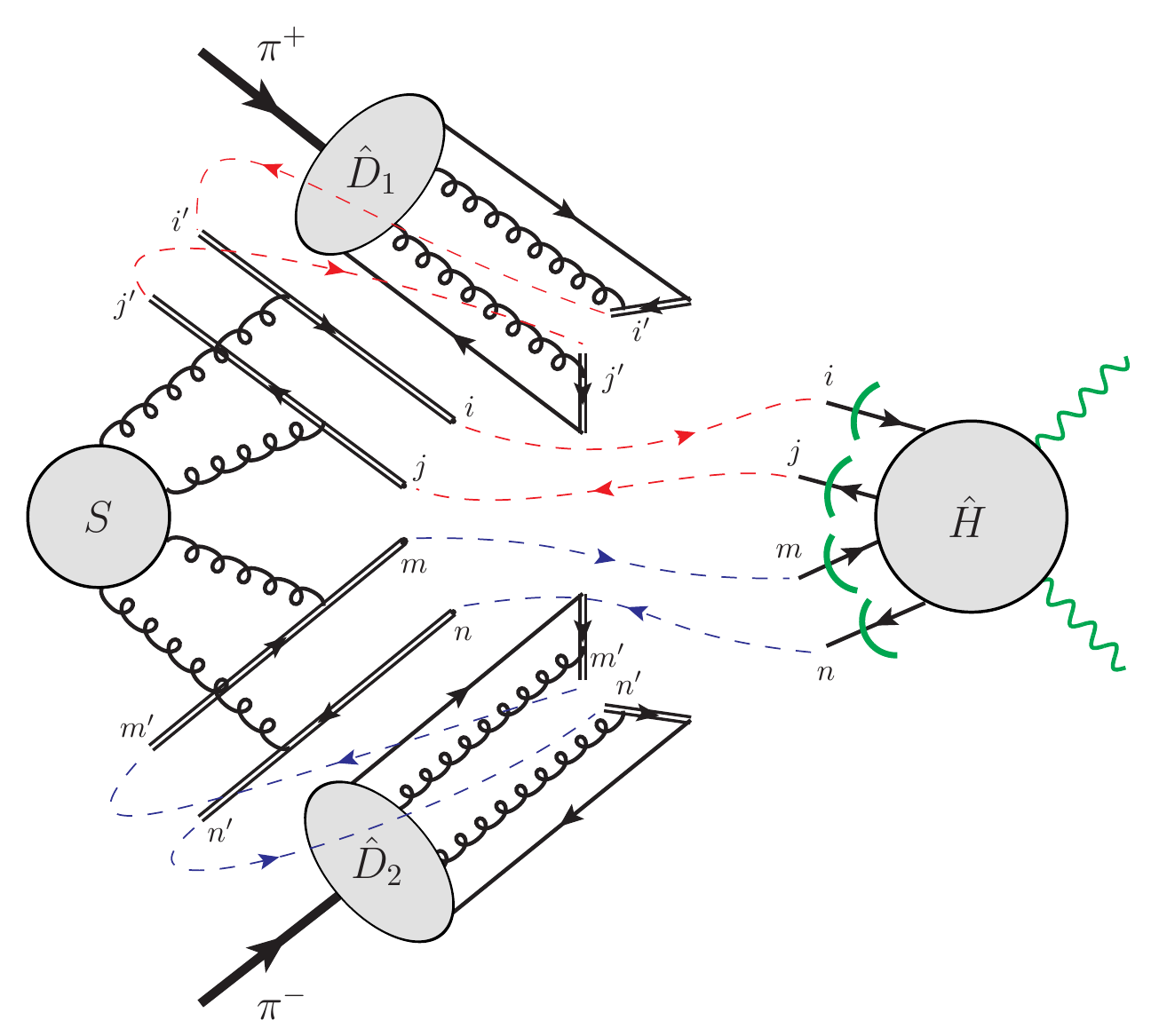}
    \caption{The result of using Ward identity for attachments of soft gluons to $\hat{D}_1$ and $\hat{D}_2$. The Wilson lines are in fundamental $\bm{3}$ or $\bar{ \bm{3} }$ color representation. The Wilson lines on $\hat{D}_1$ side come from the past along $n_1$, and those on $\hat{D}_2$ side come from the past along $n_2$. The red and blue dashed lines indicate the color flow among collinear, soft and hard subgraphs. $i,j,m,n,i',j',m',n'$ are the color indices.}
    \label{fig:ward-soft}
\end{figure}

\subsubsection{Ward identity and cancellation of the soft factor}
\label{sec:soft_cancel}

It was demonstrated in the last subsection that the collinear factors $\hat{D}_1$ and $\hat{D}_2$ can be detached from the hard part $\hat{H}$.  But, they are still connected by soft gluons from the soft factor $S$, which can communicate the colors between them. In this subsection, we use the approximations in eq.~\eqref{eq:SC}, which lead to the second and third lines of eq.~\eqref{eq:rc-app}, and the Ward identity to decouple the soft gluon attachments between $\hat{D}_1$ and $\hat{D}_2$ to achieve the factorization that we hope to derive.

As discussed in the subsection~\ref{sec:approx}, we only need to consider the ``$-$'' component of the soft momentum $k_s$ flowing from $S$ to $\hat{D}_1$, and ``$+$'' component of the soft momentum $l_s$ flowing from $S$ to $\hat{D}_2$ for the leading power contribution to the amplitude.  Similar to the collinear gluons, we can apply the Ward identity to 
$ \hat{D}_{1\,{\{\bar{\mu}\},\{\bar{\rho}\}}}(k_q, k_{\bar q}, \{k\};\{\hat{k}_{s}\}) \{\hat{k}_{s}^{\,\bar{\rho}}\}$ and 
$ \hat{D}_{2\,{\{\bar{\nu}\},\{\bar{\sigma}\}}}(l_q, l_{\bar q}, \{l\};\{\hat{l}_{s}\}) \{\hat{l}_{s}^{\,\bar{\sigma}}\} $
in the second and third lines of eq.~\eqref{eq:rc-app}, respectively, trying to detach all the soft gluons from their attachments to $\hat{D}_1$ and $\hat{D}_2$.  
However, with the Wilson lines from detaching collinear longitudinal gluons from the $\hat{H}$, the collinear subgraphs, $\hat{D}_1$ and $\hat{D}_2$, are more complicated.  
Fortunately, as in eq.~\eqref{eq:ks}, the soft momentum $k_s$ flowing into $\hat{D}_1$ is approximated by $\hat{k}_{s} \propto w_2$ and 
since the Wilson line on $\hat{D}_1$ has the vertices proportional to $w_2$, the attachment of soft gluon of momentum $\hat{k}_{s}$ to the gauge links of $\hat{D}_1$ vanishes. 
Therefore, we are allowed to sum over all possible attachments of the soft gluons to $\hat{D}_1$, including to the gauge links. Consequently, the use of Ward identity allows to detach all the soft gluons that are attached to $\hat{D}_1$ and group them into two gauge links along the direction $n_1$
in the same way as we detach collinear gluons from the $\hat{H}$.
Similarly, since $\hat{l}_{s} \propto w_1$, we can apply the Ward identify to 
$\hat{D}_{2\,{\{\bar{\nu}\},\{\bar{\sigma}\}}}(l_q, l_{\bar q}, \{l\};\{\hat{l}_{s}\}) \{\hat{l}_{s}^{\,\bar{\sigma}}\} $ to 
detach all the soft gluons that are attached to $\hat{D}_2$ and group them into two gauge links along the direction $n_2$
on the side of $\hat{D}_2$,
as shown in figure~\ref{fig:ward-soft}.

With all collinear and longitudinally polarized gluons detached from the hard part $\hat{H}$ and their impact represented by the Wilson lines to $\hat{D}_1$ and $\hat{D}_2$, as shown in figure~\ref{fig:ward-col}, and all soft gluons detached from the $\hat{D}_1$ and $\hat{D}_2$ and included into gauge links to the $S$, as shown in figure~\ref{fig:ward-soft}, 
we can express the exclusive scattering amplitude for $\pi^+\pi^-\to \gamma\gamma$ as
\begin{align}
{\cal M}^{\mu\nu}_{\pi^+\pi^- \to \gamma \gamma}=&
\int \dd z_1 \dd z_2 \ S_{ij,i'j';mn,m'n'} \,
    [{\cal P}_A\, \hat{D}_{1}(z_1,p_1)\, {\cal P}_B]_{i'j'}\,
     [{\cal P}_B\, \hat{D}_{2}(z_2,p_2)\, {\cal P}_A]_{m'n'}\,
\label{eq:color convo} \\
    &{\hskip 0.5in}\times
     \hat{H}^{\mu\nu}_{ij;mn}(k_1^+=z_1\, p_1^+; k_2^-=z_2\, p_2^-;q_T)		\, ,
     \nn	
\end{align}
where $i,j,m,n$, etc.~are color indices as labeled in figure~\ref{fig:ward-soft}, the sum of repeated color indices is understood, and $z_1\equiv k_1^+/p_1^+$ and $z_2\equiv k_2^-/p_2^-$ are momentum fractions.  In eq.~(\ref{eq:color convo}), the soft factor is given by
\begin{align}
S_{ij,i'j';mn,m'n'} = \langle 0| 
	&\Phi(0, -\infty; n_1)_{ii'} \Phi^{\dag}(0, -\infty; n_1)_{j'j}	 			
	\nn\\
	&\times
	\Phi(0, -\infty; n_2)_{mm'} \Phi^{\dag}(0, -\infty; n_2)_{n'n} |0 \rangle \, .
\end{align}
In deriving the collinear factors $\hat{D}_1$ and $\hat{D}_2$ in eq.~(\ref{eq:color convo}), we took into account the fact that at the leading power, only $\hat{k}_1$ and $\hat{k}_2$ flow into the hard part $\hat{H}^{\mu\nu}$.  For a generic pion distribution amplitude (suppressing the Wilson lines),
\begin{align}
{\cal D}(k,p)=\int \dd^4\xi\, e^{ik\cdot \xi} 
\langle 0|\overline{\psi}(0)\psi(\xi)|\pi(p)\rangle\, ,
\end{align}
we apply the following identify,
\begin{align}
& 
\int\frac{\dd^4k_1}{(2\pi)^4} {\cal D}(k_1,p_1)\, \hat{H}(\hat{k}_1;\hat{k}_2;q_T)
\nn\\
&{\hskip 0.3in}
=\int \dd z_1 \left[
\int\frac{\dd^4k_1}{(2\pi)^4} \delta\left(z_1-\frac{k_1^+}{p_1^+}\right){\cal D}(k_1,p_1)\right]
\hat{H}(\hat{k}_1=z_1\,p_1^+;\hat{k}_2;q_T)
\\
&{\hskip 0.3in}
=\int \dd z_1 \left[
\int\frac{\dd(p_1^+\xi^-)}{2\pi}\, e^{i z_1p_1^+\xi^-} 
\langle 0|\overline{\psi}(0)\psi(\xi^-)|\pi(p)\rangle|_{\xi^+=0, \, \xi_\perp=0_\perp}
\right] \hat{H}(\hat{k}_1=z_1\,p_1^+;\hat{k}_2;q_T) ,
\nn
\end{align}
to $\hat{D}_1(k_1,p_1)$ and similar identity to $\hat{D}_2(k_2,p_2)$, and obtain the collinear factors in eq.~(\ref{eq:color convo}),
\begin{align}
\hat{D}_{1}(z_1, p_1)_{i'j'}
= & \int\frac{\dd (p_1^+ \xi^-)}{2\pi} 
e^{i z_1p_1^+ \xi^-}
  	\langle 0 | T\, \bar{d}_j(0) \Phi^{\dag}(\infty, 0; w_2)_{jj'}
	\nn\\
   & {\hskip 1.2in}\times	
	\Phi(\infty, \xi^-; w_2)_{i'i} \, u_i(\xi^-) | \pi^+(p_1) \rangle \, ,
	\label{eq.da1}
\end{align}
and 
\begin{align}
\hat{D}_{2}(z_2, p_2)_{m'n'}
	= & \int\frac{\dd (p_2^- \zeta^+)}{2\pi} 
e^{i z_2p_2^- \zeta^+}
  	\langle 0 | T\, \bar{u}_n(0) \Phi^{\dag}(\infty, 0; w_1)_{nn'}	
	\nn\\
   & {\hskip 1.2in}\times
	\Phi(\infty, \zeta^+; w_1)_{m'm} \, d_m(\zeta^+) | \pi^-(p_2) \rangle \, ,
	\label{eq.da2}
\end{align}
where $T$ represents the time-ordering and $u$ and $d$ are up and down quark fields, respectively.  
In eq.~(\ref{eq:color convo}), the spinor projectors ${\cal P}_A$ and ${\cal P}_B$ are given in eqs.~(\ref{eq:spinorPA}) and (\ref{eq:spinorPB}), respectively,  and the superscripts, $\mu$ and $\nu$, are Lorentz indices of the two produced photons.  
The spinor indices in eq.~(\ref{eq:color convo}), convoluting between $\hat{H}$ and $\hat{D}$'s, are suppressed, and their factorization will be discussed in the next subsection. 

The colliding hadrons, $\pi^+$ and $\pi^-$ in our case, are color neutral.   
With all the soft gluons factored out of them, the collinear factors must be in a color singlet state,
\beq
\hat{D}_{1i'j'} \equiv  \hat{D}_{1} \frac{\delta_{i'j'}}{N_c}, \quad\quad
\hat{D}_{2m'n'} \equiv \hat{D}_{2} \frac{\delta_{m'n'}}{N_c},
\label{eq:C-singlet}
\eeq
where $\hat{D}_{1(2)} = \delta_{ij}\,\hat{D}_{1(2)ij}$. 
This color contraction connects the two Wilson lines in each collinear factor to give
\begin{align}
\hat{D}_{1}(z_1, p_1)
= 	\int\frac{\dd (p_1^+\xi^-)}{2\pi} e^{i z_1p_1^+ \xi^-} 
  	\langle 0 | T\, \bar{d}_m(0)
	\Phi(0,\xi^-; w_2)_{mn} \, u_n(\xi^-) | \pi^+(p_1) \rangle,
\label{eq:D1 factor}
\end{align}
and
\begin{align}
\hat{D}_{2}(z_2, p_2)
=  \int\frac{\dd (p_2^-\zeta^+) }{2\pi} e^{i z_2p_2^- \zeta^+} 
  	\langle 0 | T\, \bar{u}_m(0)
	\Phi(0,\zeta^+; w_1)_{mn} \, d_n(\zeta^+) | \pi^-(p_2) \rangle,
\label{eq:D2 factor}
\end{align}
where the sum of repeated color indices is understood, while the spinor indices are not summed over. 
$\Phi(0,\xi^-;w_2)$ and $\Phi(0,\zeta^+;w_1)$ are the Wilson lines in the fundamental representation, joining the $u$ and $d$ quark fields to make the DAs gauge invariant, which is a result of factorization.
Substituting eq.~\eqref{eq:C-singlet} into eq.~\eqref{eq:color convo}, we have 
\begin{align}
{\cal M}^{\mu\nu}_{\pi^+\pi^- \to \gamma \gamma}
=&
\frac{1}{N_c^2} \int \dd z_1 \dd z_2 \ S_{ij,i'i';mn,m'm'} \, 
    [{\cal P}_A\, \hat{D}_{1}(z_1,p_1)\, {\cal P}_B]\,
    [{\cal P}_B\, \hat{D}_{2}(z_2,p_2)\, {\cal P}_A]\,
\label{eq:color convo2} \\
    & \hskip 0.8in \times
    \hat{H}^{\mu\nu}_{ij;mn}(k_1^+=z_1\,p_1^+;k_2^-=z_2\, p_2^-;q_T)	\, ,
     \nn
\end{align}
where the repeated color indices, $i'$ and $m'$ are summed.  The soft factor now becomes
\begin{align}
S_{ij,i'i';mn,m'm'} &= \langle 0| 
	\left[ \Phi(0, -\infty; n_1) \Phi^{\dag}(0, -\infty; n_1) \right]_{ij}			
	\left[ \Phi(0, -\infty; n_2) \Phi^{\dag}(0, -\infty; n_2) \right]_{mn} |0 \rangle	\nn\\
	&= \delta_{ij} \delta_{mn}.
\end{align}
That is, the soft factor $S$ is in fact an identity matrix in the color space, and the exclusive scattering amplitude in eq.~\eqref{eq:color convo} is effectively factorized in color space,
\begin{align}
{\cal M}^{\mu\nu}_{\pi^+\pi^- \to \gamma \gamma}=&
\int \dd z_1 \dd z_2 \, {\rm Tr}\Big\{
    [{\cal P}_A\, \hat{D}_{1}(z_1,p_1)\, {\cal P}_B]
    [{\cal P}_B\, \hat{D}_{2}(z_2,p_2)\, {\cal P}_A]
    \nn \\
& \hskip 0.8in \times
\Big[ \frac{1}{N_c^2} \hat{H}^{\mu\nu}_{ii;mm}(k_1^+=z_1\, p_1^+;k_2^-=z_2\, p_2^-;q_T) \Big] \Big\}		\, ,
\label{eq:color convo3}
\end{align}
where the repeated color indices, $i$ and $m$ are summed, and averaged with the factor $1/N_c^2$, and the ``${\rm Tr}$'' indicates the trace over all spinor indices between $\hat{D}_1$, $\hat{D}_2$, and $\hat{H}$.
The hard part $\hat{H}$ that produces the pair of high-$q_T$ photons is given by the collision of two color-singlet, collinear, and on-shell quark-antiquark pairs, one from each colliding hadron ($\pi^+$ or $\pi^-$ in this case).

The cancellation of long-range soft gluon interactions between the colliding hadrons is essential to the factorization. It means that \textit{long-distance} connections between the two collinear subgraphs are canceled, so that the evolution of each collinear part is independent. Therefore, the collinear functions can be universal. In our case, the soft gluon cancellation happens because the active parton lines entering the hard interactions are collinear and color-neutral pairs, so that the soft gluons only see a color-neutral object from each colliding hadron, and thus there are no color correlations between the two collinear systems. This is the feature for \textit{exclusive} processes, which is also seen in the factorization of two-quarkonium exclusive production in $e^+ e^-$ annihilation~\cite{Bodwin:2010fi}.
But, this is different from the soft gluon cancellations for inclusive processes, for example in~\cite{Collins:1981ta}, where it is the unitarity (inclusiveness of the final states) that guarantees the soft cancellation.

We also stress that the above steps of factorizing collinear gluons and soft cancellation should be viewed as for a given order of perturbative diagram expansion with a given number of soft gluon lines and $\hat{D}_1$- and $\hat{D}_2$-collinear lines. Summing over different attachments of gluon lines in the same kinematic region amounts to summing over different diagrams with the same region decomposition, along with subtractions of smaller regions to avoid double counting.  Since such subtraction does not affect the used of Ward identity, after factorizing the whole amplitude into $\hat{D}_1$, $\hat{D}_2$ and $\hat{H}$, each factor should be regarded as subtracted ones. Due to the cancellation of soft gluons, the subtracted collinear factors $\hat{D}_1$ and $\hat{D}_2$ are the same as the unsubtracted ones in eqs.~\eqref{eq:D1 factor} and \eqref{eq:D2 factor}. And the subtracted hard factor $\hat{H}$ can be derived perturbatively order-by-order by using the factorization formula.

\begin{figure}[htbp]
    \centering
	\includegraphics[scale=0.85]{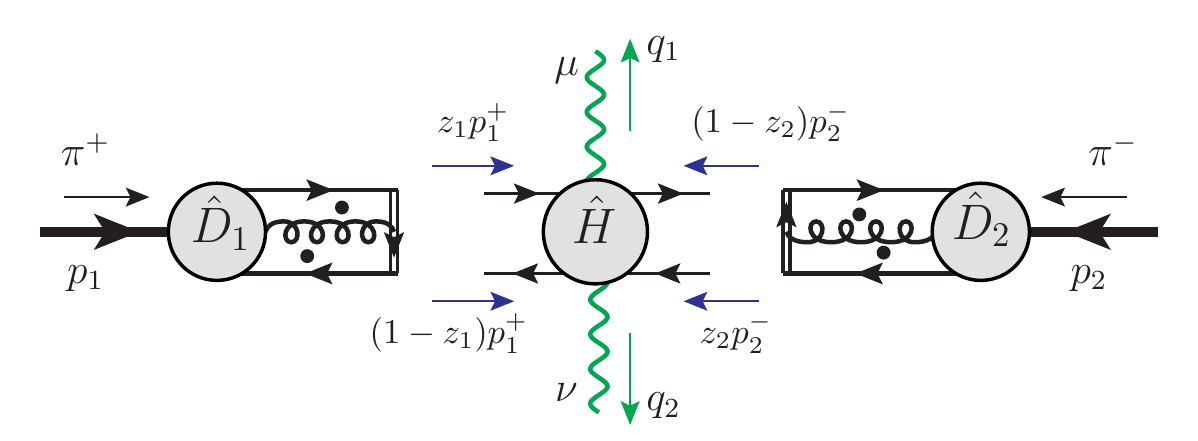}
    \caption{The factorized form for the exclusive $\pi\pi$ annihilation process.}
    \label{fig:pi-pi factorize}
\end{figure}

\subsubsection{Factorization formula}

After the cancellation of soft gluons, the leading power contributions to the exclusive scattering amplitude of $\pi^+\pi^-\to \gamma\gamma$ can be factorized into the structure shown in figure~\ref{fig:pi-pi factorize}, while the spinor indices from $\hat{D}_1$, $\hat{D}_2$ and $\hat{H}$ are still convoluted, as indicated in eq.~(\ref{eq:color convo3}), which need to be disentangled.

The factor $\hat{D}_1$ is sandwiched between $\mathcal{P}_A$ and $\mathcal{P}_B$, which indicates that only the term in $\hat{D}_1$ proportional $\gamma^-$, $\gamma_5\gamma^-$ or $\gamma_5\gamma^-\gamma_i$ with $i=1,2$ survives. Since $\pi^+$ has negative parity and zero spin, only $\gamma_5\gamma^-$ contributes. Similarly, $\hat{D}_2$ only has its $\gamma_5\gamma^+$ term that contributes. The result is 
\begin{align}
\left[
\mathcal{P}_A \hat{D}_1(z_1,p_1) \mathcal{P}_B
\right]_{\alpha\beta}
&= \frac{1}{2p_1^+}{\rm Tr}\left[ \gamma^+ \gamma_5\, \hat{D}_1(z_1,p_1) \right] 
\left[\frac{1}{2}\gamma_5 (p_1^+\gamma^-) \right]_{\alpha\beta}
\nn\\
&\equiv D_{\pi^+}(z_1)
    \left[\frac{1}{2}\gamma_5 (p_1^+\gamma^-) \right]_{\alpha\beta}  	\, ,
        	\label{eq:PA D1 PB}
\\
\left[
\mathcal{P}_B \hat{D}_2(z_2,p_2) \mathcal{P}_A
\right]_{\rho\sigma}
&= \frac{1}{2p_2^-}{\rm Tr}\left[ \gamma^- \gamma_5\, \hat{D}_2(z_2,p_2) \right] 
\left[\frac{1}{2}\gamma_5 (p_2^-\gamma^+) \right]_{\rho\sigma} 
\nn\\
&\equiv D_{\pi^-}(z_2)
    \left[\frac{1}{2}\gamma_5 (p_2^-\gamma^+) \right]_{\rho\sigma} 		\, ,
    	\label{eq:PB D2 PA}
\end{align}
where the indices $\alpha, \beta, \rho, \sigma$ here are the spinor indices, and 
the distribution amplitudes for $\pi^{\pm}$ are given by
\bse \label{eq:DA def}
\begin{align}
D_{\pi^+}(z_1)
&=
\int\frac{\dd \xi^-}{4\pi} e^{i z_1p_1^+ \xi^-}
\langle 0 | \bar{d}(0) \gamma^+ \gamma_5 \,
\Phi(0,\xi^-;w_2) u(\xi^-) | \pi^+(p_1) \rangle	\, , \\
D_{\pi^-}(z_2)
&=
\int\frac{\dd \zeta^+}{4\pi} e^{i z_2 p_2^- \zeta^+}
\langle 0 | \bar{u}(0) \gamma^- \gamma_5 \,
\Phi(0,\zeta^+;w_1) d(y^+) | \pi^-(p_1) \rangle	\, .
\end{align}
\ese 
Charge conjugation and isospin symmetry imply $D_{\pi^+}(z) = -D_{\pi^-}(z)$, following the convention of taking $(\pi^+, \pi^0, \pi^-)$ state as an isospin triplet. This particular choice does not affect our prediction for the cross section, since it is proportional to the squared amplitude.
With the separation of spinor indices, we have 
our final factorized expression for the exclusive $\pi^+\pi^-$ annihilation amplitude,
\begin{align}
\mathcal{M}^{\mu\nu}_{\pi^+\pi^- \to \gamma \gamma}=&
    \int_0^1 \dd z_1 \int_0^1 \dd z_2 \
    D_{\pi^+}(z_1)\, D_{\pi^-}(z_2)\,
    C^{\mu\nu}(z_1,z_2;p_1^+,p_2^-,q_T) \, ,
\label{eq:factorization M pi pi}
\end{align}
where the short-distance hard coefficient function $C^{\mu\nu}$ is given by
\begin{align}
C^{\mu\nu}(z_1,z_2;p_1^+,p_2^-,q_T)
&\equiv 
\left[ \frac{\gamma_5(p_1^+\gamma^-)}{2} \right]_{\alpha\beta}
\hat{H}^{\mu\nu}_{\beta\alpha; \sigma\rho}(\hat{k}_1=z_1p_1^+,\hat{k}_2=z_2p_2^-;q_T)
    \left[\frac{\gamma_5(p_2^-\gamma^+)}{2}
    \right]_{\rho\sigma} \, ,
\label{eq:hard C}
\end{align}
where a sum over repeated indices is understood, which is effectively the trace of spinor indices.
The correction to the factorized formula in eq.~(\ref{eq:factorization M pi pi}) is suppressed by powers of $m_{\pi}/q_T$.  
With the renormalization group improvement from the fact that the exclusive scattering amplitude for $\pi^+(p_1)+\pi^-(p_2)\to \gamma(q_1)+\gamma(q_2)$ should not depend on the specific hard scale (or, the factorization scale) at which we perform our factorization steps. And with the choice of this factorization scale, the pion distribution amplitudes and the short-distance coefficient function in eq.~(\ref{eq:factorization M pi pi}) become dependent on the factorization scale $\mu$.

\subsection{Gauge invariant tensor structures for the hard coefficient}

The short-distance hard coefficient, $C^{\mu\nu}(z_1, z_2; p_1^+, p_2^-, q_T)$ in eq.~\eqref{eq:hard C}, is a function of the external momenta, and its tensor structure is constrained by the symmetry of the underlying theory.  The most important constraint comes from electromagnetic gauge invariance or current conservation for the two 
external photons
\beq
q_{1\mu}C^{\mu\nu} 
= C^{\mu\nu} q_{2\nu}
= 0\, ,
\label{eq:emcurrent}
\eeq
which requires $C^{\mu\nu}$ to be expressed in terms of independent and gauge invariant tensor structures.  

Because of the explicit light-cone projection $\gamma^{\pm}$ in eq.~\eqref{eq:hard C}, we express all external momenta, $p_1, p_2, q_1, q_2$, in light-cone coordinates, 
\begin{align}
p_1^{\mu} =  p_1^+ \, \bar{n}^{\mu}\, ,
\quad
p_2^{\mu} = p_2^- \, {n}^{\mu}\, ,
\quad
q_{1,2}^{\mu}= 
q_{1,2}^+ \,\bar{n}^{\mu} + q_{1,2}^-\, n^{\mu} \mp q_T^{\mu}\, ,
\label{eq:p1p2q12}
\end{align}
as we have done for $q_{1,2}^{\mu}$ in eq.~\eqref{eq:kin q12}. 
We choose three independent vectors, $p_1^{\mu}, p_2^{\mu}$ and $q_T^{\mu}$. 
Using $q_1\cdot q_T = -q_T^\mu q_T^\nu\,g_{\mu\nu}=q_T^2$ and $q_2\cdot q_T = -q_T^2$, we can write down all the independent parity-even (P-even) current-conserving tensor structures, 
\begin{align}
\widetilde{g}_{\perp}^{\mu\nu},\quad
\widetilde{p}_{1}^{\mu}\bar{p}_{1}^{\nu}\, ,\quad
\widetilde{p}_{2}^{\mu}\bar{p}_{2}^{\nu}\, ,\quad
\widetilde{p}_{1}^{\mu}\bar{p}_{2}^{\nu}\, ,\quad
\widetilde{p}_{2}^{\mu}\bar{p}_{1}^{\nu}\, ,
\label{eq:P-even tensors}
\end{align}
where we defined
\begin{gather} 
 \widetilde{g}_{\perp}^{\mu\nu}
= g_{\perp}^{\mu\nu} + 
 \frac{q_T^{\mu} q_T^{\nu} }{q_T^2} \quad {\rm with} \quad
 g_{\perp}^{\mu\nu} 
 = g^{\mu\nu} - \bar{n}^\mu \, n^\nu - n^\mu\, \bar{n}^\nu\, ,
\nn\\
\begin{align}
 \widetilde{p}_{1}^{\mu} 
= p_1^{\mu} - \frac{p_1\cdot q_1}{q_T^2} q_{T}^{\mu}\, ,\quad
 \widetilde{p}_{2}^{\mu} 
= p_2^{\mu} - \frac{p_2\cdot q_1}{q_T^2} q_{T}^{\mu}\, ,\nn\\
 \bar{p}_{1}^{\nu}
= p_1^{\nu} + \frac{p_1\cdot q_2}{q_T^2} q_{T}^{\nu}\, ,\quad
 \bar{p}_{2}^{\nu}
= p_2^{\nu} + \frac{p_2\cdot q_2}{q_T^2} q_{T}^{\nu}\, ,
\label{eq:projected p1 p2}
\end{align}
\end{gather}
so that 
\begin{align}
q_{1\mu}\,  \widetilde{p}_{i}^{\mu} = 0 \, ,
\quad\quad
q_{2\nu}\, \bar{p}_{i}^{\nu} = 0\, ,
\quad {\rm and} \quad
q_{1\mu}\,  \widetilde{g}_{\perp}^{\mu\nu} 
=  \widetilde{g}_{\perp}^{\mu\nu} \, q_{2\nu} = 0 \, ,
\label{eq:current-conv}
\end{align}
with $i=1,2$.  Similarly, we have parity-odd (P-odd) current-conserving tensor structures 
\begin{align}
\widetilde{p}_{1}^{\mu}\varepsilon_{\perp}^{\nu\rho}q_{T\rho}\, ,\quad
\widetilde{p}_{2}^{\mu}\varepsilon_{\perp}^{\nu\rho}q_{T\rho}\, ,  \quad 
\bar{p}_{1}^{\nu}\varepsilon_{\perp}^{\mu\rho}q_{T\rho}\, ,\quad
\bar{p}_{2}^{\nu}\varepsilon_{\perp}^{\mu\rho}q_{T\rho}\, ,
\label{eq:P-odd tensors}
\end{align}
where we used the abbreviation
$
\varepsilon_{\perp}^{\mu\nu} = \varepsilon^{+-\mu\nu}\, ,
$
with the convention $\varepsilon_{\perp}^{12}=-\varepsilon_{\perp}^{21}=1$.  
One might consider 
another P-odd tensor structure $q_T^2 \varepsilon_{\perp}^{\mu\nu} + q_T^{\mu} \varepsilon_{\perp}^{\rho\nu} q_{T\rho} + q_T^{\nu} \varepsilon_{\perp}^{\mu\rho} q_{T\rho}$, which satisfies the current conservation in eq.~\eqref{eq:emcurrent}. 
But, this tensor itself vanishes for any components of $\mu$ and $\nu$.

The next constraint is from parity conservation. If we have $n$ $\gamma_5$'s in a given diagram, parity conservation requires corresponding scattering amplitude to satisfy
\beq
C^{\mu\nu}(p_{1\alpha}, p_{2\alpha}, q_{1\alpha}, q_{2\alpha})
= (-1)^n C_{\mu\nu}( p_1^\alpha,{p}_2^\alpha,{q}_1^\alpha,{q}_2^\alpha )	\, ,
\label{eq:parity}
\eeq
which holds for each individual diagram.  
For $\pi^+\pi^-$ scattering, $n=2$ and parity conservation excludes the 
tensor structures in eq.~\eqref{eq:P-odd tensors}.  In next section, we will generalize the pion-pion process to pion-nucleon scattering, for which we will have both P-even and P-odd tensor structures.

For the exclusive $\pi^+\pi^-$ scattering in this section, we can express the hard coefficient in terms of a linear combination of the P-even tensors in eq.~\eqref{eq:P-even tensors},
\begin{align}
C^{\mu\nu}
\equiv -\frac{i e^2 g^2}{2s^2}\frac{C_F}{N_c}\,\bigg(
 C_{0} \, \widetilde{g}_{\perp}^{\mu\nu}\, s  
 + C_{1} \,  \widetilde{p}_{1}^{\mu}\bar{p}_{1}^{\nu}
+ C_{2} \, \widetilde{p}_{2}^{\mu}\bar{p}_{2}^{\nu}   
  + C_{3} \,  \widetilde{p}_{1}^{\mu}\bar{p}_{2}^{\nu}
+ C_{4} \, \widetilde{p}_{2}^{\mu}\bar{p}_{1}^{\nu} \bigg)		\, ,
\label{eq:gauge-inv expansion}
\end{align}
where we introduced an overall factor $-i e^2 g^2 (C_F/N_c)/(2s^2)$ with electric charge $e$, strong coupling constant $g$, color factor $C_F/N_c$ for the leading order contribution, and collision energy  squared $s=2p_1^+p_2^-$ to make the scalar coefficients $C_i=C_i(z_1, z_2; \kappa)$ dimensionless for $i=0,\cdots,4$, with $\kappa$ introduced in eq.~\eqref{eq:kin q12}.

Substituting eq.~\eqref{eq:gauge-inv expansion} in eq.~\eqref{eq:factorization M pi pi}, we obtain 
\begin{align}
\mathcal{M}^{\mu\nu}_{\pi^+\pi^- \to \gamma \gamma}
=  & -\frac{i\, e^2 g^2 f_{\pi}^2 }{8 s^2} \frac{C_F}{N_c} \,\bigg(
 \mathcal{M}_{0} \, \widetilde{g}_{\perp}^{\mu\nu}\, s  
 + \mathcal{M}_{1} \,  \widetilde{p}_{1}^{\mu}\bar{p}_{1}^{\nu}    
+ \mathcal{M}_{2} \, \widetilde{p}_{2}^{\mu}\bar{p}_{2}^{\nu}     
+ \mathcal{M}_{3} \,  \widetilde{p}_{1}^{\mu}\bar{p}_{2}^{\nu}
+ \mathcal{M}_{4} \, \widetilde{p}_{2}^{\mu}\bar{p}_{1}^{\nu} \bigg)	\, ,
\label{eq:M result}
\end{align}
where $\mathcal{M}_i$ is the convolution of the hard scalar coefficient $C_i$ with the normalized DA $\phi(z)$
\begin{align}
\mathcal{M}_i = 
 \int_0^1 \dd z_1 \int_0^1 \dd z_2 \, \phi(z_1)\, \phi(z_2) \, C_i(z_1,z_2; \kappa)	\, ,
 \quad (i=0,\cdots, 4),
\end{align}
with $\phi(z)$ defined as
\beq
\phi(z) \equiv - \frac{2i}{f_\pi} D_{\pi^+}(z)
\,
,\quad
\int_0^1 \dd z\, \phi(z) = 1,
\eeq
and $f_{\pi}=130~{\rm MeV}$ being the pion decay constant.

To help simplify some long expressions in this paper, we introduce a notation
\def\M{\mathfrak{M}}
 \begin{align}
\mathfrak{M}[C; D_1, (z_1^m, z_1^M); D_2, (z_2^m, z_2^M)]		
 \equiv \int_{z_1^m}^{z_1^M} \dd{z_1} \int_{z_2^m}^{z_2^M} \dd{z_2}\, D_1(z_1)\, D_2(z_2)\, C(z_1, z_2).
\end{align}
We can then express our factorization formalism in eq.~\eqref{eq:factorization M pi pi} as
\beq
\mathcal{M}^{\mu\nu}_{\pi^+\pi^- \to \gamma \gamma}
= \M[C^{\mu\nu}; D_{\pi^+}, (0,1); D_{\pi^-}, (0,1)] \, ,
\eeq
and all scalar functions $\mathcal{M}_{i}$ in eq.~\eqref{eq:M result} as
\beq
\mathcal{M}_{i} = \M[C_i; \phi, (0,1); \phi, (0,1)] \, ,
\label{eq:Mi}
\eeq
with $i=0,1, ..., 4$.

\subsection{The leading-order hard coefficients}
\label{sec:hard coef}

At leading-order of $\alpha_s$, there are two types of Feynman diagrams contributing to the short-distance hard coefficients: $A$) the two observed photons are radiated from the different fermion lines, which we refer to as Type-$A$ diagrams shown in figure~\ref{fig:hard1}, and $B$)\ they are radiated from the same fermion line, which we refer to as Type-$B$ diagrams shown in figure~\ref{fig:hard2}.  With the two identical photons in the final-state, we need to consider additional diagrams that are the same as those in figure~\ref{fig:hard1} and \ref{fig:hard2}, but with the two photons switched. That is, we need to evaluate a total of 8 Type-$A$ diagrams and 12 Type-$B$ diagrams for the leading-order hard coefficients.

\begin{figure}[htbp]
\begin{center}
\begin{subfigure}{.23\textwidth}
  \centering
  \includegraphics[scale=0.65]{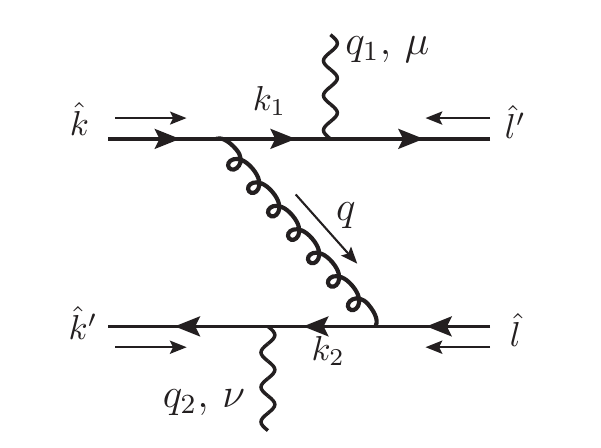}  
  \caption{}
  \label{fig:hard11}
\end{subfigure}
\begin{subfigure}{.23\textwidth}
  \centering
  \includegraphics[scale=0.65]{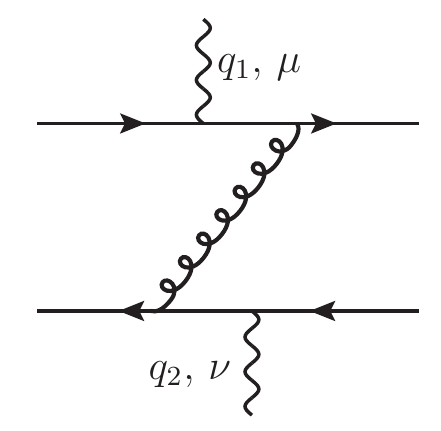}  
  \caption{}
  \label{fig:hard12}
\end{subfigure}
\begin{subfigure}{.23\textwidth}
  \centering
  \includegraphics[scale=0.65]{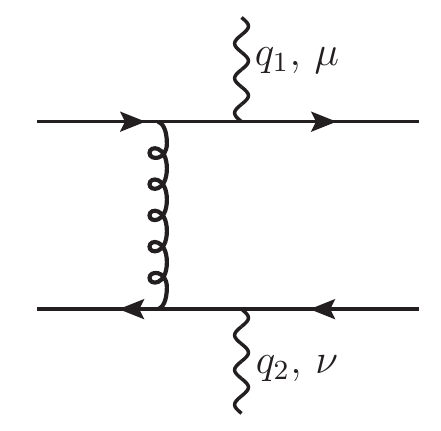}  
  \caption{}
  \label{fig:hard13}
\end{subfigure}
\begin{subfigure}{.23\textwidth}
  \centering
  \includegraphics[scale=0.65]{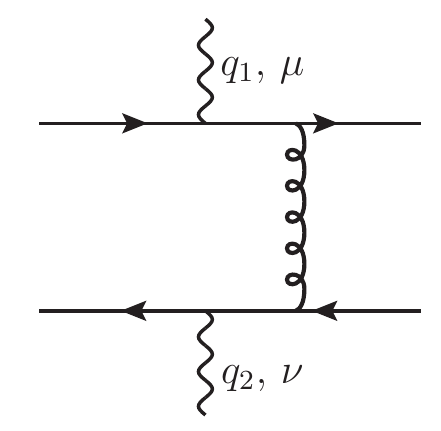}  
  \caption{}
  \label{fig:hard14}
\end{subfigure}
\caption{\label{fig:hard1}
The Type-$A$ diagrams, plus those with $q_1$ and $q_2$ switched. Relatively thicker gluon lines are used to indicate the large transverse momentum flow from one photon to another, in contrast to those in figure~\ref{fig:hard2}.}
\end{center}
\end{figure}

\begin{figure}[htbp]
\begin{center}
\begin{subfigure}{.25\textwidth}
  \centering
  \includegraphics[scale=0.65]{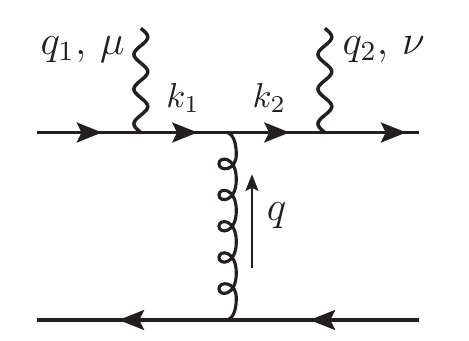}  
  \caption{}
  \label{fig:hard21}
\end{subfigure}
\begin{subfigure}{.25\textwidth}
  \centering
  \includegraphics[scale=0.65]{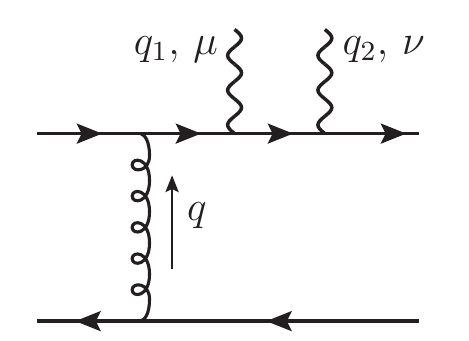}  
  \caption{}
  \label{fig:hard22}
\end{subfigure}
\begin{subfigure}{.25\textwidth}
  \centering
  \includegraphics[scale=0.65]{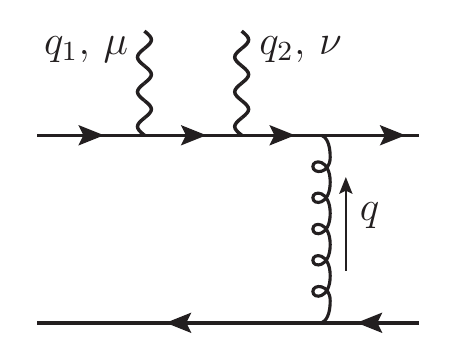}  
  \caption{}
  \label{fig:hard23}
\end{subfigure}
\\
\begin{subfigure}{.25\textwidth}
  \centering
  \includegraphics[scale=0.65]{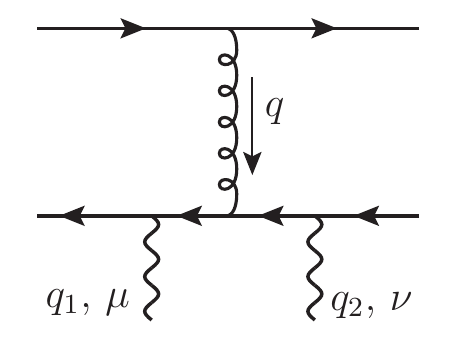}  
  \caption{}
  \label{fig:hard24}
\end{subfigure}
\begin{subfigure}{.25\textwidth}
  \centering
  \includegraphics[scale=0.65]{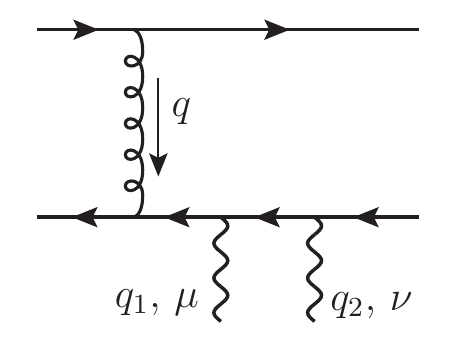}  
  \caption{}
  \label{fig:hard25}
\end{subfigure}
\begin{subfigure}{.25\textwidth}
  \centering
  \includegraphics[scale=0.65]{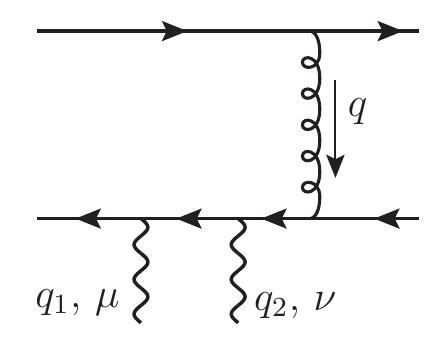}  
  \caption{}
  \label{fig:hard26}
\end{subfigure}
\caption{\label{fig:hard2}
The Type-$B$ diagrams, plus those with $q_1$ and $q_2$ switched. }
\end{center}
\end{figure}

In the CM frame of this exclusive scattering process, the large transverse momentum $q_T$ of one-photon should be balanced by that of the other photon.  This large transverse momentum $q_T$ flows through the gluon connecting the two fermion lines for all Type-$A$ diagrams, while it does not flow through the gluon for all Type-$B$ diagrams.   
Since the relative momentum of the quark and antiquark of the colliding pion, represented by the $z_1$ (or $z_2$) dependence of the pion DA in eq.~(\ref{eq:factorization M pi pi}), flows through the gluon line of the hard scattering back to the pion, the $q_T$-dependence of the gluon propagator of the Type-$A$ diagrams makes the hard coefficient, and hence, the cross section $\dd\sigma/\dd q_T^2$ of this exclusive process, be a sensitive probe for the $z_1$ (or $z_2$) dependence of pion DA.  Its sensitivity depends on the relative size of contributions to the cross section from these two-types of diagrams.  

For our calculation of the leading-order hard coefficients, we denote the diagrams in figure~\ref{fig:hard1}(a)-(d) by $A1, \cdots, A4$, sequentially, and the ones with $q_1$ and $q_2$ switched by $A1' , \cdots, A4'$, respectively. Their contributions to the hard coefficient $C^{\mu\nu}$ are denoted sequentially by $C^{\mu\nu}_{A1}, C^{\mu\nu}_{A2}$, etc. Similarly, we label the individual contribution from the Type-$B$ diagrams in a similar way.   We use $C_A^{\mu\nu}$ and $C_B^{\mu\nu}$ to represent total contribution from the Type-$A$ and Type-$B$ diagrams, respectively.

The contribution from each individual diagram in figure~\ref{fig:hard1} and \ref{fig:hard2} can be evaluated by using eq.~\eqref{eq:hard C}.  The collinear momenta of colliding partons, as labeled in figure~\ref{fig:hard1} and \ref{fig:hard2}, 
are defined as
\begin{align}
\hat{k}  = z_1\, p_1, \quad
\hat{k}' = (1-z_1) \, p_1,  \quad
\hat{l}  = z_2\, p_2,  \quad
\hat{l}' = (1-z_2) \, p_2, \quad
\end{align}
and the photon momenta are given in eq.~\eqref{eq:kin q12} and also in \eqref{eq:p1p2q12}.  
The external collinear quark and antiquark lines from $\pi^+$ on the left are 
contracted with $\gamma_5(p_1^+\gamma^-)/2$, while the collinear quark and antiquark lines from $\pi^-$ on the right are
contracted with  $\gamma_5(p_2^-\gamma^+)/2$.

At this order, all momenta of internal propagators are completely determined.  For example, for figure~\ref{fig:hard1}(a), we have
\begin{gather}
k_1  = q_1 - \hat{l}' \, ,\quad
k_2  = q_2 - \hat{k}' \, , \nn\\
q  = \hat{k} - k_1 = \hat{k} + \hat{l}' - q_1
= k_2 - \hat{l} = q_2 - \hat{l} - \hat{k}' \, ,
\end{gather}
and its contribution to $C^{\mu\nu}$ is given by
\begin{align}
 C_{A1}^{\mu\nu}
=& \, i e_u e_d \, e^2 g^2 \cdot\frac{C_F}{N_c}\cdot         
    \frac{1}{k_1^2+i\varepsilon} \cdot
	\frac{1}{k_2^2+i\varepsilon} \cdot \frac{1}{q^2+i\varepsilon}
\nn\\
& \times
	 {\rm Tr}\left[\, 
	 \left( \frac{\gamma_5}{2}(p_2^-\gamma^+)\right) \gamma^{\mu} \slash{k}_1 \gamma^{\alpha} 
	 \left( \frac{\gamma_5}{2} (p_1^+\gamma^-)\right) \gamma^{\nu} \slash{k}_2 \gamma_{\alpha} \,\right] \nn\\
=& \, -i e_u e_d \, e^2 g^2 \cdot\frac{C_F}{N_c}\cdot \frac{1}{k_1^2 + i\varepsilon} \cdot
\frac{1}{k_2^2 + i\varepsilon}\cdot
\frac{1}{q^2+i\varepsilon} 
\nn\\
&\times
	 2s \left[ 
	 g_{\perp}^{\mu\nu} p_2 \cdot q_1	
	 + \hat{l}^{\mu} q_T^{\nu} 
	 - \hat{k}^{\nu} q_T^{\mu} 
	 - q^2 n^{\mu} \bar{n}^{\nu}
	 	\right],
\label{eq:CA1}
\end{align}
where $e_u=2/3$ and $e_d=-1/3$, $s=2p_1^+p_2^-$ and the observed photon momenta are defined in eq.~\eqref{eq:p1p2q12}.
In eq.~(\ref{eq:CA1}), the momenta of internal fermion propagators are fixed as 
\bse\label{eq:A k1k2}
\begin{align}
k_1^2 & = -2 \, \hat{l}'\cdot q_1
    = -2 (1-z_2) \, (p_2\cdot q_1),    \\
k_2^2 & = -2 \, \hat{k}'\cdot q_2
    = -2 (1-z_1) \, (p_1\cdot q_2),
\end{align}
\ese
where the dependence on the parton momentum fractions $z_{1,2}$ is factored out from the external kinematic variables.  On the other hand, the exchanged gluon momentum between the two fermion lines, which is the same for the gluon propagators in all the diagrams in figure~\ref{fig:hard1}, is given by
\begin{align}
q^2 & = 2 q^+ q^- - q_T^2
= s\left[ 
\left(z_1 - \frac{q_1^+}{p_1^+} \right)
\left(1-z_2 - \frac{q_1^-}{p_2^-} \right)
- \frac{\kappa}{4} \right]  \nn\\
& = -\frac{s}{4} 
\left[ 
\left( 2z_1 - 1 \mp \sqrt{1-\kappa} \right)
\left( 2z_2 - 1 \mp \sqrt{1-\kappa} \right)
+\kappa \right] \, ,
\label{eq:A q}
\end{align}
for which the parton fractions $z_{1,2}$ cannot be factorized out of the dependence on $q_T$.  It is this entanglement of parton momentum fractions $z_{1,2}$ and external variable $q_T$ that makes Type-$A$ diagrams sensitive to the DA's functional form.

For the Type-$B$ diagrams in figure~\ref{fig:hard2}, the momenta of the internal propagators are different, although they have the same external parton momenta.  For example,  we have the contribution from figure~\ref{fig:hard2}(a), 
\begin{align}
C_{B1}^{\mu\nu}
=& \,
	i \,e_u^2 \, e^2 g^2 \cdot\frac{C_F}{N_c}
	\cdot \frac{1}{ \widetilde{k}_1^2 + i\varepsilon}
	\cdot \frac{1}{ \widetilde{k}_2^2 + i\varepsilon}
	\cdot \frac{1}{ \widetilde{q}^2 + i\varepsilon}
	\nn\\
	&\times{\rm Tr}
	\left[ 
	\left( \frac{\gamma_5}{2} (p_1^+\gamma^-) \right)
	\gamma^{\alpha} 
	\left( \frac{\gamma_5}{2} (p_2^-\gamma^+) \right)
	\gamma^{\nu} \slash{\widetilde{k}}_2 \gamma_{\alpha}	
	\slash{\widetilde{k}}_1	\gamma^{\mu}
	\right]		\nn\\
=&
	-i \,e_u^2 \, e^2 g^2 \cdot\frac{C_F}{N_c}
	\cdot \frac{1}{ \widetilde{k}_1^2 + i\varepsilon}
	\cdot \frac{1}{ \widetilde{k}_2^2 + i\varepsilon}
	\cdot \frac{1}{ \widetilde{q}^2 + i\varepsilon}
	\nn\\
	&\times
	s \left[
		 g_{\perp}^{\mu\nu} q_T^2 
		+ 2 q_T^{\mu} q_T^{\nu}
	- 4 \bar{n}^{\mu} n^{\nu} \left(\hat{k}^+ - q_1^+ \right) 
	\left(\hat{l}'^- - q_2^- \right)
	\right.	\nn\\
	&\hspace{2em}\left.
	+2 \bar{n}^{\mu} q_T^{\nu} \left(\hat{k}^+ - q_1^+ \right) 
	-2 n^{\nu} q_T^{\mu} \left(\hat{l}'^- - q_2^- \right) 
	\right] \, ,
\label{eq:CB1}
\end{align}
where the momenta of internal propagators are given by
\begin{align}
\widetilde{k}_1  = \hat{k} - q_1, \quad
\widetilde{k}_2  = q_2 - \hat{l}', \quad
\widetilde{q} = \hat{k}' + \hat{l}.
\end{align}
Similar to eq.~\eqref{eq:A k1k2}, all the three internal propagators, including the gluon propagator, 
\bse\label{eq:A k1k2q}\begin{align}
\widetilde{k}_1^2 &=-2\hat{k}\cdot q_1
= -2 z_1 (p_1\cdot q_1),\\
\widetilde{k}_2^2 &=-2\hat{l}'\cdot q_2
= -2 (1-z_2) (p_2\cdot q_2),\\
\widetilde{q}^2 &=2\hat{k}'\cdot\hat{l}
= z_2 (1-z_1) s	,
\end{align}\ese
have their dependence on parton momentum fractions $z_{1,2}$ factored out from the external kinematic variables.   
This is actually true for all the diagrams in figure~\ref{fig:hard2}. 
Consequently, when the hard coefficient $C_{B}^{\mu\nu}$ is convoluted with DAs in eq.~(\ref{eq:factorization M pi pi}), 
the measured kinematic variables are factored out of the $z_1$ and $z_2$ integration.  Therefore, for the contribution from the Type-$B$ diagrams, changing external kinematics does not directly probe the functional form of DAs.  
That is, the contribution from the Type-$B$ diagrams in figure~\ref{fig:hard2} to the factorized scattering amplitude in eq.~(\ref{eq:factorization M pi pi}) is not directly sensitive to the functional form of DAs, but rather to their integrated values or some kind of ``moments''.   

The contribution from one single diagram, such as that in eq.~\eqref{eq:CA1} or in eq.~\eqref{eq:CB1}, is not gauge invariant, and does not have the invariant form in eq.~\eqref{eq:gauge-inv expansion}, while the sum over all the diagrams at the same order should be gauge invariant and can be organized in the form in eq.~\eqref{eq:gauge-inv expansion}.  Actually, 
the sum of all Type-$A$ diagrams and the sum of all Type-$B$ diagrams are gauge invariant separately, and each of them can be organized into the form in eq.~\eqref{eq:gauge-inv expansion}.  
We can get the contribution to the scalar coefficient $C_i$ in eq.~\eqref{eq:gauge-inv expansion} from each diagram by extracting the coefficient of $g^{\mu\nu}_{\perp}$, $p_1^{\mu} p_1^{\nu}$, $p_2^{\mu} p_2^{\nu}$, $p_1^{\mu} p_2^{\nu}$ and $p_2^{\mu} p_1^{\nu}$ sequentially. The terms containing $q_T^{\mu}$ or $q_T^{\nu}$ will be naturally organized such that we have the gauge-invariant form of eq.~\eqref{eq:gauge-inv expansion}, and can be used as a cross-checking.  
For example, for the diagram in figure~\ref{fig:hard11}, we have its contribution to each scalar coefficient $C_i$,
\bse\label{eq:scalar C A1}
\begin{align}
C_{0}^{(A1)}
& = e_u e_d \cdot \frac{s^2 }{(p_1\cdot q_2) q^2} \cdot
\frac{1}{(1-z_1)(1-z_2)},   \\
C_{4}^{(A1)}
& = - e_u e_d \cdot \frac{2 s^2}{(p_1\cdot q_2)(p_2\cdot q_1)}  \cdot
\frac{1}{(1-z_1)(1-z_2)},   \\
C_{1}^{(A1)}&=C_{2}^{(A1)}=C_{3}^{(A1)}=0,
\end{align}\ese
where we have omitted the $+i\varepsilon$ since for the simple $\pi^+\pi^-$ case, those poles happen at the end points where the DA vanishes, and it does not matter to which side of the $z_{1(2)}$ integration contour the poles lie.  The complete contribution to the scalar coefficients $C_i$ in eq.~\eqref{eq:gauge-inv expansion} from all diagrams are reorganized in compact forms and given in the Appendix, where the interplay between measured $q_T$ distribution and the $z$-dependence of the DA is further discussed.

Charge conjugation sets some useful relations among the results of individual diagrams. Applying charge conjugation on a diagram effectively exchanges the $u$ and $d$ quark lines, and can be visualized by simply reversing the fermion arrows and reassigning the parton momenta such that we still have $u$ quark carrying the momentum fraction $z_1$ or $z_2$. At the level of calculating Feynman diagrams, this simply reverses the order of the $\gamma$ matrices, which does not change the value of the Dirac trace. However, for the contraction with $\gamma_5\gamma^{\mp}$ on the $\pi^{\pm}$ sides, we need to reverse $\gamma^{\mp}\gamma_5$ back to $\gamma_5\gamma^{\mp}$, so one $\gamma_5$ would bring one extra minus sign, which leads to the following relations among the diagrams,
\begin{align}
&\{ 
C_{A1}, \, C_{A1'}, \,
C_{A3}, \, C_{A4} \}^{\mu\nu}
(z_1,z_2)
=\{
C_{A2'}, \, C_{A2}, \,
C_{A3'}, \, C_{A4'}\}^{\mu\nu}(1-z_1,1-z_2),
\label{eq:C conjugation A}
\end{align}
for Type-$A$ diagrams, and 
\begin{align}
&\{
C_{B1}, \, C_{B1'}, \,
C_{B2}, \, C_{B2'}, \,
C_{B3}, \, C_{B3'}
\}^{\mu\nu}(z_1,z_2) 
\nn \\
& =
\{
C_{B4}, \, C_{B4'}, \,
C_{B5}, \, C_{B5'}, \,
C_{B6}, \, C_{B6'}
\}^{\mu\nu}(1-z_1,1-z_2)
\bigg|_{e_u\leftrightarrow e_d}
\label{eq:C conjugation B}
\end{align}
for Type-$B$ diagrams, where we also need to exchange $e_u$ and $e_d$. Consequently, we have the symmetry
\beq
C^{\mu\nu}_A(z_1,z_2)=C^{\mu\nu}_A(1-z_1, 1-z_2),
\label{eq:CA symmetry}
\eeq
for Type-$A$ diagrams, while for $C_B$ this symmetry is broken by the difference of $e_u^2$ and $e_d^2$. These relations carry through to the scalar coefficients $C_i$ ($i = 0,\cdots, 4$), which can also serve as a useful check; see the Appendix for some more discussion.

\subsection{Exclusive differential cross section}

In this paper, we focus on the exclusive production of two unpolarized back-to-back photons in $\pi^+(p_1)\pi^-(p_2)$ collisions, and derive the differential cross section as follows, 
\begin{align}
\dd\sigma=\frac{1}{2s}\, \left|\overline{ \mathcal{M}} \right|^2 \,
\frac{\dd^3\bm{q}_1}{(2\pi)^3 2|\bm{q}_1|}\frac{\dd^3\bm{q}_2}{(2\pi)^3 2|\bm{q}_2|} 
(2\pi)^4 \,\delta^{(4)}(p_1+p_2-q_1-q_2)  \, ,
\label{eq:diffsigma}
\end{align}
where $\left|\overline{ \mathcal{M}} \right|^2$ represents scattering amplitude squared, with final-state photon polarizations summed,
\begin{align}
\left|\overline{ \mathcal{M}} \right|^2 =\sum_{\lambda \lambda'}
| \mathcal{M}_{\pi^+\pi^- \to \gamma(\lambda) \gamma(\lambda')} |^2
= g_{\mu\rho} \, g_{\nu\sigma} \,  \mathcal{M}^{\mu\nu}_{\pi^+\pi^- \to \gamma \gamma} \mathcal{M}^{\rho\sigma*}_{\pi^+\pi^- \to \gamma \gamma}  \, ,
\label{eq:T}
\end{align}
where $\sum_{\lambda} \varepsilon^{\lambda \,*}_\mu\, \varepsilon^{\lambda}_{\mu'} = -g_{\mu\mu'}$ was used. 

From the factorized scattering amplitude in eq.~\eqref{eq:M result}, and using
\begin{align}
\widetilde{p}_{1}^2 
&= - \frac{(p_1\cdot q_1)^2}{q_T^2}, \quad
\widetilde{p}_{2}^2 
= - \frac{(p_2\cdot q_1)^2}{q_T^2},  \quad
\bar{p}_{1}^2 
= - \frac{(p_1\cdot q_2)^2}{q_T^2},  \quad
\bar{p}_{2}^2 
= - \frac{(p_2\cdot q_2)^2}{q_T^2}, \nn\\
&\hspace{1.1in}
\widetilde{p}_{1}\cdot \widetilde{p}_{2}
=\bar{p}_{1}\cdot \bar{p}_{2}
=\frac{1}{2} p_1\cdot p_2 \, ,
\end{align}
we obtain the scattering amplitude square as
\begin{align}
\left|\overline{ \mathcal{M}} \right|^2
= & \left(\frac{e^2\,g^2\, f_{\pi}^2}{8s}\frac{C_F}{N_c}\right)^2 
\bigg[
|\mathcal{M}_0|^2 
+
\bigg|\frac{\mathcal{M}_1 + \mathcal{M}_2}{4}
- \frac{(p_1\cdot q_1)^2 \mathcal{M}_3 + (p_1\cdot q_2)^2 \mathcal{M}_4}{s\, q_T^2}
\bigg|^2 \bigg] \, ,
\label{eq:M2}
\end{align}
where $\mathcal{M}_i$ with $i=0,1,...,4$ can be factorized if $q_T \gg \Lambda_{\rm QCD}$ and are given in eq.~\eqref{eq:Mi}.  In eq.~\eqref{eq:M2}, the terms in the bracket are dimensionless and only functions of $\kappa=4q_T^2/s$. 
Therefore, from eq.~\eqref{eq:diffsigma} with the azimuthal angle of $\bm{q}_T$ integrated, we derive the differential cross section in $q_T$,  
\begin{align}
\frac{\dd\sigma}{\dd q_T^2}
= \frac{1}{\sqrt{1- 4q_T^2/s}} \frac{ 1}{16\pi s^2} \left|\overline{ \mathcal{M}} \right|^2  \, ,
\label{eq:qT xsec}
\end{align}
with $\left|\overline{ \mathcal{M}} \right|^2$ given in eq.~\eqref{eq:M2}.  The first factor in eq.~\eqref{eq:qT xsec} is the Jacobian peak.  The differential cross section could be smoother if one changes $q_T$-distribution to $\cos\theta$-distribution with $\theta$ being the angle between the direction of one of the observed photon and the collision axis.   One should note that the value of $q_T$ alone is not enough to completely specify an event, due to the ambiguity of whether $q_1$ is in the forward or backward direction of $p_1$, corresponding to the $\pm$ solutions in eq.~\eqref{eq:kin q12}. However, since the two photons are identical, the cross section must be the same for the forward and backward configurations. Therefore, we can take the forward solution in eq.~\eqref{eq:qT xsec}, without adding a $1/2$ factor to account for the factor that two photons are identical.  For the rest of this paper, we will stick to the forward solution of eq.~\eqref{eq:kin q12}.  We can also defined the integrated ``total'' cross section for $q_T^2\ge q^2_{T{\rm min}}$ as
\begin{align}
\sigma(q^2_{T{\rm min}})
\equiv \int_{q^2_{T{\rm min}}}^{s/4}
\frac{\dd\sigma}{\dd q_T^2}\, \dd q_T^2 \, ,
\label{eq:xsec}
\end{align}
with $q^2_{T{\rm min}} \gg \Lambda^2_{\rm QCD}$ to ensure the factorization.  In our numerical estimate below, we choose 
$q^2_{T{\rm min}} =1$~GeV$^2$.

\section{Exclusive production of a pair of high transverse momentum photons in diffractive meson-baryon scatterings}
\label{s.pih2gg}

Having explained the main steps in factorizing the amplitude for the exclusive photon-pair production in the $\pi^+\pi^-$ annihilation, 
we now generalize the factorization formalism to an exclusive process involving diffractive scattering of a nucleon $N$ of momentum $p$, 
\beq
\pi(p_2) + N(p) \to \gamma(q_1)+\gamma(q_2) + N'(p'),
\label{eq:p pi process}
\eeq
where $N$ can be a neutron ($n$) or a proton ($p$) and $\pi$ can be $\pi^+$ or $\pi^-$, making up various exclusive processes, such as, $n\,\pi^+\to p\gamma\gamma$, $p\,\pi^-\to n\gamma\gamma$, $p\,\pi^-\to \Lambda^0\gamma\gamma$,   
and those that could be measured with a pion beam at J-PARC and other facilities. 
The pion beam can also be replaced by a kaon beam and make up more processes.
As shown in figure~\ref{figureempp}(c), the exclusive process, $p\,\pi^-\to n\gamma\gamma$, could be made analogous to the $\pi^+\pi^-$ collision by thinking of the $p\to n$ transition 
as taking a virtual $\pi^+$ out of the proton, carrying momentum $\Delta = p - p'$ and colliding with $\pi^-$ to produce two hard photons exclusively. 
Nevertheless, the factorization cannot be trivially adapted, because that analogy only corresponds to the ERBL region of GPD for which all the active partons from the nucleon enter into the hard part and the soft gluon momentum is not pinched in the Glauber region. Now that we are considering diffractive scattering, the presence of the spectator particles from the $N\to N'$ transition implies another kinematic region where some partons enter into the hard part and then come out to recombine with the spectators to form the diffracted $N'$. This corresponds to the DGLAP region for GPD, in which the soft gluon momentum can be partly trapped in the Glauber region. Additional argument is needed for the factorization proof, which will be given in subsection~\ref{s.proton-factorize}.

\subsection{Kinematics}
\label{sec:kin}

In the lab frame, the nucleon (pion) is moving along $+\hat{z}$ ($-\hat{z}$) direction, carrying a large plus (minus) momentum. 
Two photons with large and opposite transverse momenta are produced in the final state, together with a recoiled nucleon, or a baryon in general.  We focus on the region of phase space where $-t = -\Delta^2 \ll (q_1+q_2)^2 $.  That is, we require that the proton be recoiled in an approximately collinear direction and the invariant mass of the momentum transfer $\Delta$ be much smaller than the energy of this transfer. This is the condition that allows the scattering amplitude of the exclusive process in eq.~\eqref{eq:p pi process} to be factorized into a transition GPD of the nucleon. 

Since $\Delta$ carries a small transverse component and sufficiently large longitudinal component, it is convenient for our analysis to boost the lab frame into the CM frame of $\Delta$ and $p_2$ where $\vec{\Delta}$ is along $+\hat{z}$ direction, which is also the rest frame of $q_1$ and $q_2$, so that the $\Delta$ could mimic the momentum $p_1$ of $\pi^+$ in the Sec.~\ref{s.toy}. We denote this frame as photon frame $S_{\gamma}$, distinguished from the lab frame $S_{\rm lab}$.  The transformation from $S_{\rm lab}$ to $S_{\gamma}$ can be done by first boosting along $\vec{\Delta}_T$ such that $\vec{\Delta}$ is in parallel and head-to-head with $\vec{p}_2$, followed by a rotation in the $\vec{\Delta}_T$-$\vec{p}_2$ plane to make $\vec{\Delta}$ along $+\hat{z}$ direction.

In the $S_{\gamma}$ frame, we have the momentum conservation, 
\beq
\Delta + p_2 = q_1 + q_2 \, ,
\label{eq:mom_conv}
\eeq
and the CM collision energy square $\hat{s} \equiv (\Delta + p_2)^2 = (q_1 + q_2)^2 \gg \Lambda_{\rm QCD}^2$.   For the leading power contribution, in analog to eq.~\eqref{eq:kin}, we can parametrize $\Delta$ and $p_2$ as
\begin{align}
\Delta & = \left( \Delta^+, \, \frac{t}{2\Delta^+}, \, \bm{0}_T \right)_{\gamma}
		\simeq \left( \Delta^+, \, 0, \, \bm{0}_T \right)_{\gamma}		\, ,\nn\\
p_2 & = \left( \frac{m_{\pi}^2 }{2 p_2^-}, \, p_2^-, \bm{0}_T \right)_{\gamma}
		\simeq \left( 0, \, p_2^-, \, \bm{0}_T \right)_{\gamma}	\, ,
\label{eq:kin2}
\end{align}
where in the second step (and in the following) the ``$\simeq$'' means the neglect of small quantities suppressed by powers of $m_{\pi}^2/Q^2$ or $t/Q^2$ with the hard scale $Q\sim {\cal O}(q_T) \lesssim \sqrt{\hat{s}}$. In addition, we also implicitly take the {\it rescaled} light-like $\Delta$ and $p_2$ as the momenta entering the hard process, with
\beq
\Delta^+ = p_2^- = \sqrt{\frac{(\Delta + p_2)^2}{2}} = \sqrt{\frac{\hat{s}}{2}}\, ,
\eeq
to keep the momentum conservation manifest, which is useful for the factorization of this process.

The skewness in the lab frame is defined as 
\beq
\xi = \frac{\Delta_{\rm lab}^+}{ 2 P_{\rm lab}^+},
\eeq
where $P = (p+ p')/2$. We then have $p_{\rm lab}^+ = (1+\xi)P_{\rm lab}^+$, $p'^+_{\rm lab} = (1-\xi)P_{\rm lab}^+$, and 
\beq
\hat{s} = (\Delta + p_2)^2 
\simeq 2 \Delta^+ p_2^- 
= \frac{2\xi}{1+\xi} (2p^+ p_2^-)_{\rm lab}
\simeq \frac{2\xi}{1+\xi}\, s \, ,
\label{eq:S s}
\eeq
which defines a unique role of the skewness, quantifying the momentum flowing into the hard process from the colliding hadron of momentum $p$.

The invariant mass squared of the momentum transfer, $t = \Delta^2$, can be related to $\xi$ and the transverse component of the momentum transfer $\Delta_T$ by
\beq
t =- \left( \frac{4\xi^2}{1-\xi^2} m_p^2 + \frac{1+\xi}{1-\xi} \Delta_T^2 \right),
\label{eq:t-DeltaT}
\eeq
where $m_p$ is the proton mass and we neglect the mass difference between proton and neutron.
For a given small $t$, $\Delta_T$ is bounded to be small, and $\xi$ is effectively constrained by
\beq
0 < \xi \leq \sqrt{\frac{-t/m_p^2}{4 - t /m_p^2}} \, .
\label{eq:xi range t}
\eeq

Every event of the exclusive process in eq.~\eqref{eq:p pi process} is specified by three momenta $p'$, $q_1$ and $q_2$, which are constrained by on-shell conditions and momentum conservation, leading to $9-4=5$ degrees of freedom in kinematics. $\bm{\Delta}_T$ and $\xi$ are sufficient to specify the neutron momentum. The photon momenta are to be described by $\bm{q}_T$ in the photon frame $S_{\gamma}$, where they are back-to-back. That is, $(\bm{\Delta}_T, \xi, \bm{q}_T)$ fixes all the kinematics. Our process is insensitive to azimuthal angle in either $\bm{\Delta}_T$ or $\bm{q}_T$, and we will integrate out these angles, leaving only three degrees of freedom, $\Delta_T$, $\xi$ and $q_T$, or equivalently, $t$, $\xi$ and $q_T$ as independent variables.

\subsection{Factorization}
\label{s.proton-factorize}

\begin{figure}[htbp]
    \centering
    \includegraphics[scale=0.75]{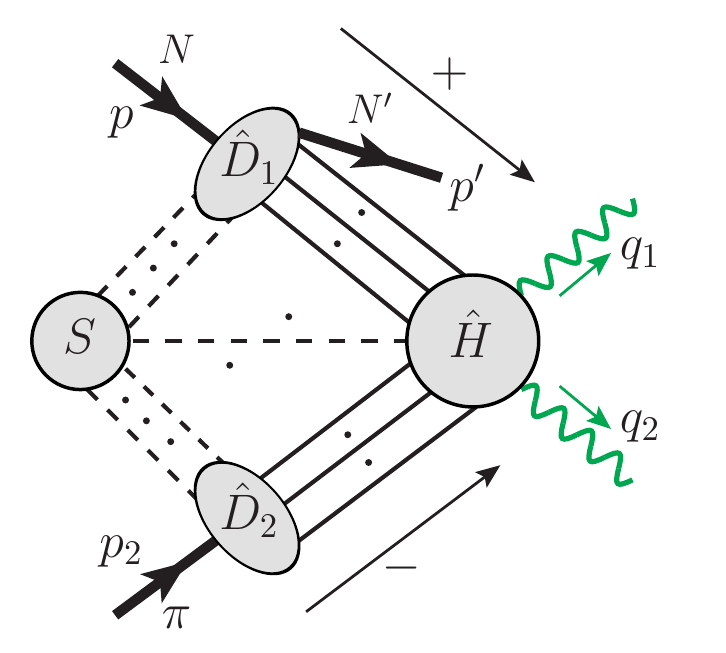}
    \caption{The general pinch-singular surface for the process \eqref{eq:p pi process}.}
    \label{fig:p-pi-PSS}
\end{figure}

We generalize the factorization formula derived in Sec.~\ref{sec:factorization1} to describe the scattering amplitude of the exclusive process in eq.~\eqref{eq:p pi process}. As indicated by the general pinch-singular surface in figure~\ref{fig:p-pi-PSS}, the initial-state nucleon momentum $p$ and slightly recoiled hadron momentum $p'$ define the direction of a collinear subgraph, $\hat{D}_1$, which is joined by a set of collinear parton lines to the hard subgraph, from which two photons with large transverse momenta are produced. 

The power counting for a pinch surface is derived in the same way as what was done in the last section. The only difference is that the dimension for the $\hat{D}_1$ is reduced by 1 because we have an extra external final-state hadron line connected to $\hat{D}_1$ in figure~\ref{fig:p-pi-PSS}. Like eq.~\eqref{eq:scale1}, we obtain the scaling behavior for corresponding reduced diagram as
\begin{align}
    {\cal M}_{N\pi\to N'\gamma\gamma}\sim
    \hat{H}\otimes \hat{D}_1\otimes \hat{D}_2\otimes  S \
    \propto\
    \lambda^{\alpha-1} 	\, ,
\label{eq:pc p pi}
\end{align}
where $\alpha$ is the same as that in eq.~\eqref{eq:pc1}.  With the minimum power $\alpha=2$, we obtain the leading pinch surfaces to the scattering amplitude of exclusive process in eq.~\eqref{eq:p pi process}, as shown in figure~\ref{fig:p pi LR}, which are slightly modified from those in figure~\ref{fig:pipi LR}.  Due to the electric charge or isospin exchange, $\hat{D}_1$ or $\hat{D} _2$ must be connected to other subdiagrams by at least two quark lines. By the same argument at the end of Sec.~\ref{sec:rd}, the pinch surface in figure~\ref{fig:p-pi LR2} is power suppressed compared to that in figure~\ref{fig:p-pi LR1}.

\begin{figure}[htbp]
\begin{center}
\begin{subfigure}{.35\textwidth}
  \centering
  \includegraphics[scale=0.7]{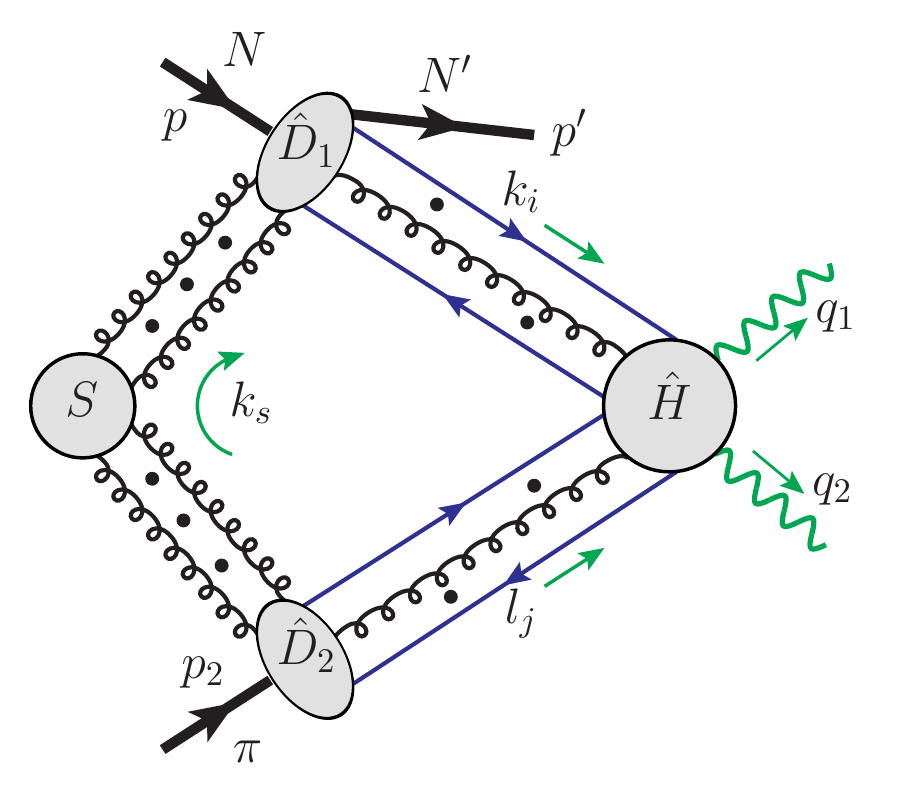}  
  \caption{}
  \label{fig:p-pi LR1}
\end{subfigure}
{\hskip 0.6in}
\begin{subfigure}{.35\textwidth}
  \centering
  \includegraphics[scale=0.7]{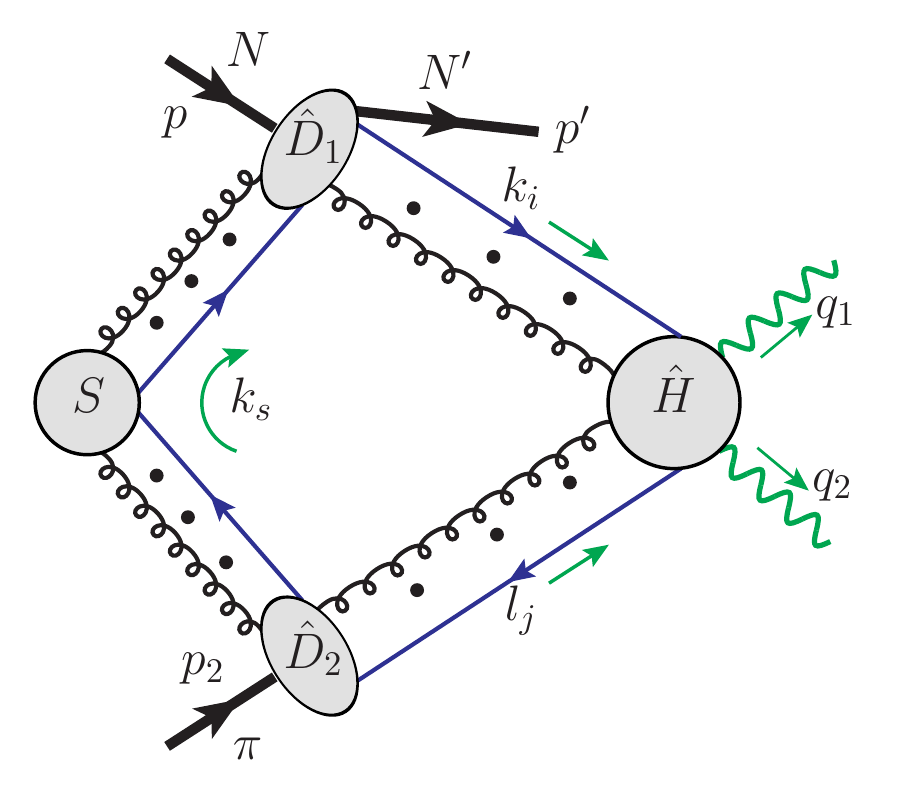}  
  \caption{}
  \label{fig:p-pi LR2}
\end{subfigure}
\caption{Two possible leading regions for the process \eqref{eq:p pi process}. Arbitrary number of gluons can connect to the collinear subgraphs $\hat{D}_1$ or $\hat{D}_2$ from $S$ or $\hat{H}$, but they have to be longitudinally polarized. }
\label{fig:p pi LR}
\end{center}
\end{figure}

\subsubsection{Deformation out of Glauber region}

Before we adopt the approximations listed in Sec.~\ref{sec:approx} to start our factorization arguments, we note one complication that distinguishes the diffractive meson-baryon process in eq.~\eqref{eq:p pi process} from the $\pi^+\pi^-$ case discussed in the last section. 

The factorization proof of $\pi^+\pi^-$ process was simplified by the fact that all the collinear parton lines go from the \textit{past} to \textit{now} when the hard collision 
takes place, without going to the future as spectators, as shown in figure~\ref{fig:soft-momentum}. All the parton lines collinear to $\pi^+$ ($\pi^-$) have positive plus (minus) momenta, and the plus/minus momenta of the soft gluons are not trapped to be much smaller than their transverse components.  We can get those soft gluons out of the Glauber region by deforming the contours of their momentum integrations, as discussed in Sec.~\ref{sec:approx}.
However, in the $\pi N$ case, or more specifically, in $p \pi^- \to n\gamma\gamma$ case, the proton-neutron transition can have either (i) all the collinear parton lines 
going from the proton \textit{into} the hard part, as shown in figure~\ref{fig:p-pi-xt1}, or (ii) some parton lines 
going from the proton \textit{into} the hard part, but others  
going to the future as spectators and  
merging with partons coming out the hard part to form a neutron, as shown in figure~\ref{fig:p-pi-xt2}. 
The type (i) corresponds to the ERBL region of GPD, and the type (ii) is for the DGLAP region. 

\begin{figure}[htbp]
\begin{center}
\begin{subfigure}{.38\textwidth}
  \centering
  \includegraphics[scale=0.65]{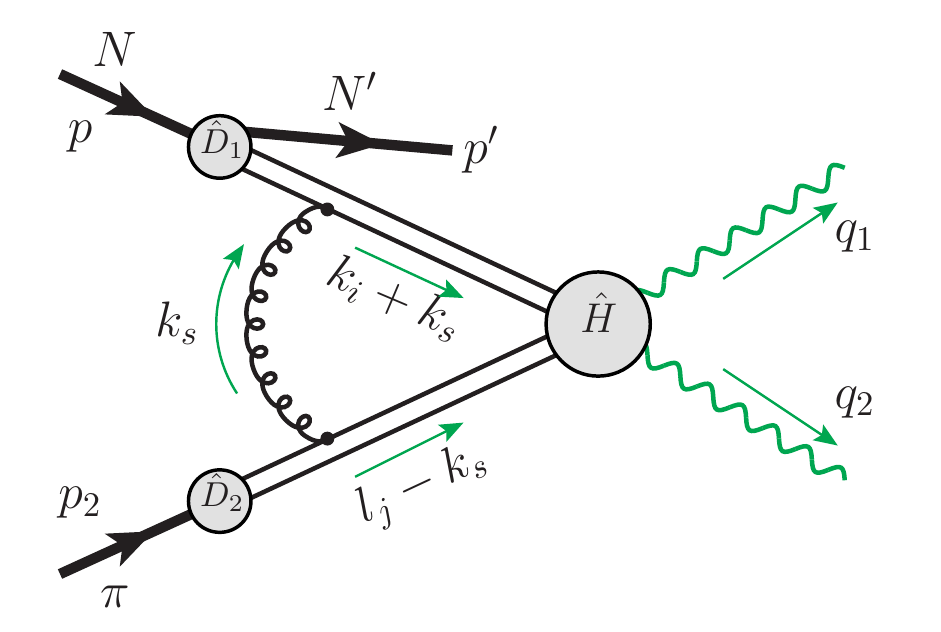}  
  \caption{}
  \label{fig:p-pi-xt1}
\end{subfigure}
{\hskip 0.3in}
\begin{subfigure}{.48\textwidth}
  \centering
  \includegraphics[scale=0.65]{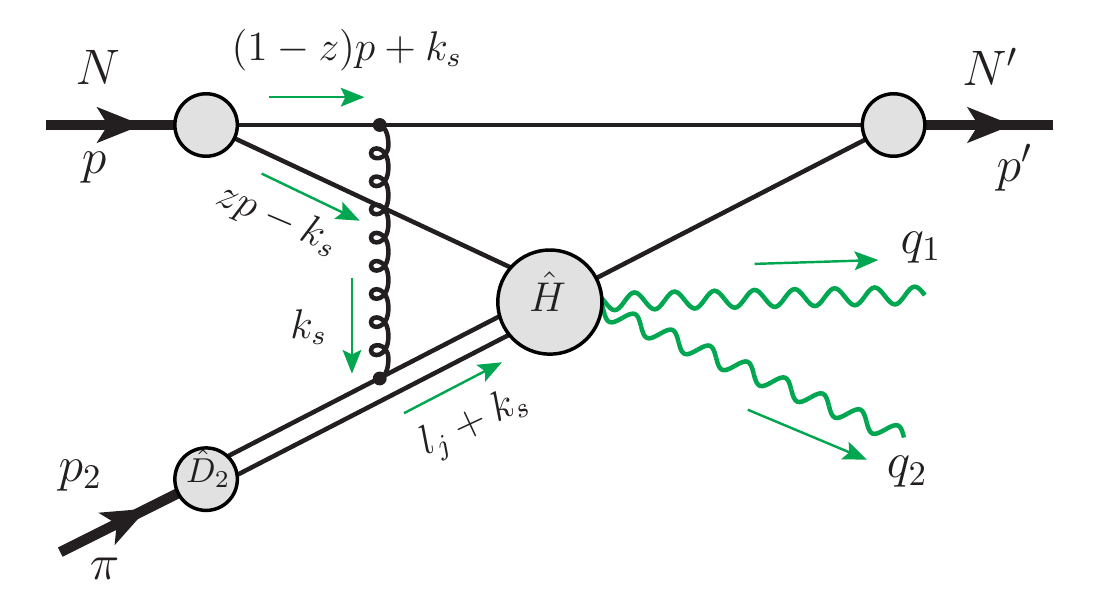}  
  \caption{}
  \label{fig:p-pi-xt2}
\end{subfigure}
\caption{
Difference in soft gluon interaction between ERBL region (a) and DGLAP region (b) in the elastic $\pi N$ process. In (a), the $k_s^-$ integration is not pinched, while in (b), the $k_s^-$ integration is pinched to be in the Glauber region.  
}
\label{fig:p-pi-xt}
\end{center}
\end{figure}

For the ERBL region, the contour deformations and approximations made to the leading regions for every possible diagram are the same as those in Sec.~\ref{sec:approx}. But for the DGLAP region, the presence of proton spectator may trap the minus momenta of soft gluons at small values.  
For example, as shown in figure~\ref{fig:p-pi-xt2}, the attachment of a soft gluon of momentum $k_s$ to a spectator of the colliding proton leads to two propagators with the denominators,
\begin{align}
((1-z)p + k_s)^2 +i\varepsilon 
\approx \
& 2(1-z)p^+ k_s^- -{\bm k}_{sT}^2+i\varepsilon \, ,
\nn \\
&\Rightarrow \quad 
k_s^- \mbox{ pole } \approx \frac{{\bm k}_{sT}^2}{2(1-z)p^+ } - i\varepsilon 
\quad  \to \quad
{\cal O}(\lambda^2 Q) - i\varepsilon \, , 
\nn \\
(z p - k_s)^2 +i\varepsilon 
\approx \ 
& 2zp^+(-k_s^-) -{\bm k}_{sT}^2+i\varepsilon \, ,
\nn \\
&\Rightarrow \quad 
k_s^- \mbox{ pole } \approx - \frac{{\bm k}_{sT}^2}{2zp^+} 
+ i \varepsilon 
\quad  \to \quad
-{\cal O}(\lambda^2 Q) + i\varepsilon \, ,
\label{eq:glauber}
\end{align}
which pinch the $k_s^-$-integration of the soft gluon momentum $k_s$ to be ${\cal O}(\lambda^2 Q)$ when $k_{sT}={\cal O}(\lambda Q)$ and trap the $k_s^-$ in the Glauber region.  The same conclusion arrives if we let $k_s$ flow through $N'(p')$ in figure~\ref{fig:p-pi-xt2}. Therefore, the argument that we used in Sec.~\ref{sec:approx} to deform the contours of plus/minus components of soft gluon momenta to get them out of the Glauber region does not work for the soft minus components in the $\pi N$ case when the nucleon $N$ is moving in the ``+'' direction.

Luckily, the poles for the plus components of the soft momenta are solely provided by the collinear lines from the $\pi$, which all go \textit{into} the hard part with positive minus momenta.  All the poles from $l_j + k_s$ lie on the same half plane, and therefore, we can deform $k_s^+$ as 
\beq
k_s^+ \to k_s^+ + i\mathcal{O}(p^+),
\eeq
when it lies in the Glauber region flowing into the $\hat{D}_2$ subgraph. This is the maximal extent that we can deform $k_s^+$, which leads the soft momentum $k_s$ all the way into $\hat{D}_1$-collinear region.  That is, the soft Glauber mode is deformed to be a collinear mode, which is only possible when all the collinear lines from the $\pi$ flow into the hard part $\hat{H}$. 
Had we considered an exclusive double diffractive process: $pn\to pn\gamma\gamma$, with a pair of back-to-back high transverse momentum photons produced while the nucleons are slightly diffracted, we would have both plus and minus components of soft momenta pinched in the Glauber region, which forbids the double diffractive processes, like $pn\to pn\gamma\gamma$, 
$p\bar{p}\to p\bar{p}+{\rm jet}+{\rm jet}$, etc.,  
to be factorized into two GPDs and a hard part~\cite{Soper:1997gj}, 
even though there is indeed a hard scale provided by the transverse momenta of the photons or the jets.

After the deformation of Glauber gluons, we can apply all the approximations in Sec.~\ref{sec:approx}. Since we will not deform $k_s^-$ in DGLAP region, it does not matter what $i\varepsilon$ prescription we assign to $k_s^-$. We choose the same convention as in Sec.~\ref{sec:approx} to be compatible with ERBL region, for which we do need to deform $k_s^-$.

\subsubsection{Soft cancellation and factorization}

We first use the same arguments presented in the last section to factorize the collinear subgraph $\hat{D}_2$ from the hard part $\hat{H}$ and the soft factor $S$.  The approximation in eqs.~\eqref{eq:k2H} and \eqref{eq:HC2} allows us to use Ward identity 
to detach all longitudinally polarized collinear gluons of $\hat{D}_2$ from the hard part $\hat{H}$, and factorize them into Wilson lines along $w_1$, as shown in figure~\ref{fig:p-pi-ward-col}. Like the $\pi^+\pi^-\to \gamma\gamma$ case in the last section, the Wilson lines connected to $\hat{D}_2$ point to the past due to the choice of $i\varepsilon$ in eq.~\eqref{eq:HC2}.

\begin{figure}[htbp]
    \centering
	\includegraphics[scale=0.65]{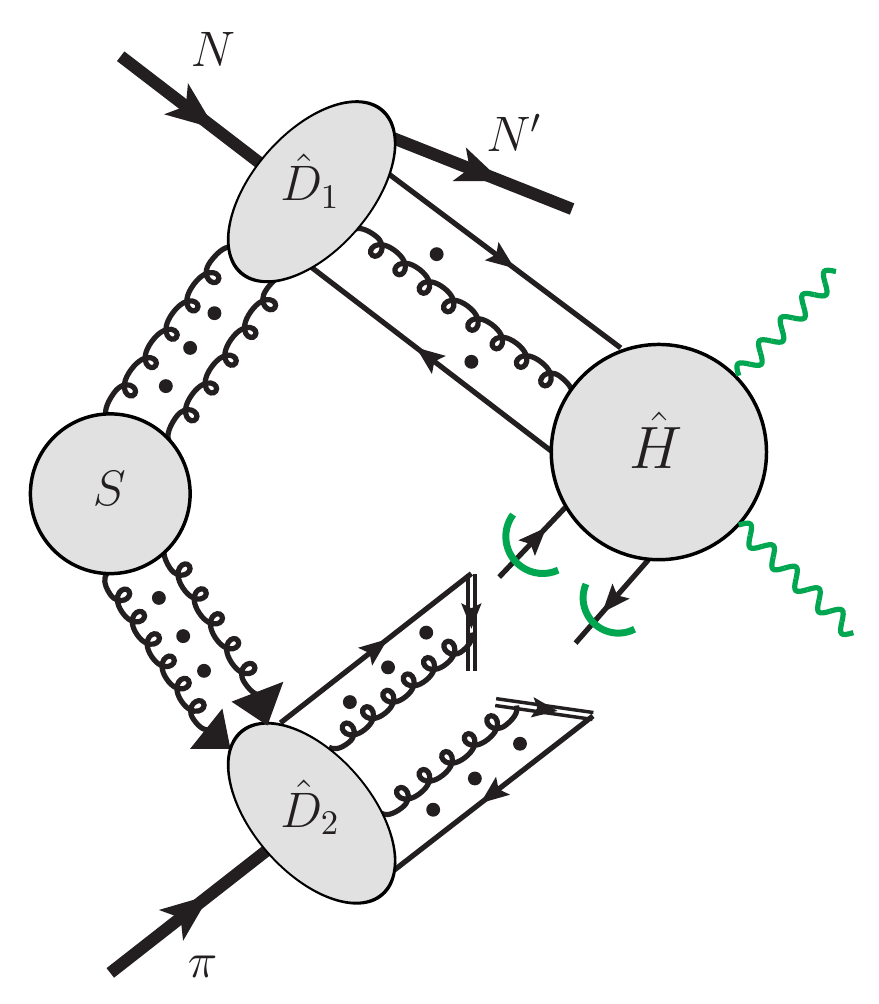}
    \caption{The result of using Ward identity for $\hat{D}_2$-collinear gluons. The Wilson lines point along $w_1$ to the past. The notations are similar to figure~\ref{fig:ward-col}.}
    \label{fig:p-pi-ward-col}
\end{figure}

\begin{figure}[htbp]
    \centering
	\includegraphics[scale=0.65]{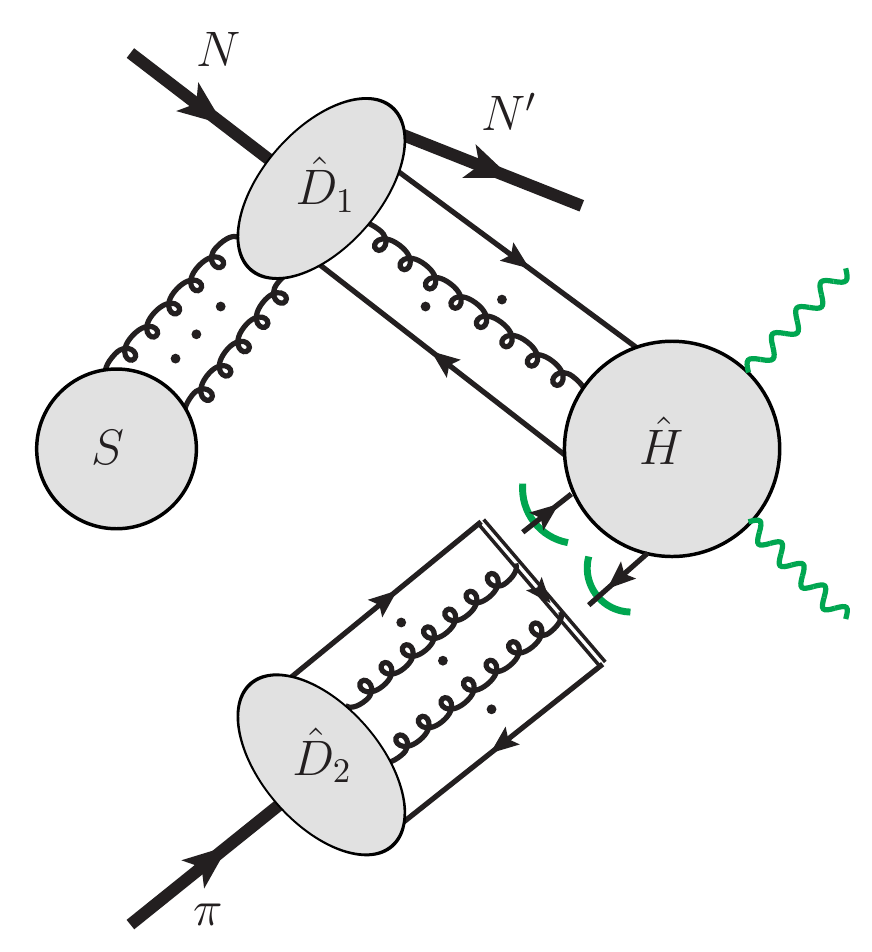}
    \caption{The result of using Ward identity for soft gluons coupling to $\hat{D}_2$. Those gluons cancel. The Wilson line is along $w_1$.}
    \label{fig:p-pi-ward-soft}
\end{figure}

\begin{figure}[htbp]
    \centering
	\includegraphics[scale=0.85]{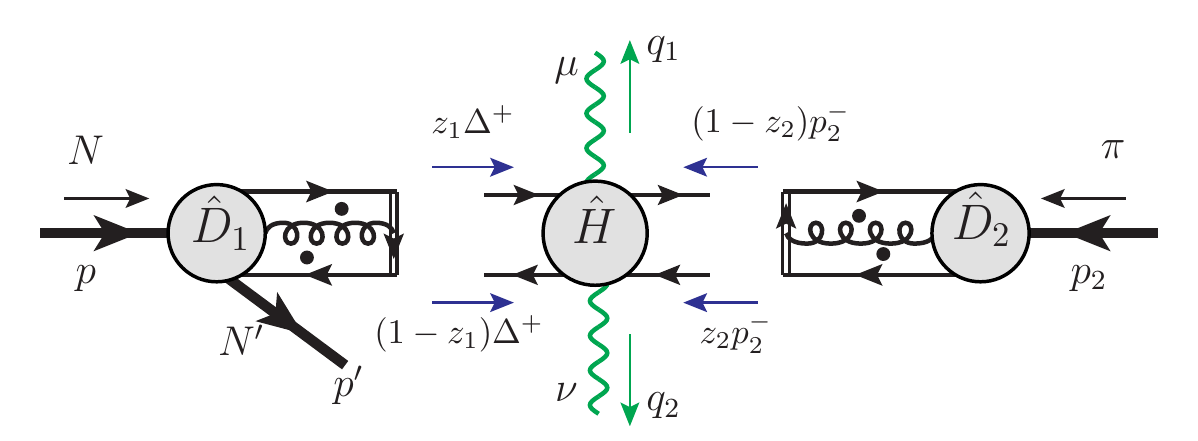}
    \caption{The factorized form for the process \eqref{eq:p pi process}.}
    \label{fig:p-pi factorize}
\end{figure}

Next, having eqs.~\eqref{eq:ks2} and \eqref{eq:SC2}, we can use Ward identity to factorize soft gluons out of the collinear factor $\hat{D}_2$. This leaves the collinear factor $\hat{D}_2$ uncoupled to $\hat{D}_1$, so that $\hat{D}_2$ ends up being color singlet. By the same method of Sec.~\ref{sec:soft_cancel}, the soft gluons coupling to $\hat{D}_2$ cancel. The rest of the soft gluons only couple to $\hat{D}_1$, as in figure~\ref{fig:p-pi-ward-soft}, and can be grouped into $\hat{D}_1$.

We can then use eqs.~\eqref{eq:k1H} and \eqref{eq:HC1}, and the Ward identity to factorize all longitudinally polarized collinear gluons of $\hat{D}_1$ out of the hard part $\hat{H}$. This step is similar to that of the $\pi^+\pi^-$ case, since the soft gluon connection to $\hat{D}_2$ has been canceled, which would have pinched the minus components of soft gluon momenta into the Glauber region.  After factorizing the longitudinally polarized collinear gluons from the $\hat{H}$ into Wilson line, we get a color singlet $\hat{D}_1$.  Therefore, we complete the factorization arguments and have a factorized result, as shown in figure~\ref{fig:p-pi factorize}. 
The color structure of the hard part takes the same form as in eq.~\eqref{eq:color convo3}. But, the spinor indices are still convoluted between $\hat{D}_1$ and $\hat{H}$, as well as between $\hat{D}_2$ and $\hat{H}$, and will be dealt with in next subsection.

\subsubsection{Factorization formula}

Similar to eq.~\eqref{eq:color convo3}, we derived the factorized formalism for the scattering amplitude of the exclusive process in \eqref{eq:p pi process}, corresponding to the factorized diagram in figure~\ref{fig:p-pi factorize},
\begin{align}
    \mathcal{M}^{\mu\nu}_{N\pi\to N'\gamma\gamma}
    =&
    \int \dd z_1 \dd z_2 \,{\rm Tr}\Big\{ 
    [\mathcal{P}_A\, \hat{D}_1(z_1,p,p')\, \mathcal{P}_B]
    [\mathcal{P}_B\, \hat{D}_2(z_2,p_2)\, \mathcal{P}_A]
    \nn\\
&\hspace{0.7in}\times
\Big[ \frac{1}{N_c^2} \hat{H}^{\mu\nu}_{ii;mm}(k_1^+=z_1\,  \Delta^+;k_2^-=z_2\, p_2^-;q_1,q_2;\mu) \Big] \Big\}   ,
\label{eq:M p pi}
\end{align}
where the repeated color indices, $i$ and $m$ are summed, and averaged with the factor $1/N_c^2$, and the ``${\rm Tr}$'' indicates the trace over all spinor indices between $\hat{D}_1$, $\hat{D}_2$, and $\hat{H}$.  In eq.~\eqref{eq:M p pi},
$\mathcal{P}_A$, $\mathcal{P}_B$, and $\hat{D}_2(z_2,p_2)$ are the same as those in eq.~\eqref{eq:color convo3},
but $\hat{D}_1(z_1,p,p')$ is different, which now represents the transition GPD of the nucleon $N$,
\begin{align}
&\hat{D}_{1}(z_1,p,p')_{\alpha\beta}
= \int\frac{\dd (\Delta^+y^-)}{2\pi} e^{iz_1\Delta^+ y^-}
\langle N'(p') | \bar{d}_{\beta}(0) 
\Phi(0,y^-;w_2)\, u_{\alpha}(y^-) | N(p) \rangle
 \\
&\quad = \int\frac{\dd (\Delta^+y^-)}{2\pi} e^{i(2z_1-1)\xi P^+ y^-}
\langle N'(p') | \bar{d}_{\beta} \left(-\frac{y^-}{2}\right) 
\Phi\left(-\frac{y^-}{2},\frac{y^-}{2};w_2\right)\, u_{\alpha}\left(\frac{y^-}{2}\right) | N(p) \rangle	\, ,
\nn
\end{align}
where $\alpha,\beta$ are spinor indices, $w_2$ is as in eq.~\eqref{eq:aux vec}, color indices have been implicitly summed, and in the second line, we shifted the position of the operator to be consistent with the convention in~\cite{Diehl:2003ny}. Now $\mathcal{P}_A$ and $\mathcal{P}_B$ sandwiching $\hat{D}_1$ picks out only the term proportional $\gamma^-$, $\gamma^-\gamma_5$ or $\gamma^-\gamma_5\gamma_i$. Because of helicity conservation, the transversity GPD associated with $\gamma^-\gamma_5\gamma_i$ does not contribute at leading power. Effectively, we have 
\begin{align}
&\left[\mathcal{P}_A \hat{D}_1(z_1,p, p') \mathcal{P}_B\right]_{\alpha\beta} 	
\nn \\
& {\hskip 0.2in} = 
\frac{1}{2\Delta^+}{\rm Tr}\left[ \gamma^+ \hat{D}_1 \right] 
\left[\frac{1}{2}(\Delta^+\gamma^-) \right]_{\alpha\beta}
+
\frac{1}{2\Delta^+}{\rm Tr}\left[ \gamma^+ \gamma_5 \hat{D}_1 \right] 
\left[\frac{1}{2}\gamma_5(\Delta^+\gamma^-) \right]_{\alpha\beta}
\nn \\
& {\hskip 0.2in} \equiv 
{\cal F}^{ud}_{NN'}(z_1, \xi, t)\left[ \frac{1}{2}(\Delta^+\gamma^-) \right]_{\alpha\beta}
	+ \widetilde{\cal F}^{ud}_{NN'}(z_1, \xi, t) \left[\frac{1}{2}\gamma_5(\Delta^+\gamma^-) \right]_{\alpha\beta}  \, ,
\label{eq:PA C1 PB}   
\end{align}
where ${\cal F}^{ud}_{NN'}(z_1, \xi, t)$ and $\widetilde{\cal F}^{ud}_{NN'}(z_1, \xi, t)$ are 
GPDs with different chirality
characterizing the amplitude for the transition of hadron $N$ to $N'$,
\bse \label{eq:GPD def}
\begin{align}
&{\cal F}^{ud}_{NN'}(z_1, \xi, t)
=
\int\frac{\dd y^-}{4\pi} e^{iz_1\Delta^+ y^-}
\langle N'(p') | \bar{d}(0) \gamma^+ 
\Phi(0, y^-;w_2)\, u(y^-) | N(p) \rangle 
\\
&\hspace{2em}
=
\int\frac{\dd y^-}{4\pi} e^{i(2z_1-1)\xi P^+ y^-}
\langle N'(p') | \bar{d}\left(-\frac{y^-}{2}\right) \gamma^+
\Phi \left(-\frac{y^-}{2},\frac{y^-}{2};w_2\right)\, u\left(\frac{y^-}{2}\right)  | N(p) \rangle 
\nn \\
&\hspace{2em}= F^{ud}_{NN'}(x=(2z_1-1)\xi, \xi, t)\, ,
\label{eq:gpd_z2x}\\
&\widetilde{\cal F}^{ud}_{NN'}(z_1, \xi, t)
=
\int\frac{\dd y^-}{4\pi} e^{iz_1\Delta^+ y^-}
\langle N'(p') | \bar{d}(0) \gamma^+\gamma_5 
\Phi(0,y^-;w_2)\, u(y^-) | N(p) \rangle 
\\
&\hspace{2em}
=
\int\frac{\dd y^-}{4\pi} e^{i(2z_1-1)\xi P^+ y^-}
\langle N'(p') | \bar{d}\left(-\frac{y^-}{2}\right) \gamma^+\gamma_5 
\Phi\left(-\frac{y^-}{2},\frac{y^-}{2};w_2\right)\, u\left(\frac{y^-}{2}\right)  | N(p) \rangle 
\nn \\
&\hspace{2em}
= \widetilde{F}^{ud}_{NN'}(x=(2z_1-1)\xi, \xi, t)\, ,
\label{eq:gpd_z2xa}
\end{align}
\ese
where $F^{ud}_{NN'}(x, \xi, t)$ and $\widetilde{F}^{ud}_{NN'}(x, \xi, t)$ are the GPDs defined with the convention in Ref.~\cite{Diehl:2003ny}.
Note that we are using an unusual variable $z_1$ to label the momentum fraction of an active parton ($u$ quark here), as indicated in figure~\ref{fig:p-pi factorize}, in order to have a direct analogy to the $\pi^+\pi^-$ process that we studied in the last section.  As clearly indicated in eqs.~\eqref{eq:gpd_z2x} and \eqref{eq:gpd_z2xa}, the momentum fraction $z_1$ is closely related to the common variables of GPDs, such as $x$ and $\xi$, 
\beq
z_1 = \frac{x+\xi}{2\xi}.
\label{eq:z_1}
\eeq
Consequently, the range of $z_1$ is different from $[0,1]$ for the nucleon side, as opposed to $z_2$ for the $\pi$, and is given by
\beq
z_m \equiv \frac{-1+\xi}{2\xi} \leq z_1 \leq \frac{1+\xi}{2\xi} \equiv z_M.
\label{eq:z1limits}
\eeq
The choice of $z_1$ parameter highlights the so-called ERBL region, which lies between $-\xi < x < \xi$, and is now given by $0<z_1<1$. In this region, a pair of quark and antiquark with positive momentum fractions enters the hard scattering.  On the other hand, one of the DGLAP regions $\xi < x < 1$ with a quark scattering configuration corresponds to $1<z_1<(1+\xi)/2\xi$, while the other DGLAP region $-1 < x < -\xi$ with an antiquark scattering configuration becomes $-(1-\xi)/2\xi < z_1 < 0$.

Inserting eqs.~\eqref{eq:PA C1 PB} and \eqref{eq:PB D2 PA} into eq.~\eqref{eq:M p pi} we obtain the factorized scattering amplitude for the elastic process in eq.~\eqref{eq:p pi process}
\begin{align}
\mathcal{M}^{\mu\nu}
=& 
    \int_{z_m}^{ z_M} \dd z_1 \int_0^1 \dd z_2 \,
    \Big[ \widetilde{\cal F}^{ud}_{NN'}(z_1,\xi,t) D_{\pi^-}(z_2) C^{\mu\nu}(z_1,z_2) 
        \nn\\
 & {\hskip 1.1in}
 + {\cal F}^{ud}_{NN'}(z_1,\xi,t) D_{\pi^-}(z_2) \widetilde{C}^{\mu\nu}(z_1,z_2) \Big]	\, ,
\label{eq:factorization M p pi}
\end{align}
where $C^{\mu\nu}$ is the same as that in eq.~\eqref{eq:hard C} with $p_1^+$ replaced by $\Delta^+$, which has $\gamma_5$ attached on both proton and pion sides so is chiral even, while $\widetilde{C}^{\mu\nu}$ is given by
\beq
\widetilde{C}^{\mu\nu}(z_1,z_2)
\equiv 
{\rm Tr}\left[ \frac{\Delta^+\gamma^-}{2} H^{\mu\nu}(\hat{k}_1,\hat{k}_2;q_1,q_2;\mu)
    \frac{\gamma_5(p_2^-\gamma^+)}{2}
    \right] \, ,
\label{eq:hard C odd}
\eeq
which only has one $\gamma_5$ on the pion side and is referred as chiral odd. The correction to the factorized scattering amplitude in eq.~\eqref{eq:factorization M p pi} is suppressed by an inverse power of the high transverse momentum of observed photon $q_T$ in $S_\gamma$. 

The hard coefficients $C^{\mu\nu}$ and $\widetilde{C}^{\mu\nu}$, and the factorized formalism in eq.~\eqref{eq:factorization M p pi} are manifestly invariant under a boost along $\hat{z}$. Since the transformation from $S_{\rm lab}$ to $S_{\gamma}$ is only by a boost along $\hat{z}$, up to a boost and rotation characterized by $\Delta_T$, which is neglected at leading power, the factorization formula \eqref{eq:factorization M p pi} takes the same form in the $S_{\gamma}$ frame, and the hard coefficients $C^{\mu\nu}$ and $\widetilde{C}^{\mu\nu}$ can be calculated in $S_{\gamma}$, in the same way as for $\pi^+\pi^-$ case.

If $N={\rm proton}$ and $N'={\rm neutron}$, these transition GPDs can be related to the nucleon GPDs by isospin symmetry~\cite{Mankiewicz:1997aa}
\begin{align}
{\cal F}^{ud}_{pn}(z_1, \xi, t) &= {\cal F}^u_{p}(z_1, \xi, t) - {\cal F}^u_{n}(z_1, \xi, t), \nn\\
\widetilde{{\cal F}}^{ud}_{pn}(z_1, \xi, t) &= \widetilde{{\cal F}}^u_{p}(z_1, \xi, t) - 
	\widetilde{{\cal F}}^{u}_{n}(z_1, \xi, t).
\label{eq:isospin Fud}
\end{align}

\subsection{The leading-order hard coefficients}
\label{sec:C C-odd}

The leading-order diagrams are the same as those in figure~\ref{fig:hard1} and \ref{fig:hard2}, except that now we have two sets of hard coefficients, obtained with different spinor projectors on the nucleon side.
The calculation of the chiral-even coefficients is the same as $\pi^+\pi^-$ case, and the results are reorganized in a compact form in the Appendix with $z_1$ taking the value within $[z_m, z_M]$. From the parity constraint \eqref{eq:parity}, the chiral-odd coefficient $\widetilde{C}^{\mu\nu}$ can be expanded into the P-odd gauge invariant tensor structures in eq.~\eqref{eq:P-odd tensors}, with $p_1$ replaced by $\Delta$. Similarly to eq.~\eqref{eq:gauge-inv expansion}, we have 
\begin{align}
\widetilde{C}^{\mu\nu}
= -\frac{e^2g^2}{2\,\hat{s}^{2}}\frac{C_F}{N_c}\bigg(
&
  \widetilde{C}_{1} \,  
  \widetilde{\Delta}^{\mu} \varepsilon_{\perp}^{\nu\rho} q_{T\rho} 
+ \widetilde{C}_{2} \,
  \widetilde{p}_{2}^{\mu}
  \varepsilon_{\perp}^{\nu\rho} q_{T\rho}   
+ \widetilde{C}_{3} \,  
  \bar{\Delta}^{\nu}
  \varepsilon_{\perp}^{\mu\rho} q_{T\rho} 
+ \widetilde{C}_{4} \, 
  \bar{p}_{2}^{\nu} 
  \varepsilon_{\perp}^{\mu\rho} q_{T\rho}  \bigg) \, ,
\label{eq:gauge-inv expansion odd}
\end{align}
where $\widetilde{\Delta}^{\mu}$ and $\bar{\Delta}^{\nu}$ are defined in the same way as $\widetilde{p}_1^{\mu}$ and $\bar{p}_1^{\nu}$ in eq.~\eqref{eq:projected p1 p2}, respectively. The dimensionless scalar coefficients $\widetilde{C}_1$ to $\widetilde{C}_4$ can be extracted from the calculated result of each diagram by using eq.~\eqref{eq:gauge-inv expansion odd}, and isolating the coefficient of the term proportional to $\Delta^{\mu}$, $p_2^{\mu}$, $\Delta^{\nu}$ and $p_2^{\nu}$ sequentially. The results are collected in the Appendix.

Following the discussion above eqs.~\eqref{eq:C conjugation A} and \eqref{eq:C conjugation B}, charge conjugation implies similar relations for the chiral-odd coefficients, but with a minus sign, i.e.,
\begin{align}
&\{ 
\widetilde{C}_{A1}, \, \widetilde{C}_{A1'}, \,
\widetilde{C}_{A3}, \, \widetilde{C}_{A4} \}^{\mu\nu}
(z_1,z_2)
=
-\{
\widetilde{C}_{A2'}, \, \widetilde{C}_{A2}, \,
\widetilde{C}_{A3'}, \, \widetilde{C}_{A4'}\}^{\mu\nu}(1-z_1,1-z_2)
\label{eq:C conjugation A tilde}
\end{align}
for Type-$A$ diagrams, and 
\begin{align}
&\{
\widetilde{C}_{B1}, \, \widetilde{C}_{B1'}, \,
\widetilde{C}_{B2}, \, \widetilde{C}_{B2'}, \,
\widetilde{C}_{B3}, \, \widetilde{C}_{B3'}
\}^{\mu\nu}(z_1,z_2)\bigg|_{e_u\leftrightarrow e_d}
\nn\\
&\hspace{2em} =
-\{
\widetilde{C}_{B4}, \, \widetilde{C}_{B4'}, \,
\widetilde{C}_{B5}, \, \widetilde{C}_{B5'}, \,
\widetilde{C}_{B6}, \, \widetilde{C}_{B6'}
\}^{\mu\nu}(1-z_1,1-z_2)
\label{eq:C conjugation B tilde}
\end{align}
for Type-$B$ diagrams.
These relations carry through to each scalar coefficient $\widetilde{C}_1,\cdots, \widetilde{C}_4$, which has been checked in the calculations. 
Similar to the symmetric relation in eq.~\eqref{eq:CA symmetry}, we obtain 
an antisymmetric relation for $\widetilde{C}^{\mu\nu}$,
\beq
\widetilde{C}^{\mu\nu}_A(z_1,z_2)=-\widetilde{C}^{\mu\nu}_A(1-z_1, 1-z_2) ,
\label{eq:CA  O symmetry}
\eeq
for Type-$A$ diagrams, while for $\widetilde{C}_B$ this antisymmetry is broken by the difference of $e_u^2$ and $e_d^2$.

\subsection{Cross section}

Using eqs.~\eqref{eq:gauge-inv expansion}, \eqref{eq:gauge-inv expansion odd} and \eqref{eq:factorization M p pi}, we obtain the factorized scattering amplitude $\mathcal{M}^{\mu\nu}$ as
\begin{align}
\mathcal{M}^{\mu\nu}_{N\pi\to N'\gamma\gamma}
&= \frac{i e^2 g^2f_{\pi}}{4\hat{s}^2} \frac{C_F}{N_c} 
\nn\\
& \times 
	\bigg[ 
		i \left( \mathcal{M}_{0} \, \widetilde{g}_{\perp}^{\mu\nu}\, \hat{s}
 	+ \mathcal{M}_{1} \,  \widetilde{\Delta}^{\mu}\bar{\Delta}^{\nu} 
	+ \mathcal{M}_{2} \, 
			\widetilde{p}_{2}^{\mu}\bar{p}_{2}^{\nu}   
		+ \mathcal{M}_{3} \,  
			\widetilde{\Delta}^{\mu}\bar{p}_{2}^{\nu}
		+ \mathcal{M}_{4} \, 
			\widetilde{p}_{2}^{\mu}\bar{\Delta}^{\nu} \right)
	\nn\\ 
	&  \quad
	+ \left(
  \widetilde{\mathcal{M}}_{1} \,  
  \widetilde{\Delta}^{\mu} \varepsilon_{\perp}^{\nu\rho} q_{T\rho} 
+ \widetilde{\mathcal{M}}_{2} \,
  \widetilde{p}_{2}^{\mu}
  \varepsilon_{\perp}^{\nu\rho} q_{T\rho}    
+ \widetilde{\mathcal{M}}_{3} \,  
  \bar{\Delta}^{\nu}
  \varepsilon_{\perp}^{\mu\rho} q_{T\rho} 
+ \widetilde{\mathcal{M}}_{4} \, 
  \bar{p}_{2}^{\nu} 
  \varepsilon_{\perp}^{\mu\rho} q_{T\rho}  \right)
		\bigg] \, ,
\end{align}
where 
\begin{align}
\mathcal{M}_i 
    &= 
	\int_{z_m}^{z_M} \dd z_1 \int_0^1 \dd z_2 \,
    \widetilde{\cal F}^{ud}_{NN'}(z_1,\xi,t) \, \phi(z_2) \, C_i(z_1,z_2) 
    \nn\\
    & {\hskip 0.2in}   
      = \M[C_i; \widetilde{\cal F}^{ud}_{NN'}, (z_m,z_M); \phi, (0,1)] 
    ,	\nn\\
\widetilde{\mathcal{M}}_i 
    & = 
	\int_{ z_m}^{ z_M } \dd z_1 \int_0^1 \dd z_2 \,
     {\cal F}^{ud}_{NN'}(z_1,\xi,t) \, \phi(z_2) \, \widetilde{C}_i(z_1,z_2)
     \nn\\
    & {\hskip 0.2in}  
     = \M[\widetilde{C}_i; {\cal F}^{ud}_{NN'}, (z_m,z_M); \phi, (0,1)] \, ,
\label{eq:Mi F Ft}
\end{align}
with $i = 0, \cdots, 4$ for $\mathcal{M}_i$ or $1, \cdots, 4$ for $\widetilde{\mathcal{M}}_i$.
Like eq.~\eqref{eq:M2}, we have the full scattering amplitude squared, summing over the photon polarizations,
\begin{align}
\left|\overline{\mathcal{M}}\right|^2
&= \left( \frac{e^2 g^2 f_{\pi}}{4\hat{s}} \frac{C_F}{N_c} \right)^2
	\Bigg[ \bigg(
|\mathcal{M}_0|^2 
+\bigg|\frac{\mathcal{M}_1 + \mathcal{M}_2}{4}
- \frac{(\Delta\cdot q_1)^2 \mathcal{M}_3 + (\Delta\cdot q_2)^2 \mathcal{M}_4}{ \hat{s} \, q_T^2}
\bigg|^2 \bigg) 
\nn\\
&  \hspace{0.8in}+
\bigg| \frac{(\Delta\cdot q_1) \widetilde{\mathcal{M}}_1 - (\Delta\cdot q_2) \widetilde{\mathcal{M}}_2}{\hat{s}}
\bigg|^2		
+ \bigg| \frac{(\Delta\cdot q_2) \widetilde{\mathcal{M}}_3 - (\Delta\cdot q_1) \widetilde{\mathcal{M}}_4}{\hat{s}}
\bigg|^2
\Bigg] \, ,
\label{eq:T p pi}
\end{align}
where the average (sum) over the spins of initial-state nucleon $N$ (final-state $N'$) is included in $|\mathcal{M}_i|^2$ and $|\widetilde{\mathcal{M}}_i|^2$.

Instead of summing (or averaging) over all nucleon spins, we can introduce GPDs sensitive to the hadron spin by expressing the matrix elements of nucleon states in eq.~\eqref{eq:GPD def} in terms of independent combinations of nucleon spinors and corresponding ``form factors'' or spin dependent GPDs,
\begin{align}
{\cal F}^{ud}_{NN'}(z_1,\xi,t)=&\frac{1}{2P^+}
		\bigg[\, {\cal H}^{ud}_{NN'}(z_1,\xi,t) \,\bar{u}(p')\gamma^+ u(p) 
		\nn \\
		& {\hskip 0.4in}
		- {\cal E}^{ud}_{NN'}(z_1,\xi,t) \,\bar{u}(p')\frac{i\sigma^{+\alpha}
		\Delta_{\alpha}}{2m_p} u(p)
		\bigg],	\\
\widetilde{\cal F}^{ud}_{NN'}(z_1,\xi,t)=&\frac{1}{2P^+}
		\bigg[\, \widetilde{\cal H}^{ud}_{NN'}(z_1,\xi,t) \,\bar{u}(p')\gamma^+\gamma_5 u(p) 
		 \nn \\
		&  {\hskip 0.4in}
		- \widetilde{\cal E}^{ud}_{NN'}(z_1,\xi,t) \,
		\bar{u}(p')\frac{i\gamma_5\sigma^{+\alpha}\Delta_\alpha}{2m_p} u(p)
		\bigg] .
\end{align}
Consequently, all scattering amplitudes corresponding to independent tensor structures, $\mathcal{M}_i$ and $\widetilde{\mathcal{M}}_i$ in eq.~\eqref{eq:T p pi} can be expressed in terms of the spin dependent GPDs, 
\begin{align}
\mathcal{M}_i = &
\frac{1}{2P^+}
    \left[\, \mathcal{M}_i^{[\widetilde{\cal H}]}\,
    \bar{u}(p')\gamma^+\gamma_5 u(p) 
	- \mathcal{M}_i^{[\widetilde{\cal E}]} \,
	\bar{u}(p')\frac{i\gamma_5\sigma^{+\alpha}\Delta_\alpha}{2m_p} u(p)
	\right]	\, ,\nn\\
\widetilde{\mathcal{M}}_i = &
\frac{1}{2P^+}
    \left[\, \widetilde{\mathcal{M}}_i^{[{\cal H}]}\,
    \bar{u}(p')\gamma^+ u(p) 
	- \widetilde{\mathcal{M}}_i^{[{\cal E}]} \,
	\bar{u}(p')\frac{i\sigma^{+\alpha}
		\Delta_{\alpha}}{2m_p} u(p)
	\right]	\, ,
\end{align}
where the superscript ``$[{\cal H}]$'' means to replace the corresponding ${\cal F}^{ud}_{NN'}$ in eq.~\eqref{eq:Mi F Ft} by ${\cal H}^{ud}_{NN'}$, etc. Multiplied by their complex conjugate with the spin of $N$ ($N'$) averaged (summed), we have
\bse\label{eq:M2 spin sum}
\begin{align}
\left| \mathcal{M}_i \right|^2  = &
(1-\xi^2)\, 
\left|\mathcal{M}_i^{[\widetilde{\cal H}]}\right|^2
- 2\xi^2\, {\rm Re} \left(
    \mathcal{M}_i^{[\widetilde{\cal H}]}{}^*     \mathcal{M}_i^{[\widetilde{\cal E}]} 
\right) 
- \frac{\xi^2 t}{4m^2_p} 
\left|\mathcal{M}_i^{[\widetilde{\cal E}]} \right|^2 	\, ,
	\\
\left|\widetilde{\mathcal{M}}_i \right|^2  
= & 
(1-\xi^2)\, \left|\widetilde{\mathcal{M}}_i^{[{\cal H}]} \right|^2
- 2\xi^2\, {\rm Re} \left(         
    \widetilde{\mathcal{M}}_i^{[{\cal H}]}{}^* \widetilde{\mathcal{M}}_i^{[{\cal E}]} 
\right)   
-   \left( \frac{t}{4m^2_p} + \xi^2 \right) 
 \left| \widetilde{\mathcal{M}}_i^{[{\cal E}]} \right|^2 	\, ,
\end{align}\ese
where the factor $1/2$ for the spin average has been included. 

In our numerical analysis in the next section, we take $|t| \leq 0.2~{\rm GeV}^2$, which constrains $\xi$ to be $\xi \leq 0.23 $ by eq.~\eqref{eq:xi range t}. Then the terms containing $\mathcal{M}_i^{[\widetilde{\cal E}]}$ or $\widetilde{\mathcal{M}}_i^{[{\cal E}]}$ are suppressed by a factor of about $0.1$ or smaller, compared to the terms containing $|\mathcal{M}_i^{[\widetilde{\cal H}]} |^2$ or $|\widetilde{\mathcal{M}}_i^{[{\cal H}]} |^2$ . We can thus neglect them for a rough estimate.
Using eq.~\eqref{eq:M2 spin sum}, we can rewrite eq.~\eqref{eq:T p pi} as
\begin{align}
\left|\overline{\mathcal{M}}\right|^2
&\approx (1-\xi^2) \left( \frac{e^2 g^2 f_{\pi}}{4\hat{s}}  \frac{C_F}{N_c} \right)^2
 \nn\\
& \hspace{0.5em} \times
\Bigg[ \bigg(
|\mathcal{M}_0^{[\widetilde{\cal H}]}|^2 
+\bigg|\frac{\mathcal{M}_1^{[\widetilde{\cal H}]} + \mathcal{M}_2^{[\widetilde{\cal H}]} }{4}
- \frac{(\Delta\cdot q_1)^2 \mathcal{M}_3^{[\widetilde{\cal H}]} + (\Delta\cdot q_2)^2 \mathcal{M}_4^{[\widetilde{\cal H}]} }{ \hat{s} \, q_T^2}
\bigg|^2 \bigg)
\nn\\
& \hspace{1.5em}
+
\bigg| \frac{(\Delta\cdot q_1) \widetilde{\mathcal{M}}_1^{[{\cal H}]} - (\Delta\cdot q_2) \widetilde{\mathcal{M}}_2^{[{\cal H}]}}{\hat{s}}
\bigg|^2		
+ \bigg| \frac{(\Delta\cdot q_2) \widetilde{\mathcal{M}}_3^{[{\cal H}]} - (\Delta\cdot q_1) \widetilde{\mathcal{M}}_4^{[{\cal H}]}}{\hat{s}}
\bigg|^2
\Bigg]	\, .
\label{eq:T p pi 2}
\end{align}

As discussed in Sec.~\ref{sec:kin}, we can specify an event by $\bm{\Delta}_T, \xi$ and $\bm{q}_T$, with $\bm{q}_T$ being the transverse momentum of the photons in the photon frame $S_{\gamma}$. This gives
\begin{align}
\dd\sigma = \frac{1}{2s} 
			\frac{\dd\xi \dd^2\bm{\Delta}_T}{(1-\xi^2) (2\pi)^3}
			\frac{\dd^2\bm{q}_T}{8\pi^2 \hat{s}}
			\frac{ \left|\overline{\mathcal{M}}\right|^2 }{ \sqrt{ 1-\hat{\kappa} } } \, ,
\label{eq:xsec formula0}
\end{align}
where $\left|\overline{\mathcal{M}}\right|^2$ is given in eq.~\eqref{eq:T p pi 2} and $\hat{\kappa} = 4q_T^2/\hat{s} \leq 1$ is the analog of $\kappa$ (defined below eq.~\eqref{eq:kin q12}) for the photon system in the $S_{\gamma}$ frame. The direction of $\bm{q}_T$ can be defined with respect to the $N-N'$ plane, or $\bm{p}$-$\bm{\Delta}_T$ plane. But since $\left|\overline{\mathcal{M}}\right|^2$ is 
for unpolarized scattering,
it does not depend on the azimuthal angles of $\bm{q}_T$ and $\bm{\Delta}_T$, so we can integrate them out. That allows us to only use three scalars $\Delta_T$, $\xi$ and $q_T$ to describe the events, which by eq.~\eqref{eq:t-DeltaT} can be transformed to the three scalar variables $(t, \xi, q_T)$, and corresponding differential cross section,
\begin{align}
\frac{\dd\sigma}{\dd |t|\, \dd \xi\, \dd q_T^2} =
  \frac{\pi}{64} 
  \left(\alpha_e \alpha_s \frac{f_{\pi}}{s^2} \frac{C_F}{N_c} \right)^2
  \frac{(1-\xi^2)}{\xi^2}
  \frac{(1+\xi)}{\xi}
  \frac{\mathcal{B} }{\sqrt{1-\hat{\kappa}}} \, ,
\label{eq:xsec formula}
\end{align}
where $\mathcal{B}$ stands for the big square bracket in eq.~\eqref{eq:T p pi 2}, which is dimensionless and can be evaluated numerically once we know the pion DAs and nucleon's GPDs. In eq.~\eqref{eq:xsec formula}, we have separated the $\xi$ dependent factor into two parts, in which the second part, $(1+\xi)/\xi$, is canceled when we integrate over $q_T^2$ from $q_{T {\rm min}}^2$ to $\hat{s}/4 = \xi/(1+\xi) (s/2)$.

\section{Numerical results}
\label{s.numerical}

In this section, we evaluate the cross sections for producing a pair of high transverse momentum photons in exclusive pion-pion and pion-nucleon scattering and test their sensitivity to the shape of DAs and GPDs in terms of active parton's momentum fraction.   

\subsection{End-point sensitivity and improvement from Sudakov suppression}
\label{sec:Sudakov}

Before we introduce our choices of DAs and GPDs to evaluate the factorized cross sections, we discuss the well-known ``end-point'' sensitivity associated with perturbative evaluation of factorized elastic scattering processes, and its impact on the new type of exclusive processes introduced in this paper.  

\begin{figure}[htbp]
    \centering
    \includegraphics[scale=0.6]{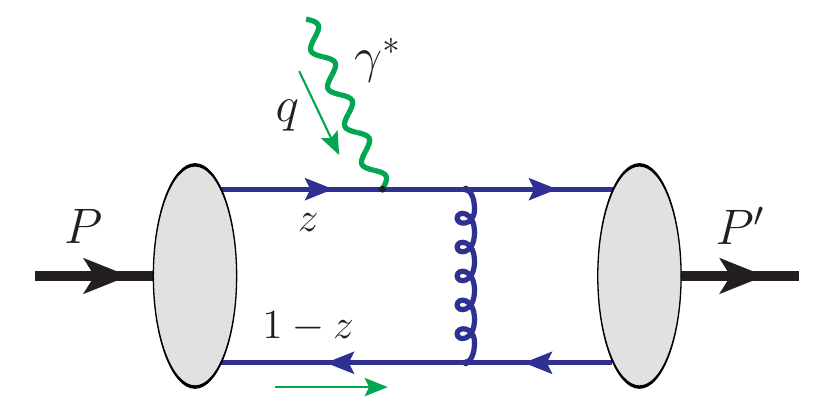} {\hskip 0.15\textwidth}
        \includegraphics[scale=0.6]{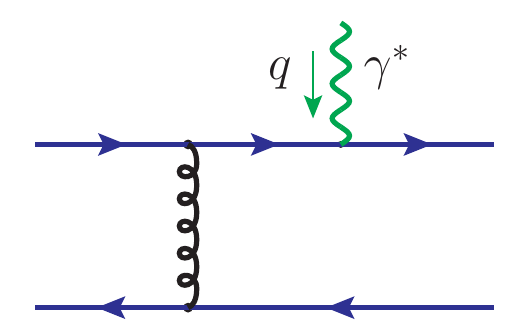}\quad
            \includegraphics[scale=0.6]{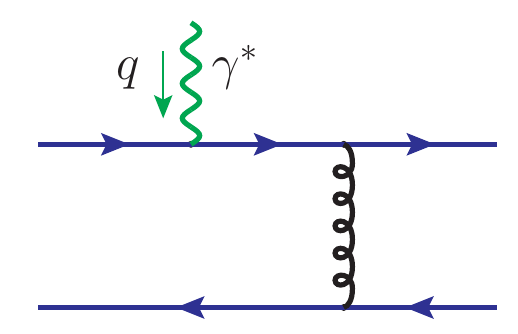}\quad \\
            (a) \hskip 0.5\textwidth (b)
\caption{(a) Sketch for pion Form Factor; (b) Leading order Feynman diagrams for the partonic hard part of the factorized pion Form Factor.}
\label{fig:pionFF}
\end{figure}

For the comparison, we consider the well-known perturbative calculation of pion Form Factor $F_{\pi}(Q^2)$, as sketched in figure~\ref{fig:pionFF}(a), which can be extracted from elastic electron-pion scattering: $e(\ell)+\pi(p_\pi) \to e(\ell')+\pi(p_{\pi}')$.  When the momentum transfer $q = \ell-\ell'$ has a high virtuality, with $Q^2\equiv -q^2 \gg \Lambda_{\rm QCD}^2$, the pion Form Factor takes the factorized form as~\cite{Lepage:1979zb},
\begin{equation}
F_\pi(Q^2) \approx \int_0^1 \dd z_1 \int_0^1 \dd z_2\,  \phi(z_1) T_B(z_1,z_2,Q^2)\, \phi(z_2) \, ,
\label{eq:pionFF}
\end{equation}
where $\phi$ is pion DA,  $T_B(z_1,z_2,Q^2)$ represents the hard scattering, and the factorization scale dependence is suppressed.  With the leading order diagrams in figure~\ref{fig:pionFF}(b), the short-distance hard part is given by~\cite{Lepage:1979zb}  
\begin{equation}
T_B(z_1,z_2,Q^2) \approx 16\pi C_F\, \frac{\alpha_s(Q^2)}{z_1 z_2 Q^2} \, ,
\label{eq:H4pionFF}
\end{equation}
with color factor $C_F=4/3$ for SU(3) color.  By substituting this lowest order hard part in eq.~\eqref{eq:H4pionFF} into eq.~\eqref{eq:pionFF}, it is clear that the pion Form Factor measurement is only sensitive to the ``moment'' of pion DA, $\int_0^1 dz z^{-1} \phi(z)$, not the detailed shape of $\phi(z)$, even when the probing scale $Q^2$ varies.  Although the ``moment'' $\int_0^1 dz z^{-1} \phi(z)$ is expected to be finite since $\phi(z)\to 0 $ as $z\to 0$, the short-distance hard part  in eq.~\eqref{eq:H4pionFF} is actually singular as $z_1$ (and/or $z_2$) $\to 0$, corresponding to the situation when the virtuality of the exchanged gluon in figure~\ref{fig:pionFF} goes to zero and the ``hard'' scattering is actually not taking place at a ``short-distance''.   The reliability of this perturbative fixed-order calculation near the ``end-point'' region when $z_1$ (and/or $z_2$) $\to 0$ could be improved by taking into account the ``Sudakov suppression'' from resumming high order Sudakov logarithmic contributions.  For example, the leading order perturbatively calculated hard part in eq.~\eqref{eq:H4pionFF} could be improved as~\cite{Li:1992nu}
\begin{equation}
T_B(z_1,z_2,Q^2) \approx 16\pi C_F
\int_0^{\infty}  \dd b\,
\alpha_s(t)\, b\, K_0(\sqrt{z_1z_2}Q\, b)
\, e^{-{\cal S}(z_1,z_2,b,Q)} \, ,
\label{eq:ModifiedH}
\end{equation}
where the running coupling constant $\alpha_s$ is evaluated at $t={\rm max}(\sqrt{z_1 z_2}Q, 1/b)$, $K_0$ is the modified Bessel function of order zero and the Sudakov factor ${\cal S}(z_1,z_2,b,Q)$ is given in eq.~(14) of Ref.~\cite{Li:1992nu}.  In keeping the same factorized form in eq.~\eqref{eq:pionFF} with the modified hard part in eq.~\eqref{eq:ModifiedH}, an evolution of pion $\phi(z)$'s factorization scale from $1/b$ to the hard scale $Q$ was neglected.  With the Sudakov suppression, the perturbative hard part $T_B(z_1,z_2,Q^2)$ in eq.~\eqref{eq:ModifiedH} is no longer singular as $z_1$ (and/or $z_2$) goes to zero.

Like the pion Form Factor, the perturbative hard part calculated from the Type-$B$ diagrams in figure~\ref{fig:hard2} is also singular in the ``end-point'' region when $z_1$ (and/or $z_2$) $\to 0$ or $1$, as clearly evident from the behavior of the three propagators in eq.~\eqref{eq:A k1k2q}.  In addition, like the hard part of pion Form Factor in eq.~\eqref{eq:H4pionFF}, the dependence on active parton momentum fractions $z_1$ and $z_2$ in eq.~\eqref{eq:A k1k2q} is completely decoupled from the external kinematic variables, and consequently, the contribution from the Type-$B$ diagrams to the exclusive cross section is only sensitive to the ``moment'' of pion DA.  

On the other hand, the three propagators for the Type-$A$ diagrams in figure~\ref{fig:hard1}, as shown in eqs.~\eqref{eq:A k1k2} and \eqref{eq:A q}, have slightly different features.  The contribution from the Type-$A$ diagrams is less singular in the ``end-point'' region when $z_1$ or $z_2$ goes to zero.  The dependence on active parton momentum fractions $z_1$ and $z_2$ cannot be completely decoupled from the external kinematic variables.  As shown in eq.~\eqref{eq:A q}, $z_1$ and $z_2$ are entangled with externally measured photon transverse momentum $q_T$.  It is this entanglement that makes the $q_T$-distribution of this exclusive cross section to be sensitive to the shape of the $z$-dependence of pion DA, or GPDs in pion-baryon scattering.

\subsection{Enhanced sensitivity to the shape of pion DAs}
\label{sec:DA shape}

To demonstrate that the differential cross section $\dd\sigma/\dd q_T^2$ for exclusive $\pi^+\pi^- \to \gamma\gamma$ process is sensitive to both the ``moment'' as well as the detailed shape of pion DA, we introduce a power-form parametrization for the normalized pion DA,
\beq
\phi_{\alpha}(z) = \frac{ z^{\alpha}(1-z)^{\alpha} }{ \B(1+\alpha, 1+\alpha) }\, ,
\label{eq:DA power}
\eeq
with $\alpha > 0$ so that the ``moment'' $\int_0^1 \dd z z^{-1} \phi_{\alpha}(z)$ is finite. When $\alpha=1$, this normalized pion DA is effectively the same as the so-called asymptotic form of pion DA when factorization scale $\mu\to\infty$~\cite{Lepage:1980fj}.  In this subsection, we vary the power $\alpha$ to show how $\dd\sigma/\dd q_T^2$ changes. In the following numerical calculation, we use fixed electromagnetic coupling $\alpha_e = 1/137$ and the one-loop running strong coupling constant $\alpha_s(\mu)$ evaluated at the scale $\mu = q_T$.  For exclusive $\pi^+\pi^- \to \gamma\gamma$, which could be a Sullivan-type process as a part of the $p\pi^-\to n \gamma\gamma$ diffractive scattering when the $|t|$ is small, we choose the collision energy $\sqrt{s} = 3-6$~GeV, and 
require $q_T$ to be greater than $1~{\rm GeV}$.  

\begin{figure}[htbp]
    \centering
    \includegraphics[scale=0.55]{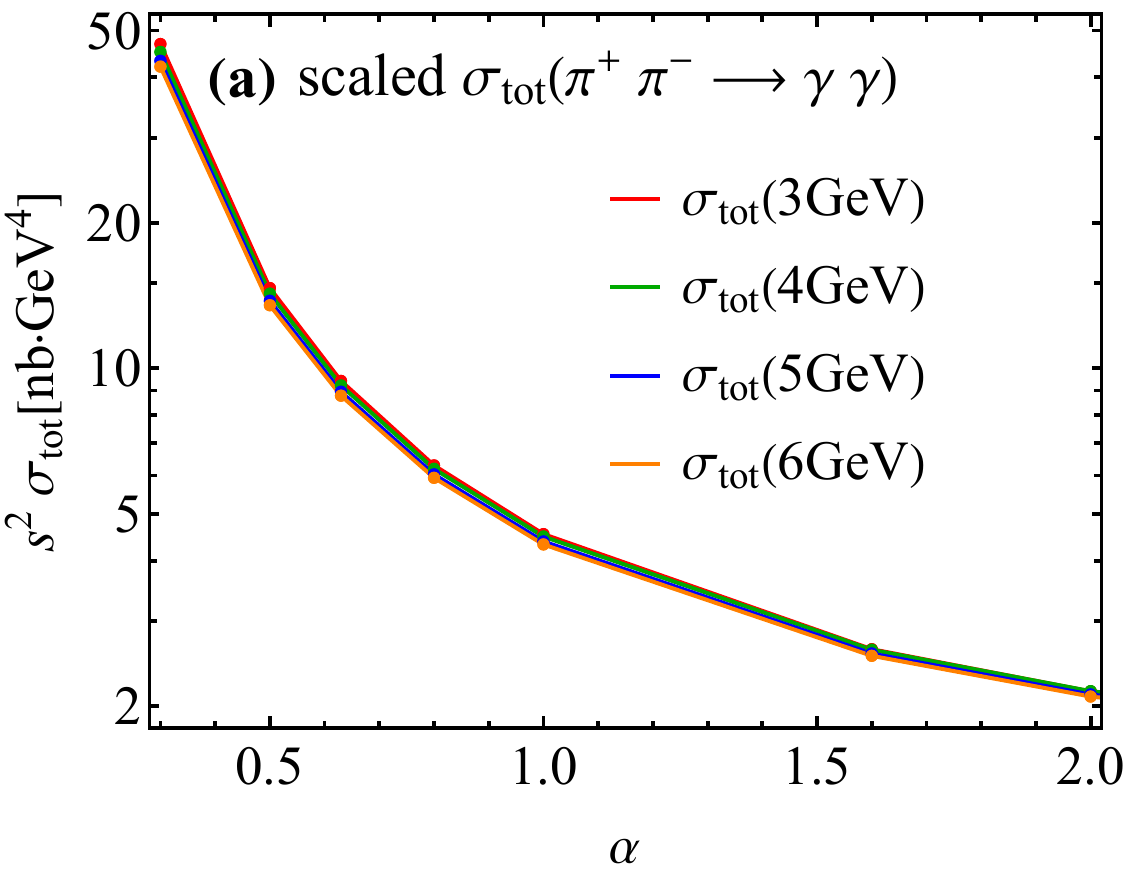}\quad
    \includegraphics[scale=0.55]{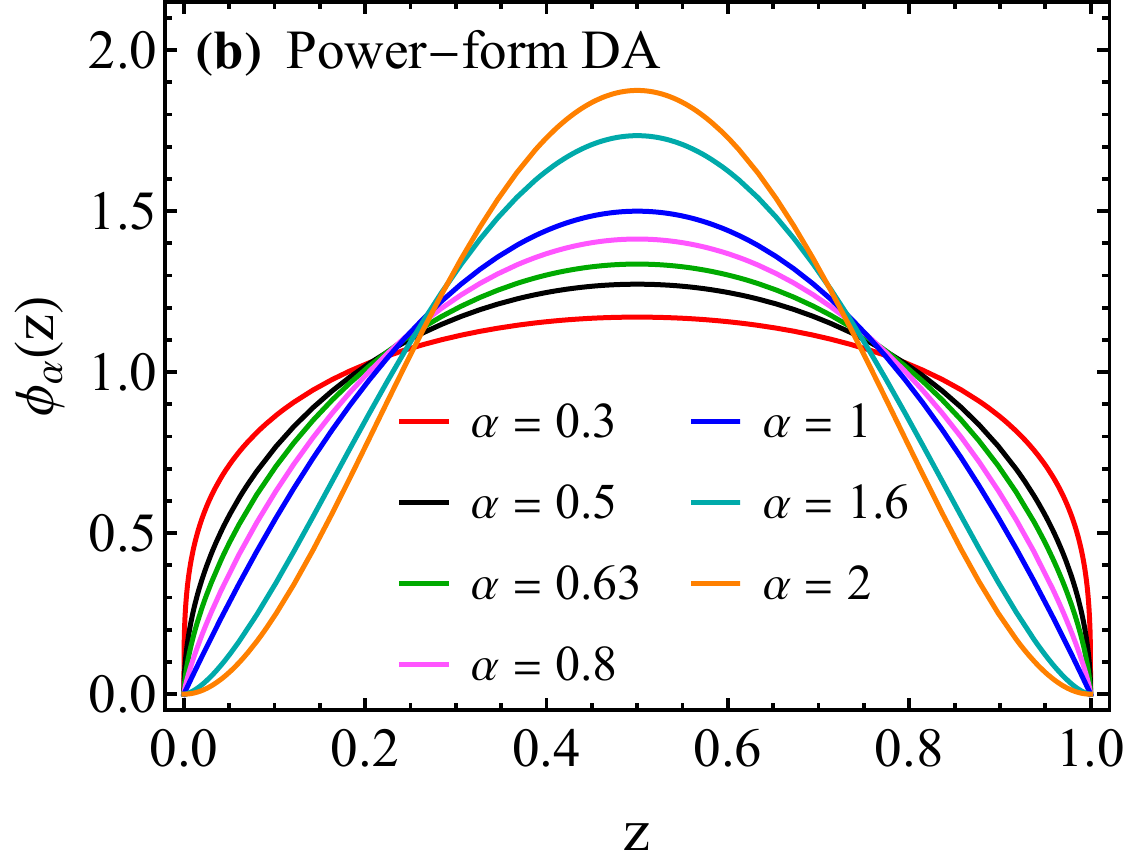}\quad
\caption{The total cross section (in (a)) for different DAs (shown in (b)) and different CM energies, where a scaling factor $s^2$ has been multiplied. The total cross section is obtained by integrating over $q_T$ from $1~{\rm GeV}$ to $\sqrt{s}/2$. 
    The dots on the curves are the points that were explicitly calculated.}
    \label{fig:xsec pi pi plot}
\end{figure}

In figure~\ref{fig:xsec pi pi plot}(a), we plot the ``total'' cross section defined in eq.~\eqref{eq:xsec} with $q_{T{\rm min}}=1$~GeV as a function of the power $\alpha$ of the normalized pion DA for various collision energies.  Corresponding shapes of the normalized pion DAs are shown in figure~\ref{fig:xsec pi pi plot}(b).  To minimize its dependence on the collision energy, we multiplied a scaling factor $s^2$ to the cross section, which effectively puts all the curves with four different collision energies on top of each other.  However, as shown in figure~\ref{fig:xsec pi pi plot}(a), the scaled cross section shows a very strong dependence on the value of $\alpha$, which is not because the partonic hard part is a good probe of the shape of DAs.  Instead, such a strong dependence on $\alpha$ is caused by the ``end-point'' sensitivity of the perturbatively calculated partonic hard part as discussed in the last subsection, and the fact, as shown in figure~\ref{fig:xsec pi pi plot}(b), that the value of pion DAs at different $\alpha$ have very different values near the ``end-point''.

Like the ``Sudakov'' suppression treatment for the ``end-point'' region of the pion Form Factor, an improvement of the ``end-point'' sensitivity is also needed to improve the reliability of perturbative calculation of the factorized hard parts for this new type of exclusive processes, which is beyond the scope of the current paper.

As pointed out in Sec.~\ref{sec:Sudakov}, the propagator of the gluon has a very different momentum structure for Type-$A$ diagrams from those of the Type-$B$ diagrams.  The entanglement of momentum fraction $z_1$, $z_2$ and the observed $q_T$ in the Type-$A$ diagrams makes the $q_T$ distribution sensitive to the $z$-dependence of the pion DAs.

\begin{figure}[htbp]
    \centering
    \includegraphics[scale=0.5]{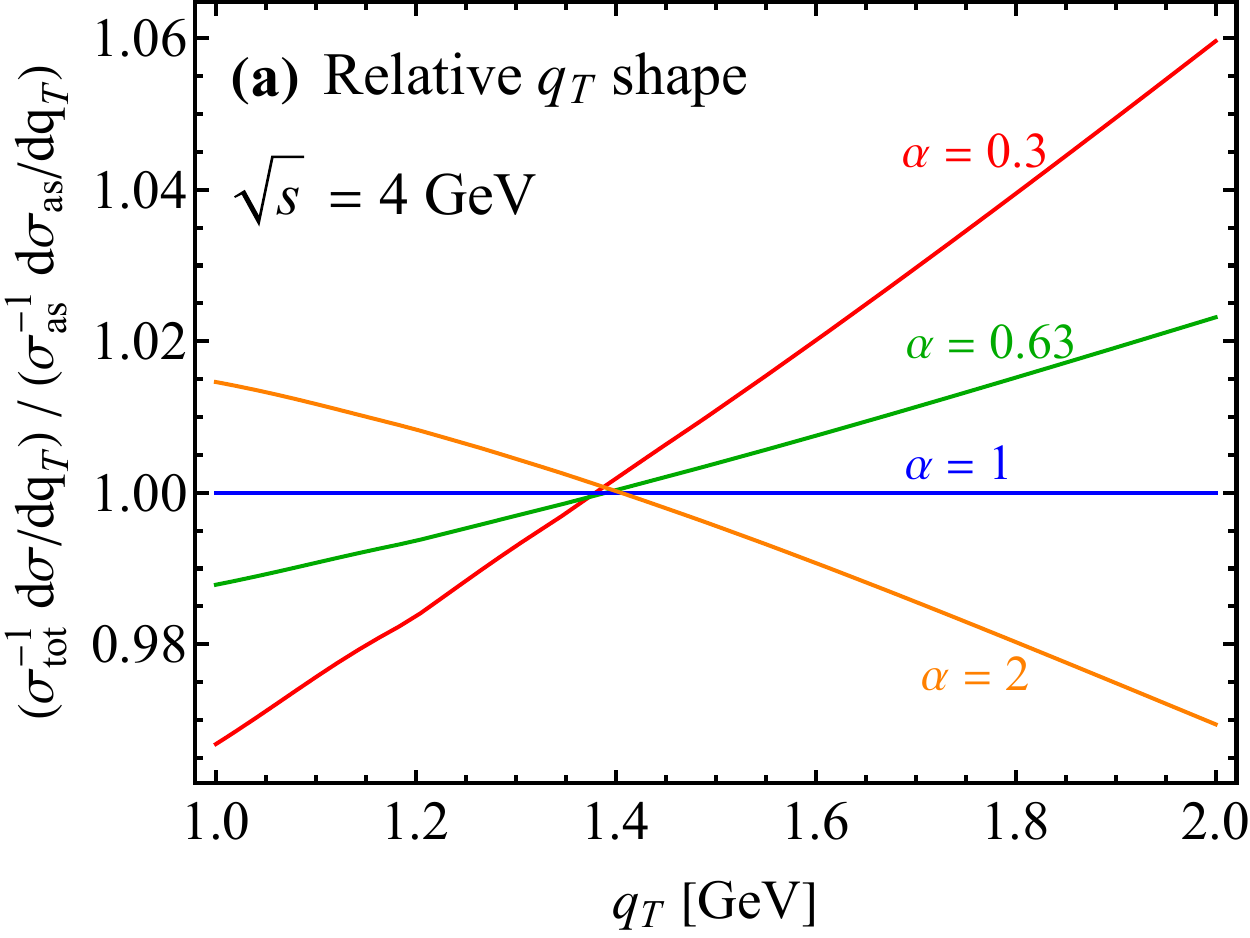} \quad
    \includegraphics[scale=0.5]{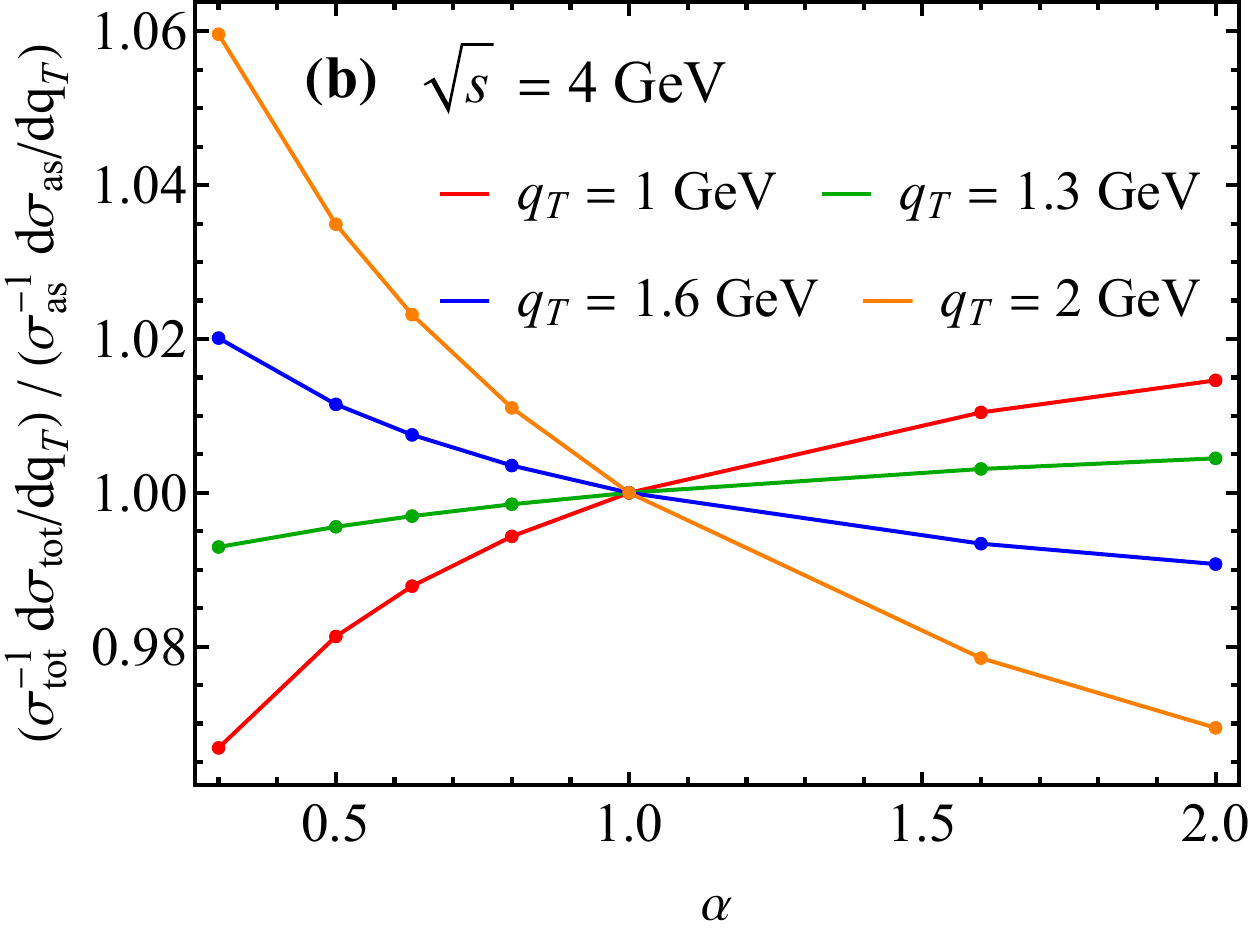}\quad
    \caption{(a) The relative $q_T$ shape for a few choices of power-form DAs with different values of $\alpha$. The relative $q_T$ shape is obtained by dividing the normalized $q_T$ distribution by the one with the asymptotic DA form. (b) The same normalized $q_T$ distribution as a function of $\alpha$ of the power-form DAs.}
    \label{fig:relative qT}
\end{figure}

In figure~\ref{fig:relative qT}(a), we plot the normalized $q_T$ distribution, defined as $\dd \sigma/\dd q_T$ divided by the total cross section $\sigma_{\rm tot}\equiv \sigma(q_{T{\rm min}}=1~{\rm GeV})$ as defined in eq.~\eqref{eq:xsec}, with respect to the same normalized $q_T$ distribution evaluated with asymptotic pion DA ($\alpha=1$).   Corresponding normalized pion DAs are plotted in figure~\ref{fig:xsec pi pi plot}(b).  In figure~\ref{fig:relative qT}(b), we plot the same normalized $q_T$ distribution as a function of the power $\alpha$ at different values of $q_T$.  The normalized $q_T$ distribution at different $q_T$ values have very different dependence on the $\alpha$. 
Naively, from figure~\ref{fig:relative qT}, it seems that the $q_T$-dependence provides additional 10\% sensitivity on the shape of the pion DA.  Actually, the $q_T$-dependence should have provided a much stronger sensitivity to the shape of pion DAs, if the ``end-point'' sensitivity of the perturbatively calculated partonic hard parts are better controlled.  As pointed out in Sec.~\ref{sec:Sudakov}, the Type-$B$ diagrams have a much stronger singular behavior at the ``end-point'' than that of Type-$A$ diagrams.   Consequently, the Type-$B$ diagrams give a much bigger fraction of $\dd\sigma/\dd q_T$ from the ``end-point'' region of the pion DAs than what the Type-$A$ diagrams can give, while the Type-$A$ diagrams are more sensitive to the shape of pion DAs.  If we can improve the reliability of perturbatively calculated partonic hard cross section near the ``end-point'' for both Type-$A$ and Type-$B$ diagrams, the Type-$A$ diagrams would contribute a much bigger fraction to the differential cross section $\dd\sigma/\dd q_T$, making the measurement of $\dd\sigma/\dd q_T$ more sensitive to the shape of the $z$-dependence of the pion DAs.

\subsection{Enhanced sensitivity to the shape of GPDs}
\label{sec:GPD shape}

In this subsection, we try to demonstrate that the photon $q_T$ distribution of exclusive meson-baryon scattering process is sensitive to the functional shapes of nucleon GPD and pion DA.  The dependence on pion DA was discussed in Sec.~\ref{sec:DA shape} along with the exclusive $\pi\pi$ annihilation process.  We now focus on the sensitivity to the shape of nucleon GPD, and fix pion DA to the power-form in eq.~\eqref{eq:DA power} with $\alpha = 0.63$, which is the value compatible with the Lattice QCD calculation of the second moment of DA~\cite{Bali:2019dqc}.   

As discussed in Sec.~\ref{s.pih2gg}, the integration range of active momentum fraction $z_1$ of GPDs is extended from $(0, 1)$ to $((\xi-1)/2\xi, (\xi+1)/2\xi)$, as shown in eq.~\eqref{eq:z1limits}, and consequently, the propagators in partonic diagrams could be on-shell leading to poles along the integration contour of $z_1$. As discussed in Sec.~\ref{sec:factorization1}, the reduced diagram analysis ensures that the only perturbative pinch singularity at leading power is on the lines collinear to the external hadrons, which are systematically removed from the hard part of partonic scattering and absorbed into universal long-distance DAs or GPDs. The only possible singularities of the perturbatively calculated partonic hard part could appear at the ``end-point" of the integration, and need to be suppressed by the behavior of non-perturbative DAs and/or GPDs, or by improving high order perturbative calculations as discussed in Sec.~\ref{sec:Sudakov}.  When the non-pinched pole of $z_1$ locates along the contour, we use the distribution identity
\beq
\frac{1}{z_1 - a \pm i\varepsilon} = P\frac{1}{z_1 - a} \mp i\pi \delta(z_1 - a)
\label{eq:pole avoid}
\eeq
as a practical method to deform the contour~\cite{Qiu:1991wg}, where $P$ means the principal-value integration. Our numerical integration strategy is to individually separate each pole and use Eq.~\eqref{eq:pole avoid} to deal with the poles on the integration contour of $z_1$. In this approach, the non-pinched poles lead to imaginary parts to the scalar coefficients $\mathcal{M}_i$ and $\widetilde{\mathcal{M}}_i$ of the factorized scattering amplitude, and both their real and imaginary parts contribute to the exclusive cross section through the absolute values in eq.~\eqref{eq:T p pi 2}.

For our numerical analysis below, we use the kinematics of J-PARC~\cite{Aoki:2021cqa} with a charged pion beam of energy around $20$~GeV, as well as that of AMBER~\cite{Adams:2018pwt} with a pion beam of energy $150~{\rm GeV}$ hitting on fixed targets. 
For nucleon GPD, we choose the GK parametrization~\cite{Goloskokov:2005sd,Goloskokov:2007nt,Goloskokov:2009ia,Kroll:2012sm}, which models the GPD using double distribution,
\begin{align}
H_i(x, \xi, t) = \int_{-1}^1 \dd\beta \int_{-1+|\beta|}^{1-|\beta|}\dd\alpha \,
	\delta(x-\beta-\xi \alpha) \,  f_i(\beta, \alpha, t)	\,	,
\label{eq:DD int}
\end{align}
where the subscript $i$ refers to the choice of parton flavor and nucleon GPDs $H_i(x,\xi,t)$ are defined with the convention in Ref.~\cite{Diehl:2003ny}, as specified in eq.~\eqref{eq:GPD def}. The double distribution $ f_i(\beta, \alpha, t)$ is parametrized as
\begin{align}
f_i(\beta, \alpha, t)
= e^{\left( b_i + \alpha_i' \ln|\beta|^{-1} \right) t } \cdot h_i(\beta)\cdot w_i(\beta, \alpha)	\, , 
\label{eq:DD ansatz}
\end{align}
where $h_i(\beta)$ is the forward PDF of flavor $i$, and $w_i$ is a weight function
\begin{align}
w_i(\beta, \alpha) = \frac{\Gamma(2n_i+2)}{2^{2n_i+1} \Gamma^2(n_i+1)}
	\frac{\left[(1-|\beta|)^2 - \alpha^2 \right]^{n_i} }{(1-|\beta|)^{2n_i+1}}		\, ,
\label{eq:weight}
\end{align}
which characterizes the $\xi$ dependence of GPD $H_i(x, \xi, t)$ and is normalized as
\begin{align}
\int_{-1+|\beta|}^{1-|\beta|}\dd\alpha \, w_i(\beta, \alpha)  = 1 \, .
\end{align}
The larger the power $n_i$ is, the less dependent $H_i(x, \xi, t)$ is on $\xi$. In the limit that $n_i \to \infty$, $w_i \to \delta(\alpha)$, and we have
\begin{align}
H_i(x, \xi, t) = e^{\left( b_i + \alpha_i' \ln|x|^{-1} \right) t } h_i(x)	\, ,
\end{align}
which has no dependence on $\xi$ at all. 

\begin{figure}[h!]
    \centering
    \includegraphics[scale=0.65]{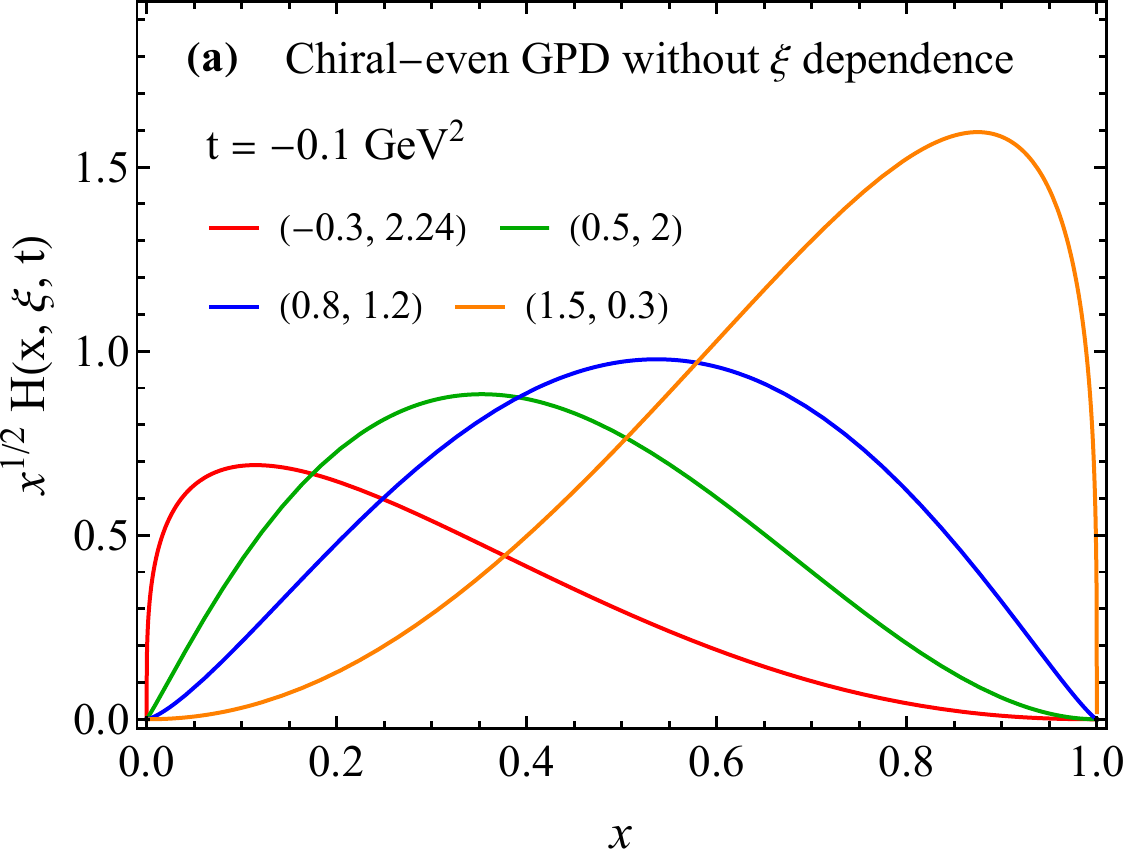}
    \includegraphics[scale=0.65]{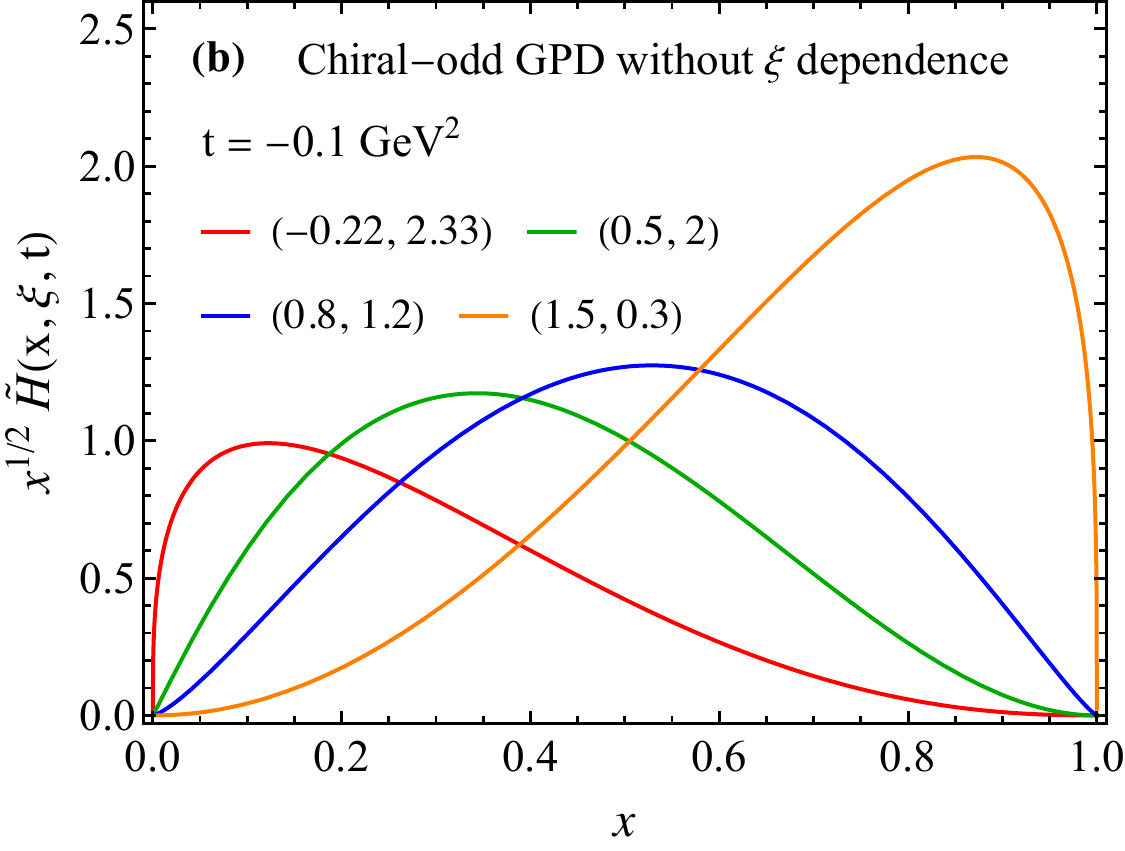}
    \caption{Chiral-even GPD $H(x,\xi,t)$ (a) and chiral-odd GPD $\widetilde{H}(x,\xi,t)$ (b) for $t= -0.1~{\rm GeV}^2$ and different power parameters $(\rho, \tau)$.}
    \label{fig:gpd_model}
\end{figure}

The quark double distribution is decomposed into valence and sea components, and sea quark components are taken to be the same for $u_{\rm sea}$ and $d_{\rm sea}$. Since our process is only sensitive to $H_u - H_d$ (or $\widetilde{H}_u - \widetilde{H}_d$) (see eq.~\eqref{eq:isospin Fud}), the sea components cancel, and only valence components contribute,\footnote{This is also the reason that we neglected the so-called $D$-term in eq.~\eqref{eq:DD ansatz} since it only appears for gluon and sea quarks.}
for which we have
\begin{align}
f^q_{\rm val}(\beta, \alpha, t) = 
\left[ f^q(\beta, \alpha, t) + \varepsilon_f f^q( - \beta, \alpha, t) \right] \theta(\beta) \, ,
\label{eq:DD val}
\end{align}
where $\varepsilon = +1$ for $H$ and $-1$ for $\widetilde{H}$. The condition $\theta(\beta)$ means that $H^q_{\rm val}(x,\xi,t)\neq 0$ only when $-\xi < x \leq 1$.

In the GK model, $b_{\rm val} = 0$ for both $H$ and $\widetilde{H}$, and $\alpha'_{\rm val} = 0.9~\rm{GeV}^{-2}$ for $H$ and $0.45~\rm{GeV}^{-2}$ for $\widetilde{H}$. The forward parton density $h_i(\beta)$ is parametrized as a ``power series" of $\beta$, fitted to global-fit PDFs. It is not our purpose to use a realistic GPD, but instead we want to see how different forms of GPDs affect the $q_T$ distribution, so it is convenient to use a simple functional form for $h(\beta)$, for which we choose 
\begin{align}
h_{ud}(\beta) = h_{u_{\rm v}}(\beta) - h_{d_{\rm v}}(\beta) = N \frac{\beta^{\rho} (1-\beta)^{\tau} }{B(1+\rho, 1+ \tau)}	\, ,
\label{eq:h(beta)}
\end{align}
which is similar to eq.~\eqref{eq:DA power} but with possibly different powers $\rho$ and $\tau$. The normalization factor is $N = 1$ for $H$ and $N = \eta_u - \eta_d = 1.267$ for $\widetilde{H}$~\cite{Kroll:2012sm}. The parameters $\rho$ and $\tau$ are fitted to the GK model at $\mu = 2~{\rm GeV}$, and we have the best fit
\begin{align}
(\rho_0, \tau_0) &= (-0.30, 2.24) \mbox{ for }  H		\, , \nn\\
(\rho_0, \tau_0) &= (-0.22, 2.33) \mbox{ for }  \widetilde{H}		\, .
\label{eq:best-fit}
\end{align}
This gives a $h(\beta)$ peaked near $\beta = 0$. We will vary the powers $(\rho, \tau)$ around the best-fit values and compare the change of observables.

\begin{figure*}[htbp]
    \centering
    \includegraphics[scale=0.6]{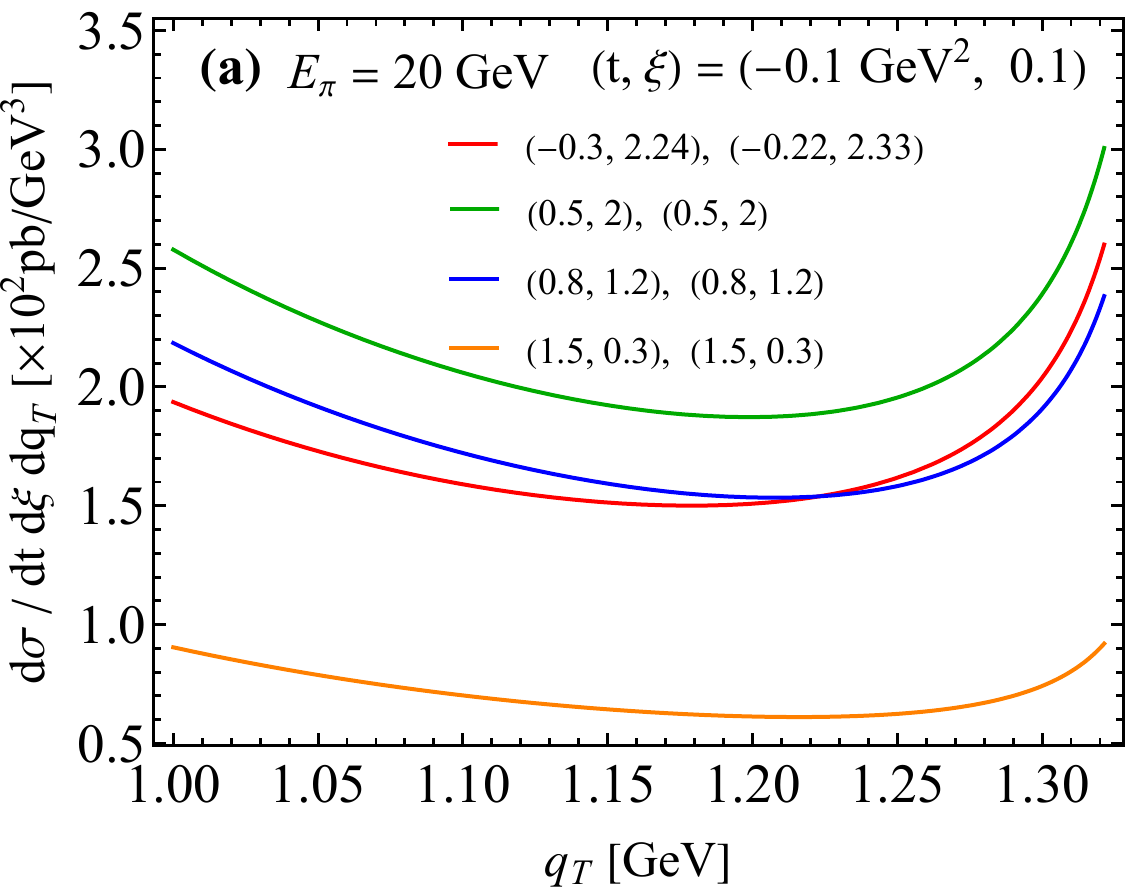}\quad
    \includegraphics[scale=0.58]{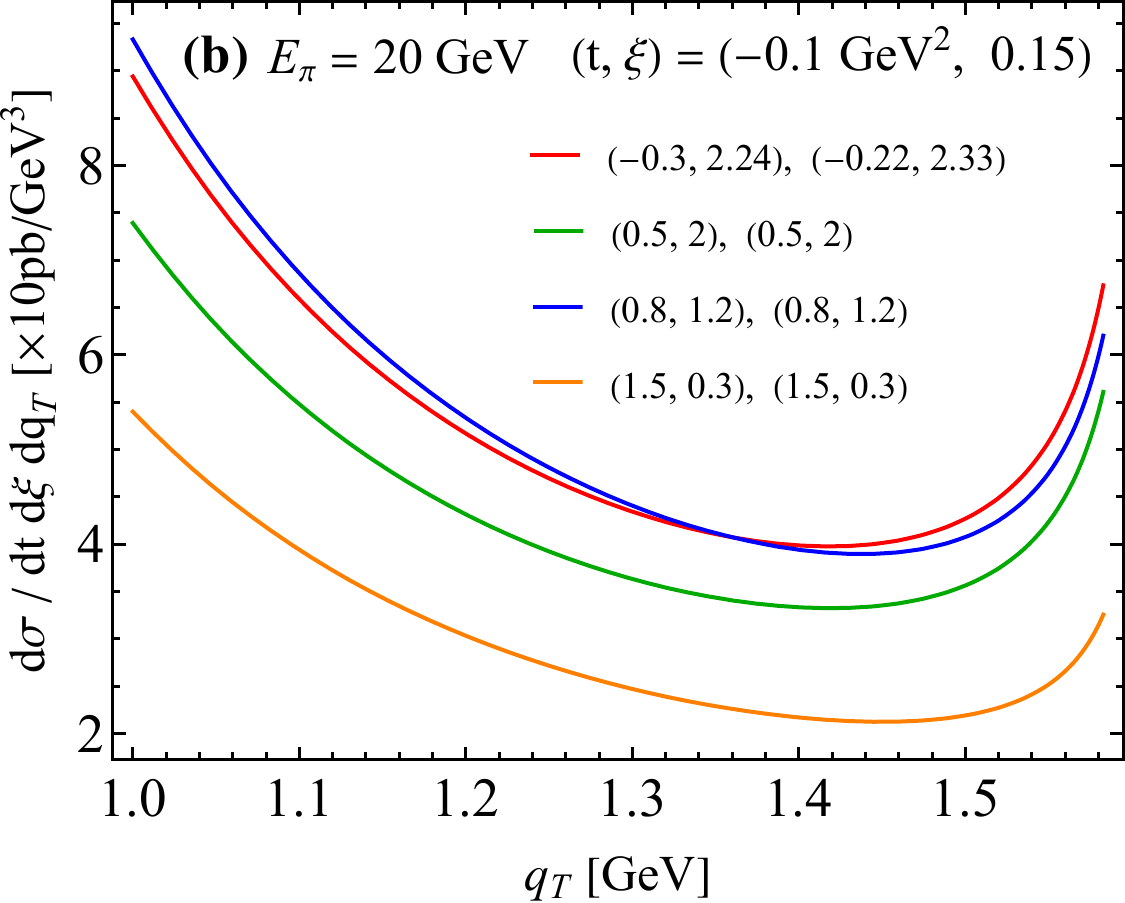}\\
    \vspace{1em}
    \includegraphics[scale=0.6]{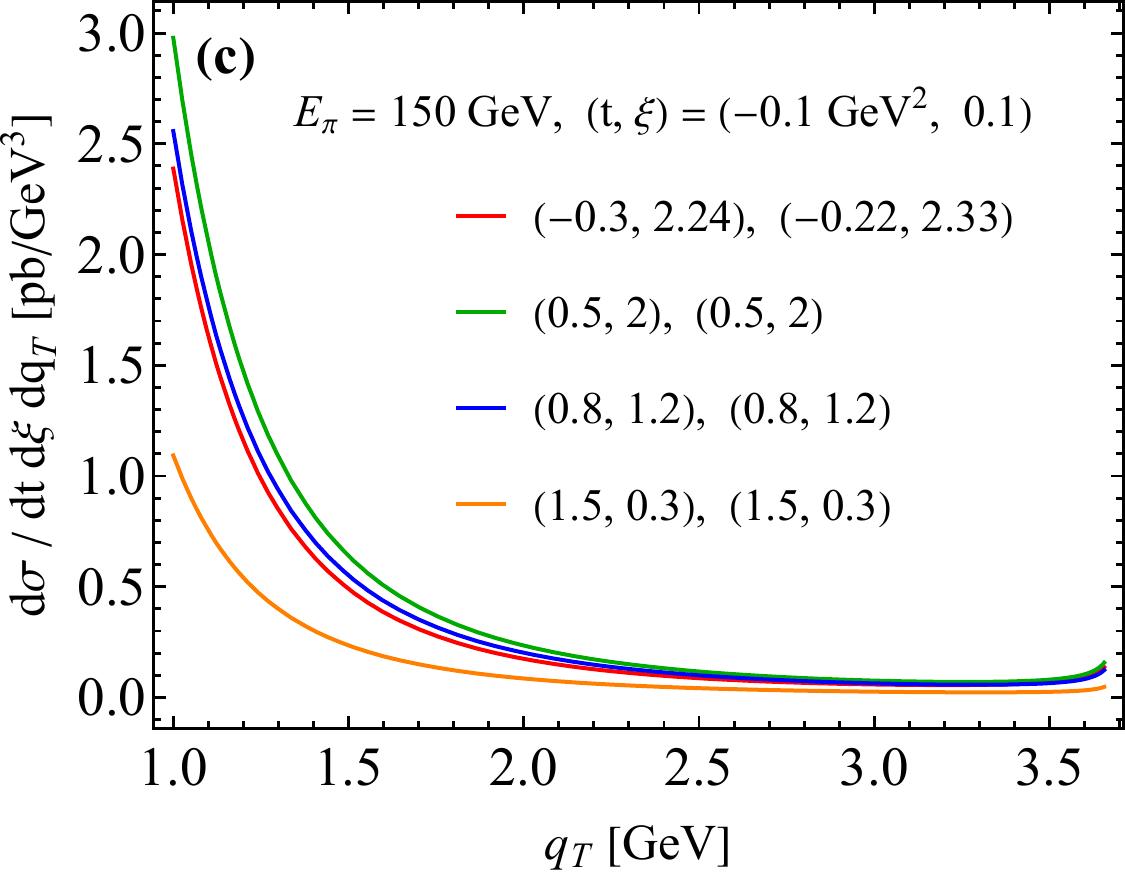}\quad
    \includegraphics[scale=0.6]{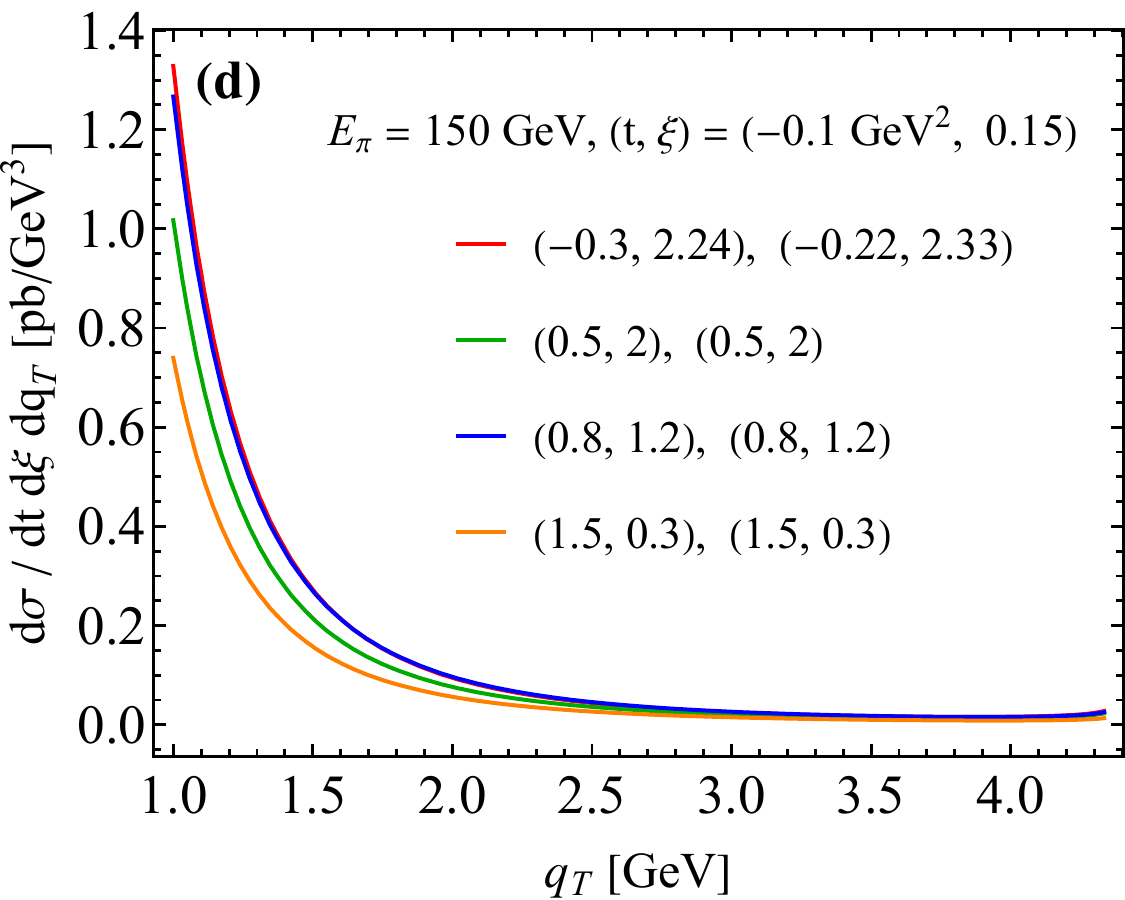}
    \caption{Differential cross section in eq.~\eqref{eq:xsec formula} as a function of photon $q_T$ for two choices of pion beam energies, along with two sets of $(t,  \, \xi)$ values.  Different curves correspond to different $(\rho, \tau)$ parameters for the GPD models of the chiral-even GPDs followed by that of the chiral-odd GPDs.  The rise at large $q_T$ is due to the Jacobian peak of the differential cross section.}
    \label{fig:pi-p absolute qt}
\end{figure*}

\begin{figure*}[htbp]
    \centering
    \includegraphics[scale=0.6]{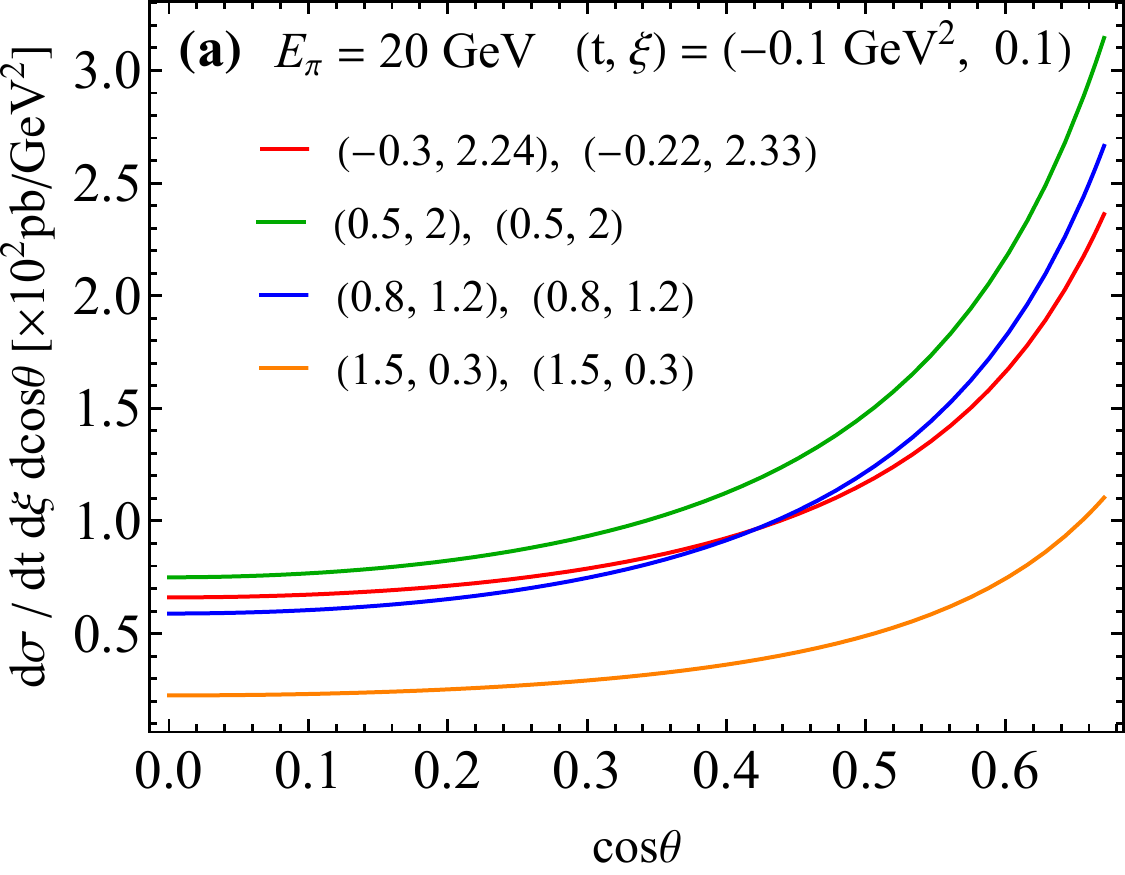}\quad
    \includegraphics[scale=0.6]{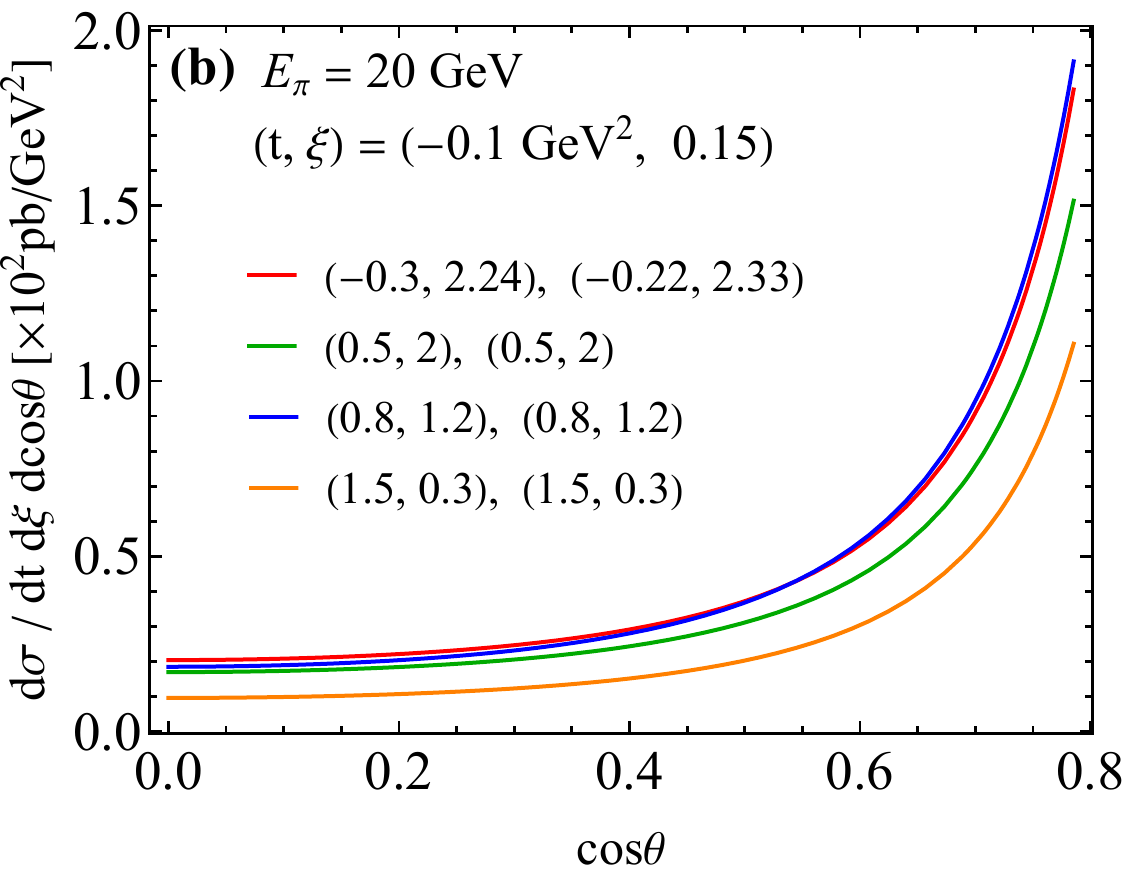}\\
    \vspace{1em}
    \includegraphics[scale=0.6]{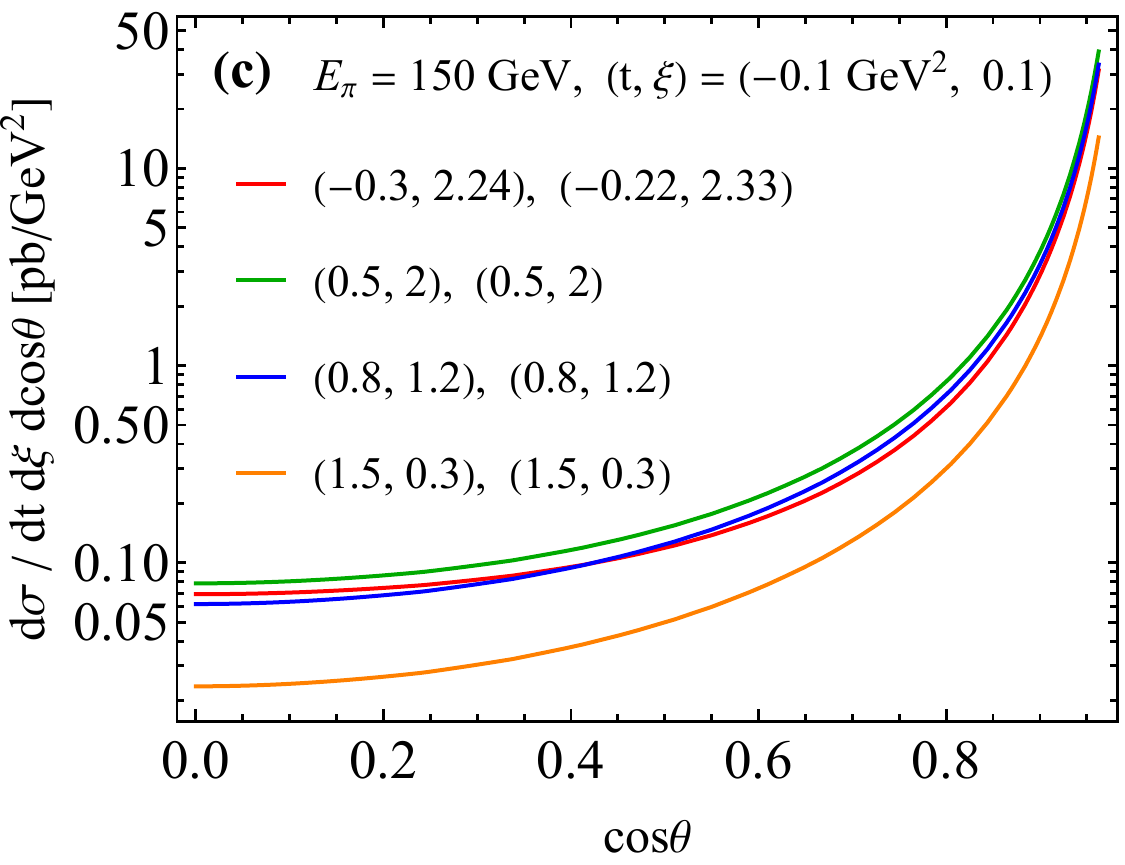}\quad
    \includegraphics[scale=0.6]{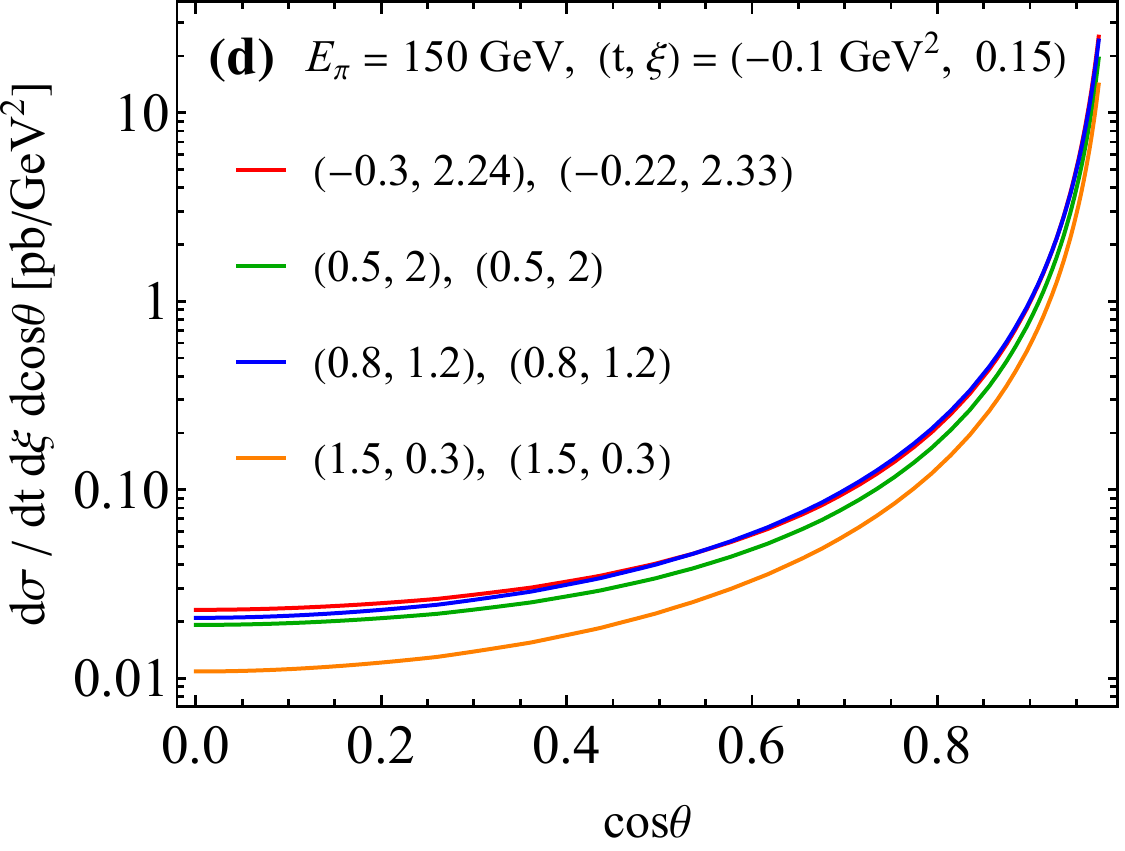}
    \caption{Differential cross section in eq.~\eqref{eq:xsec formula} as a function of $\cos\theta$ of the observed photon with all parameters chosen to be the same as that in figure~\ref{fig:pi-p absolute qt}.
    }
     \label{fig:pi-p absolute c}
\end{figure*}

\subsubsection{Sensitivity to GPD's $x$ dependence}
\label{sec:x-dependence GPD}

First, we examine the sensitivity of measured photon $q_T$ distribution to the $x$-dependence of nucleon GPDs. For simplicity we take $n_i \to \infty$ in eq.~\eqref{eq:weight} to remove the $\xi$ dependence for both $H$ and $\widetilde{H}$, and have a simplified model for nucleon transition GPDs
\begin{align}
H^{ud}_{pn}(x, \xi, t) &= \theta(x) x^{-0.9\, t/{\rm GeV}^2} 
			\frac{x^{\rho} (1-x)^{\tau} }{B(1+\rho, 1+ \tau)}		\, , \nn\\
\widetilde{H}^{ud}_{pn}(x, \xi, t) &= \theta(x) x^{-0.45\, t/{\rm GeV}^2} 
			\frac{1.267 \, x^{\rho} (1-x)^{\tau} }{B(1+\rho, 1+ \tau)}	\, .
\label{eq:H no-xi}
\end{align}
Apart from the best-fit parameters in eq.~\eqref{eq:best-fit}, we choose an additional set of parameters,
\beq
(\rho,\tau) = (0.5, 2),\, (0.8, 1.2),\, (1.5, 0.3),
\eeq
for both $H^{ud}_{pn}(x, \xi, t)$ and $\widetilde{H}^{ud}_{pn}(x, \xi, t)$. This gives a set of GPDs with their $x$-dependence peaked between $x=0$ and $1$, as shown in Figs.~\ref{fig:gpd_model} for $t = -0.1~{\rm GeV}^2$. Although there is no explicit $\xi$ dependence in eq.~\eqref{eq:H no-xi},  
the hard-part integration in \eqref{eq:Mi F Ft} still knows about $\xi$ since $z_1$ is a function of $x$ and $\xi$ as defined in eq.~\eqref{eq:z_1}.
Moreover, $\xi$ characterizes the CM energy of the hard collision (eq.~\eqref{eq:S s}) and thus the range of $q_T$. Therefore, the integration of $q_T$ also differs for different $\xi$. As a result there will still be substantial $\xi$ dependence of the cross section.

\begin{figure*}[htbp]
    \centering
    \includegraphics[scale=0.6]{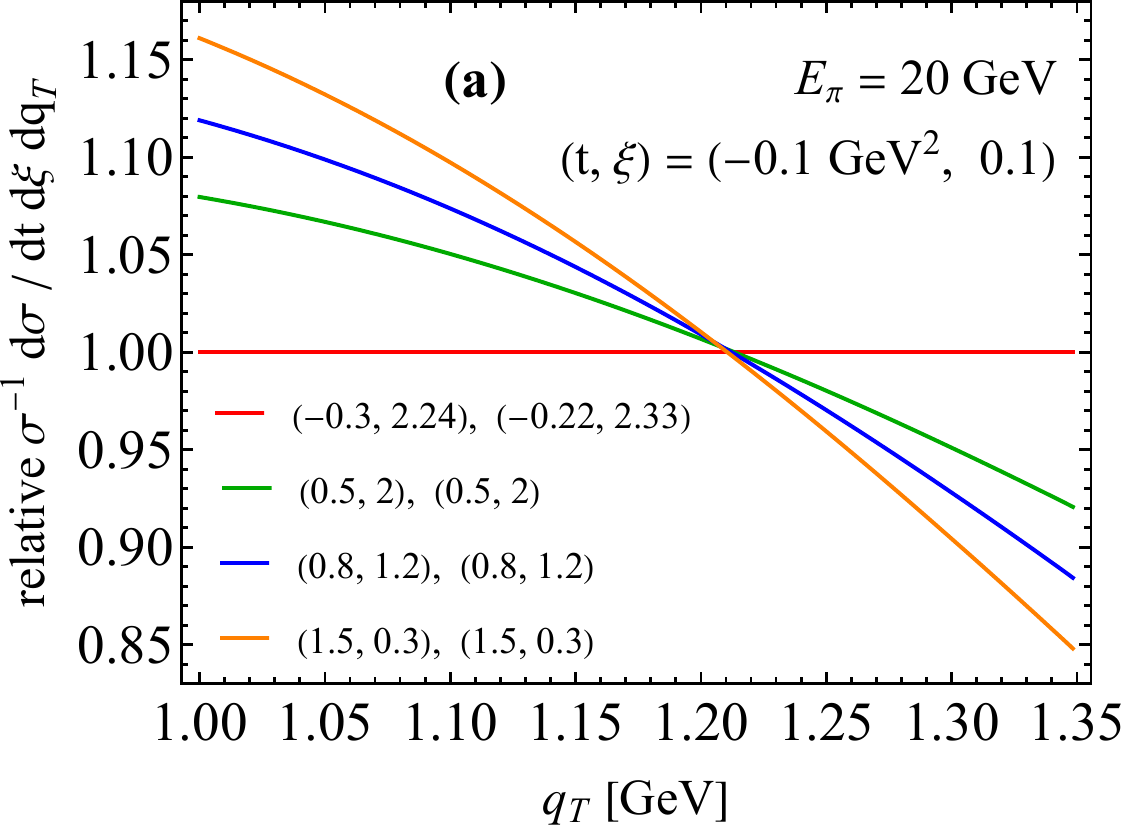}\quad
    \includegraphics[scale=0.6]{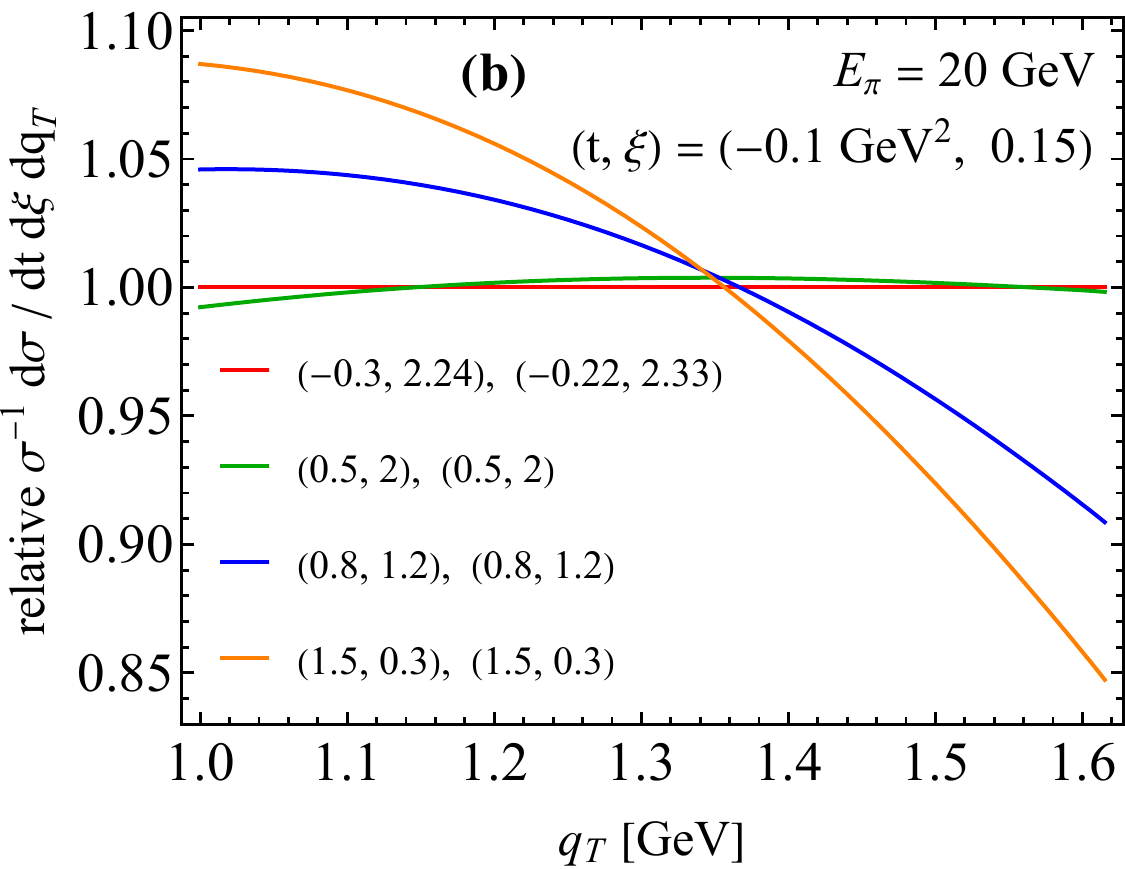}\\
    \vspace{1em}
    \includegraphics[scale=0.6]{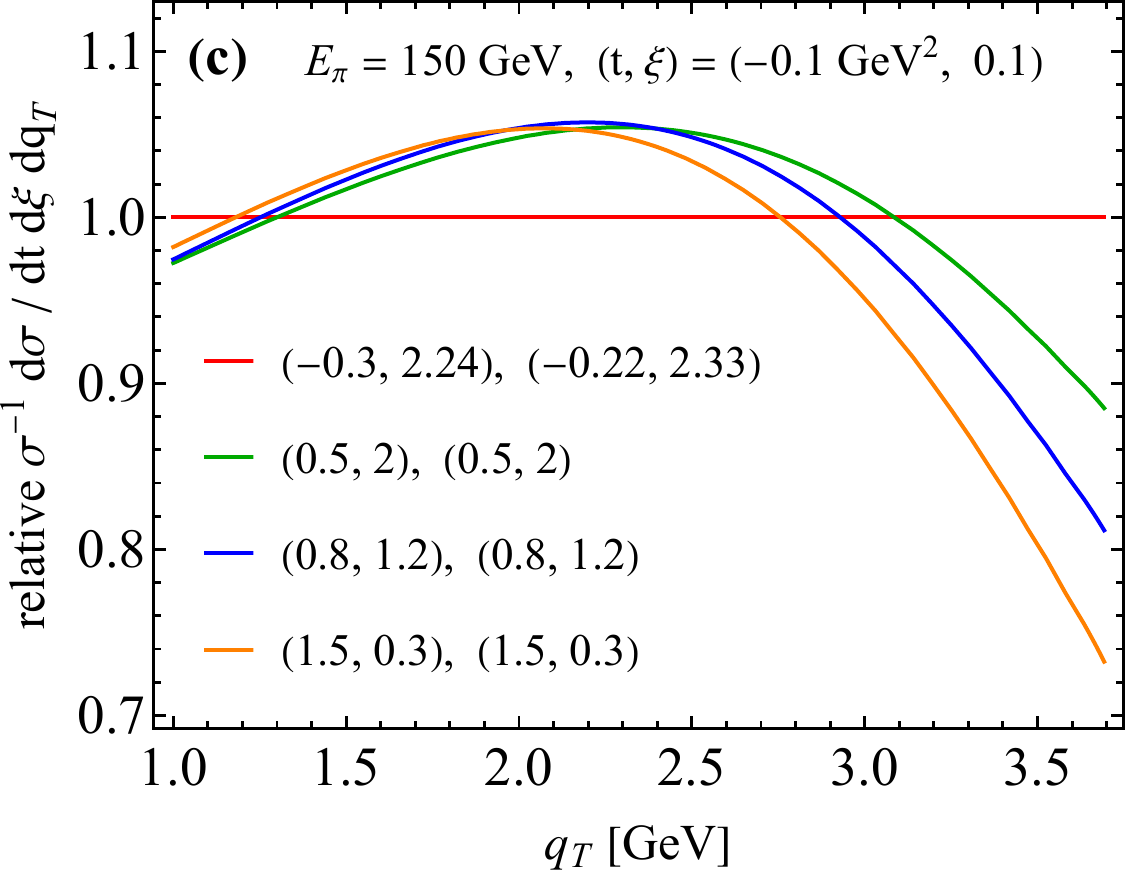}\quad
    \includegraphics[scale=0.6]{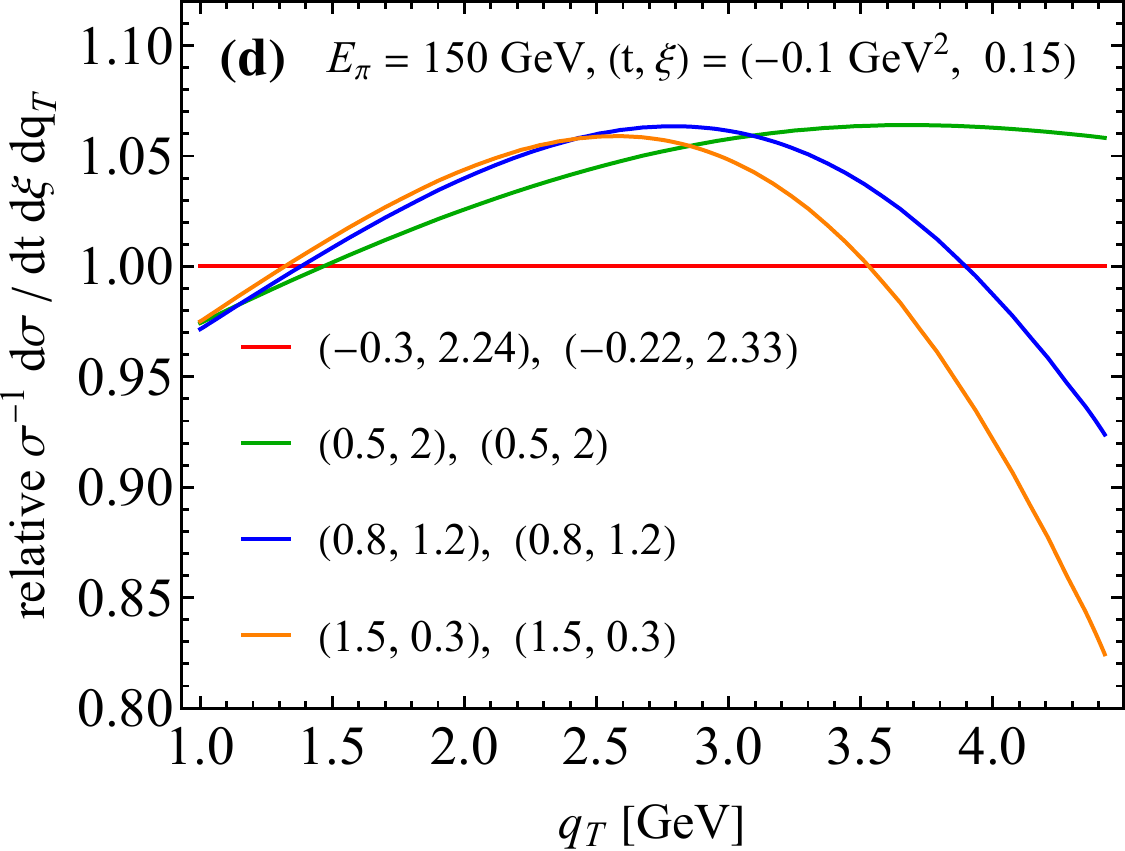}
    \caption{Ratio of normalized differential cross sections $\sigma^{-1} \dd \sigma/\dd t\, \dd\xi \, \dd q_T$ as a function of observed photon $q_T$ evaluated with the GPD model in eq.~\eqref{eq:H no-xi}. Different curves correspond to different parameter sets of the GPD model in eq.~\eqref{eq:H no-xi}.}
    \label{fig:pi-p relative qt}
\end{figure*}

With the model nucleon GPDs in figure~\ref{fig:gpd_model}, we plot in figure~\ref{fig:pi-p absolute qt} the absolute differential cross section in eq.~\eqref{eq:xsec formula} as a function of measured photon $q_T$ at both J-PARC and AMBER pion beam energies, along with two choices of $(t,  \, \xi)$ values.  We have restricted $q_T \ge 1$~GeV to ensure that power correction to the factorization formalism is sufficiently small.  The upper bound of $q_T$ depends on the collision energy and $\xi$.  Different curves correspond to different choices of $(\rho, \tau)$ parameters for the GPD models, which are chosen to be the same for both chiral-even and chiral-odd GPDs.  The rise at large $q_T$ is due to the Jacobian peak of the differential cross section.  We can avoid the Jacobian peak by plotting the differential cross sections with respect to $\cos\theta = \sqrt{1-4q_T^2/\hat{s}}$ with $\theta$ being the angle between the observed photon and collision $\hat{z}$-axis, instead of $q_T$, as shown in figure~\ref{fig:pi-p absolute c}. By comparing plots on the left and right --- with different $\xi$, and plots on the top and bottom --- with increase of collision energy $\sqrt{s}$, the $q_T$ distribution becomes more and more dominated by small $q_T$.  As $\sqrt{s}$ and $\xi$ (or $\sqrt{\hat{s}}$) increase, more phase space opens up for the production of the two back-to-back photons. As $q_T$ decreases, the virtualities of the quark propagators in the leading-order diagrams in figure~\ref{fig:hard1} and \ref{fig:hard2} decrease, leading to the enhancement of differential cross sections.

To make the difference of $q_T$ shapes more manifest to better visualize the sensitivity of measured $q_T$ distribution to the $x$-dependence of nucleon GPDs, we plot in figure~\ref{fig:pi-p relative qt} the {\it ratio} of the normalized differential cross sections as a function of $q_T$ for two different collision energies, like what we plotted in figure~\ref{fig:relative qT}.  The normalized cross sections are defined by dividing the differential cross sections by $\sigma(q_{T{\rm min}}=1~{\rm GeV})$. The ratio of the normalized differential cross sections is defined by dividing by the one evaluated with the best-fit GPD model parameters in eq.~\eqref{eq:best-fit} --- the red curve.  
Taking the ratio of normalized differential cross sections effectively removes the huge variation of the absolute values of the cross sections and enhances the dependence on the parameters of GPD models, as clearly shown in figure~\ref{fig:pi-p relative qt}.
It is evident that as the peak in $x$-distribution of GPD model in figure~\ref{fig:gpd_model} shifts from $0$ to $1$, the $q_T$ shape differs by around $10\%$ to $20\%$, without even considering the possible improvement from better control of the ``end-point'' sensitivity as discussed in Sec.~\ref{sec:Sudakov}. And by comparing figure~\ref{fig:relative qT} and \ref{fig:pi-p relative qt}, we find more sensitivity to the shape of GPD than that of DA, which means the sensitivity comes more from the DGLAP region than the ERBL region. Hence, we can conclude that the \textit{shape} of $q_T$ distribution has significant sensitivity to the $x$-dependence (or equivalently, the $z_1$-dependence) of GPDs.

\subsubsection{Sensitivity to GPD's $\xi$ dependence}

In contrast to the $x$-dependence of GPDs, which is proportional to the relative momentum of the active quark-antiquark pair from the diffractive nucleon, 
the $\xi$ and $t$ are direct kinematic observables once we measure the momentum of the diffracted nucleon in an event. So, in principle, getting information on $\xi$ and $t$ is much more direct than getting the $x$-dependence.  However, since GPDs are collective functions of $(x, \xi, t)$, extracting the $(\xi,t)$ dependence of GPDs from measured $(\xi, t)$ dependence of exclusive cross sections depends on how $x$-dependence is entangled with $\xi$- and $t$-dependence in GPDs, and also, in practice, how GPDs are parametrized in terms of their $(x,\xi,t)$-dependence.

The measured $\xi$-dependence of this new type of exclusive processes has three major sources: 
(1) $\xi$-dependence of GPDs, e.g., the parameter $n_i$-dependence of the GK model in eq.~\eqref{eq:weight};
(2) $\xi$-dependence from the factorized scattering amplitudes, i.e., the convolution in eq.~\eqref{eq:Mi F Ft}); and
(3) kinematic effect from the fact that $\xi$ characterizes the CM energy of the hard collision when cross section is expressed in terms of $(t, \xi, q_T^2)$.
The kinematic effect is reflected by the $(1-\xi^2)/\xi^2$ factor in eq.~\eqref{eq:xsec formula} and is independent of (1) and (2).  In principle, it is not possible to separate the $x$-dependence from the $\xi$-dependence of GPD because of (2), i.e., the convolution of GPD and hard coefficient depends on $\xi$. 
In this subsection, we try to explore to what extent the cross section depends on how $\xi$ is parametrized in the GPD.

To focus on the $x$-dependence, we set $n_i\to \infty$ in our GPD model in eq.~\eqref{eq:weight} in our discussion in last subsection, which led to a model of GPDs that has no dependence on $\xi$ as shown in eq.~\eqref{eq:H no-xi}.  To test the sensitivity to $\xi$-dependence, we choose $n_i = 0$ and $n_i = 1$ as two additional model GPDs.  We still keep the same parametrization of $h_i(\beta)$ in eq.~\eqref{eq:h(beta)}. The advantage of using small integers for $n_i$ is that we can analytically integrate out eq.~\eqref{eq:DD int} and express GPD in terms of special functions. Since our proposed process is only sensitive to the valence region, letting $n_i\to n_{\rm val}$, and combining eq.~\eqref{eq:DD int} with eqs.~\eqref{eq:DD ansatz}, \eqref{eq:weight}, \eqref{eq:DD val}, and \eqref{eq:h(beta)} gives us the GPD model, 
\begin{align}
{\rm (GPD)}^{ud}_{pn}(x,\xi, t) 	 = N 
\begin{dcases}
	\frac{ \B_{x_1}(1+\rho - \alpha_v' t, \tau) - 
	\B_{x_2}(1+\rho - \alpha_v' t, \tau) }{ 2 \, \xi \, B(1+\rho,\, 1+\tau)} 
		& x\geq \xi \, ,\\
	\frac{ \B_{x_1}(1+\rho - \alpha_v' t, \tau) }{ 
		2 \, \xi \, B(1+\rho,\, 1+\tau)} 
		& - \xi \leq x < \xi  \, ,\\
	\hspace{4em} 0 
		& x < -\xi  \, ,
\end{dcases}
\label{eq:n=0}
\end{align}
for $n_{\rm val} = 0$, and
\begin{align}
&{\rm (GPD)}^{ud}_{pn}(x,\xi, t) = N \nn\\
& \times	
\begin{dcases}
	\frac{3(1-\xi^2)}{4\xi^3 \B(1+\rho,\, 1+\tau)} \times \bigg[	
	 - \B_{x_1}(3 + \rho - \alpha_v' t, \tau - 2)  
	 + \B_{x_2}(3 + \rho - \alpha_v' t, \tau - 2) \\
	\hspace{2em} + (x_1+x_2) \bigg( \B_{x_1}(2 + \rho - \alpha_v' t, \tau - 2)
	-  \B_{x_2}(2 + \rho - \alpha_v' t, \tau - 2)	 \bigg) \\
	\hspace{2em} - x_1 x_2 \bigg( \B_{x_1}(1 + \rho - \alpha_v' t, \tau - 2)  
	- \B_{x_2}(1 + \rho - \alpha_v' t, \tau - 2)  \bigg) \bigg]
		& {\hskip -0.1in}  x\geq \xi \, , \\
	\frac{3(1-\xi^2)}{4\xi^3 \B(1+\rho,\, 1+\tau)} \times \bigg[	
	- \B_{x_1}(3 + \rho - \alpha_v' t, \tau - 2)  \\
	\hspace{2em} + (x_1+x_2) \B_{x_1}(2 + \rho - \alpha_v' t, \tau - 2)	
	- x_1 x_2  \B_{x_1}(1 + \rho - \alpha_v' t, \tau - 2) 
	\bigg]
		&{\hskip -0.3in}  - \xi \leq x < \xi \, , \\
	\hspace{4em} 0 
		& {\hskip -0.1in} x < -\xi \, ,
\end{dcases}
\label{eq:n=1}
\end{align}
for $n_{\rm val} = 1$,
where 
\beq
x_1 = \frac{x+\xi}{1+\xi}, \quad
x_2 = \frac{x-\xi}{1-\xi} \, ,
\eeq
and
\beq
\B_x(a, \, b) = \int_0^x \dd y \, y^{a-1} \, (1-y)^{b-1}
\eeq
is the incomplete Beta function. The parameter $b_v$ in eq.~\eqref{eq:DD ansatz} has been set to $0$. $\alpha_v'$ and $N$ are taken unchanged from $n_i = \infty$.

\begin{figure}[htbp]
    \centering
    \includegraphics[scale=0.38]{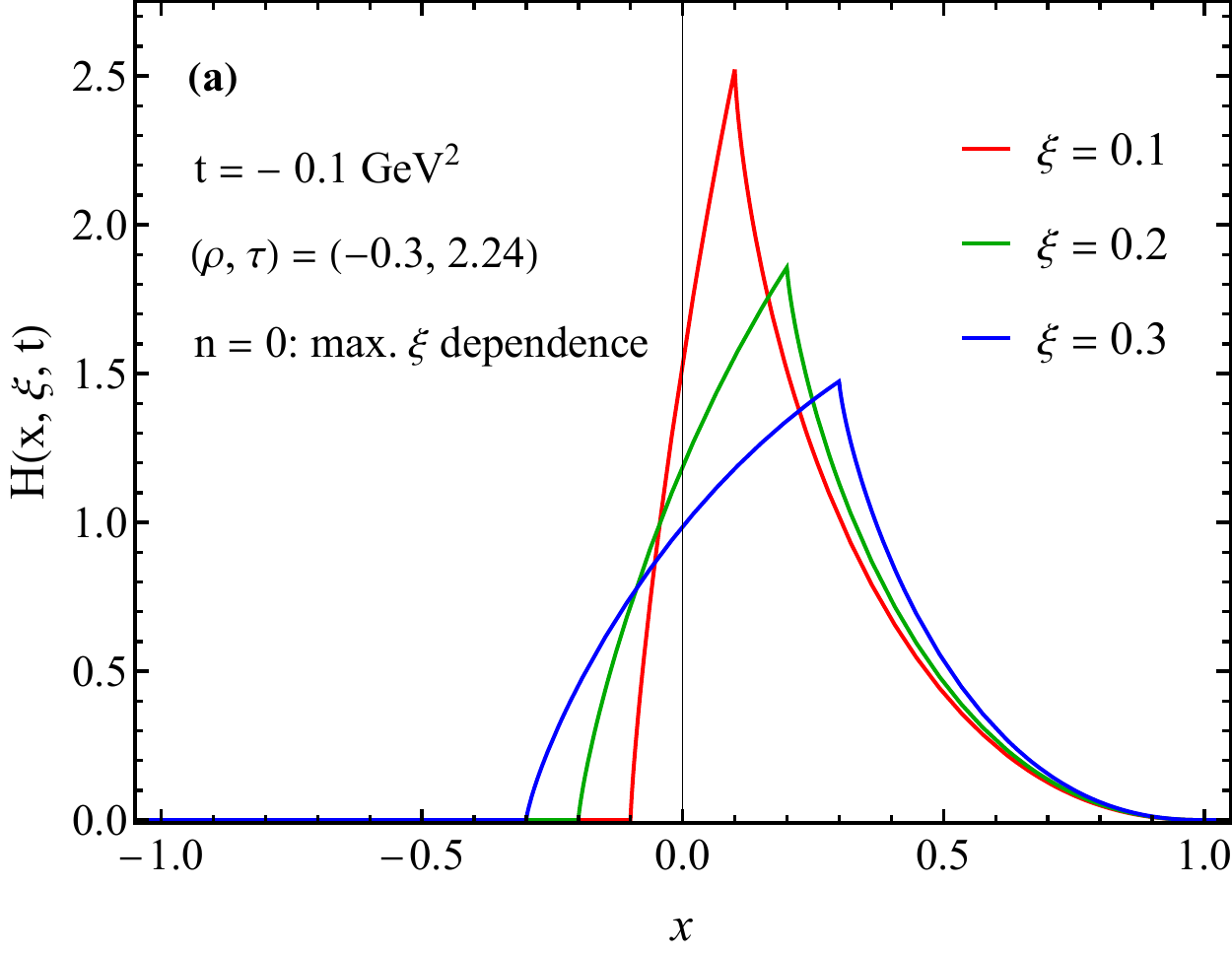}	\,
    \includegraphics[scale=0.38]{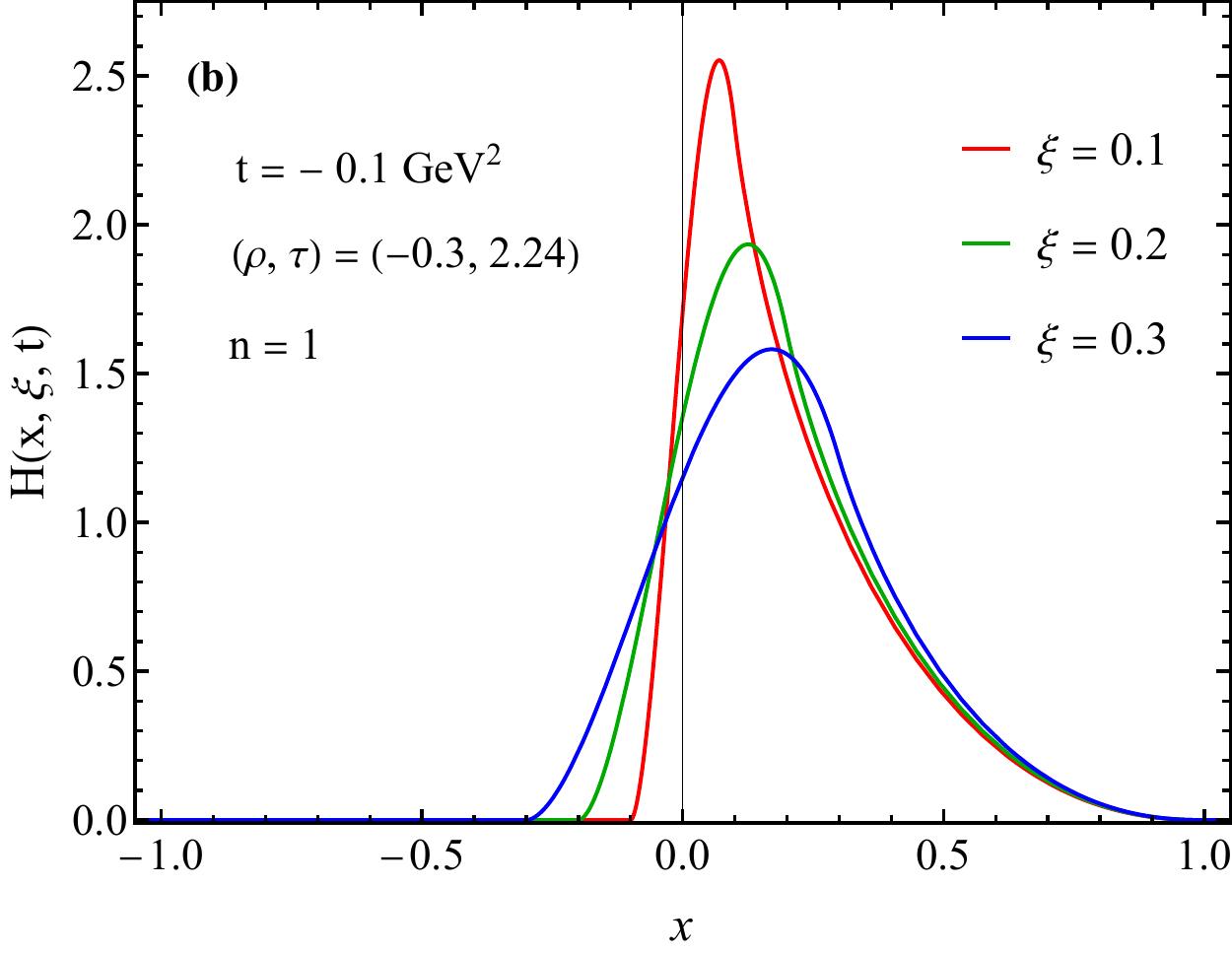}	\,
	\includegraphics[scale=0.38]{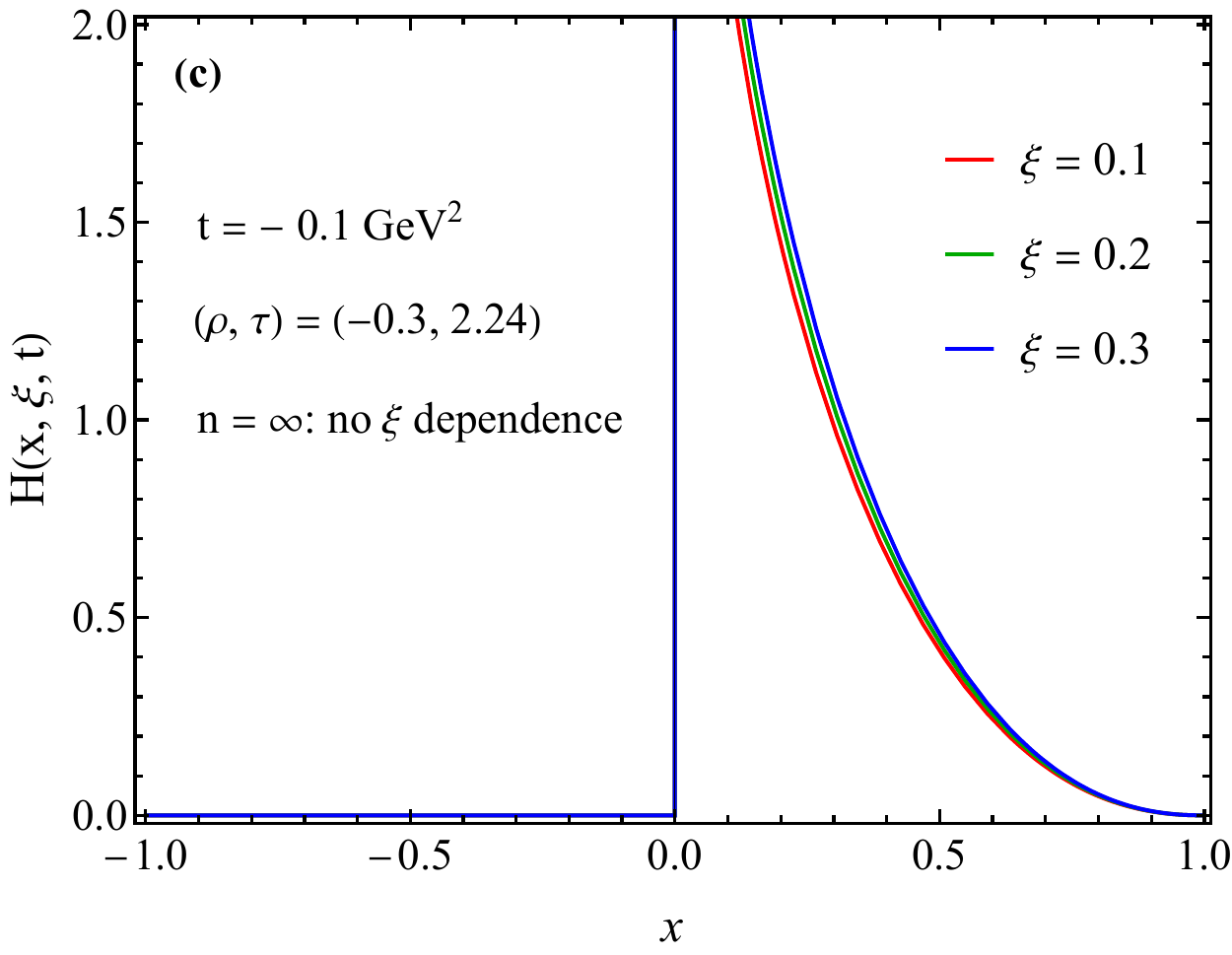}
    \caption{Chiral-even GPD $H(x,\xi,t)$ as a distribution of $x$ for three different $\xi$ and three different values of $n_{\rm val}$, which controls the GPD $\xi$ dependence in the GK model.  }
    \label{fig:gpd xi}
\end{figure}

\begin{figure}[htbp]
    \centering
    \includegraphics[scale=0.5]{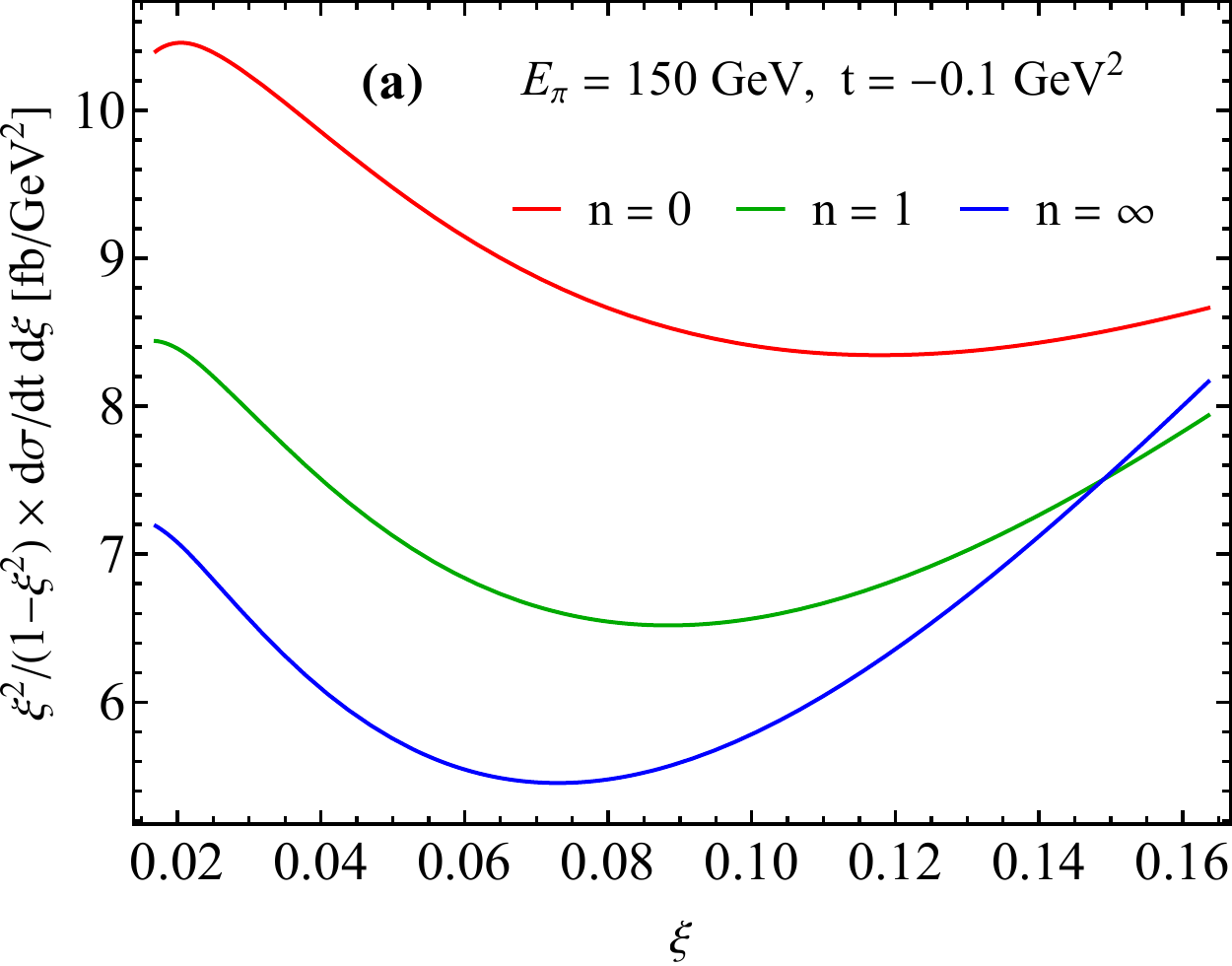}	\hspace{2em}
    \includegraphics[scale=0.5]{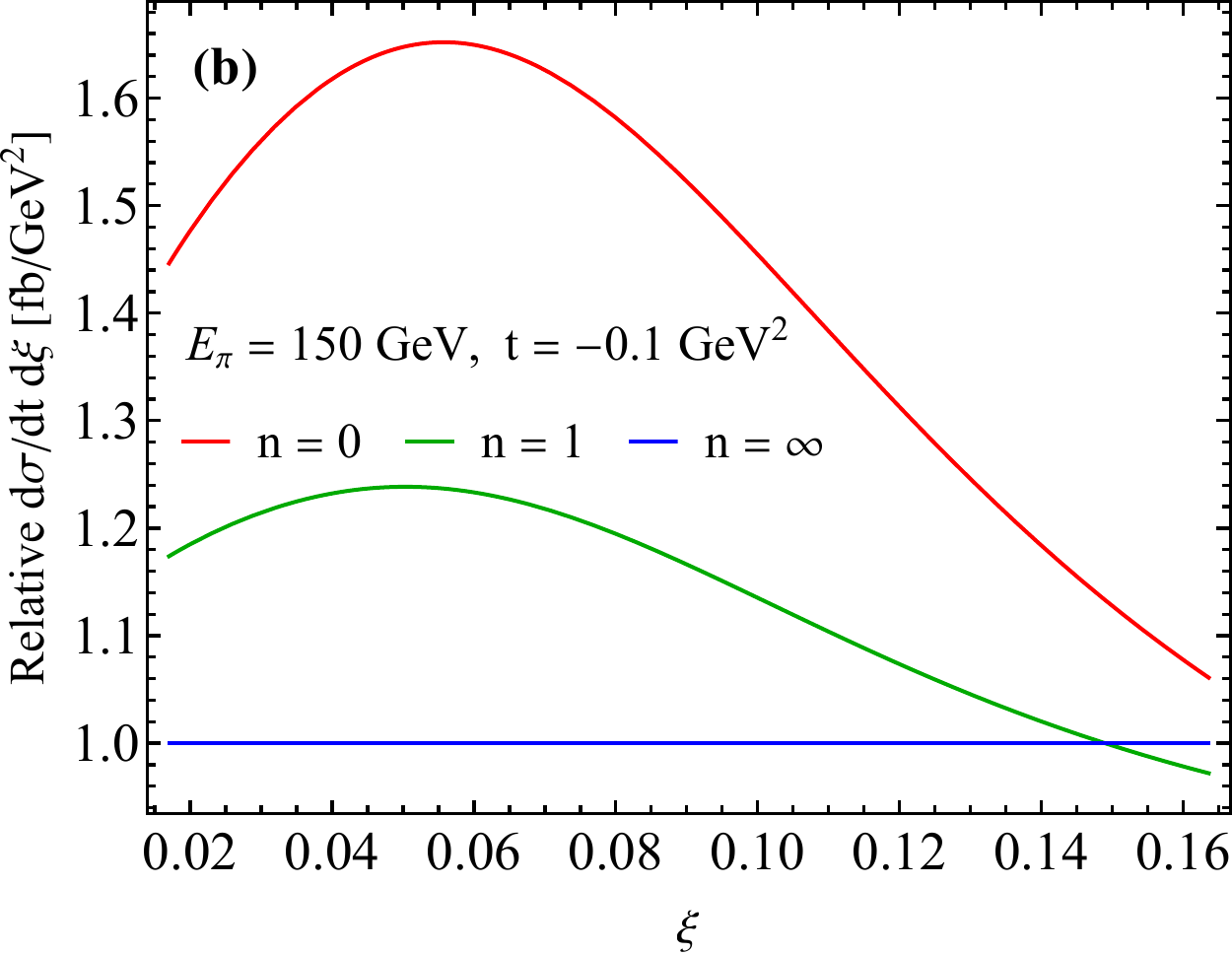}
    \caption{(a) Absolute and (b) relative distributions of $\xi$ at $t = -0.1~{\rm GeV}^2$ for $150~{\rm GeV}$ pion beam, for the three different GPD models shown in figure~\ref{fig:gpd xi}. The relative distribution in (b) is obtained by dividing each curve in (a) by the one with $n=\infty$.}
    \label{fig:gpd xi xsec}
\end{figure}

In figure~\ref{fig:gpd xi}, we plot our models for chiral-even GPD $H(x,\xi,t)$ as functions of $x$ for three values of $n_{\rm val} = 0, 1, \infty$.  Our model GPDs for $n_{\rm val} = 0$ and $1$ are given in eqs.~\eqref{eq:n=0} and \eqref{eq:n=1}, respectively.  For $n_{\rm val} = \infty$, the GPDs are given in eq.~\eqref{eq:H no-xi}.  We fix $(\rho, \tau)$ to be the best-fit values~\eqref{eq:best-fit}, $t=-0.1~{\rm GeV}^2$ and show GPDs for three values of $\xi (=0.1,0.2,0.3)$.  GPDs with $n_{\rm val} = 0$ have the maximum $\xi$ dependence while those with $n_{\rm val} = \infty$ have no $\xi$ dependence, which is clearly evident from the examples of chiral-even GPD $H(x,\xi,t)$ in figure~\ref{fig:gpd xi}. 

By integrating out $q_T$, we plot the cross section as a distribution of $\xi$ in figure~\ref{fig:gpd xi xsec} for the AMBER energy $E_{\pi} = 150~{\rm GeV}$, where the kinematic factor $(1-\xi^2)/\xi^2$ has been divided out. We see that different $\xi$ parametrizations do reflect themselves in the $\xi$-distribution of differential cross sections. Their relative differences are better seen by taking ratios to the one with $n=\infty$, as seen in figure~\ref{fig:gpd xi xsec}(b). Comparing $n=\infty$ with $n=1$, we see that introducing some $\xi$ dependence to GPDs through $n=1$ leads to $20\%$ change to the $\xi$ distribution of the cross sections. Then increasing the $\xi$ dependence from $n=1$ to $n=0$ leads to a further $20\% \sim 40\%$ change.

\subsubsection{Sensitivity to GPD's $t$ dependence}

Same as the $\xi$-dependence, the $t$-dependence of the diffractive cross section is experimentally determined.  On the other hand, the $t$-dependence of theoretically factorized cross section comes from (1) the $t$ dependence of GPD and (2) kinematic effect of hard process.  As shown in eq.~\eqref{eq:xi range t}
the value of $t$ actually constrains the available range of the $\xi$.

It is worth emphasizing that $t$ does not enter the hard process directly as an immediate consequence of the leading-power factorization, which is accurate up to power corrections of $|t|/Q^2$. However, the information of $t$ is not lost, but is captured by GPDs.  The Fourier transform of GPDs with respect to the transverse component of $t$ leads to transverse spatial density distributions of quarks and gluons inside a bound hadron, which could reveal very valuable information on how quarks and gluons are distributed in an environment of a confined hadron.
Comparing with the $x$-dependence, $t$-dependence is more visible in a physical process and will not be explored in more details in this work.

\section{Discussion and Outlook}
\label{s.outlook}

Exclusive processes provide valuable information that is different from and complementary to inclusive processes. Without breaking the hadron, exclusive diffractive processes that can be factorized into DAs and GPDs provide not only one hard scale to characterize the particle nature of quarks and gluons inside the hadron, but also a secondary soft scale $t$ that allows us to probe into the transverse structure of the hadron to explore much needed information on spatial distributions of quarks and gluons in a bound hadron.  However, the hard scale for many existing exclusive processes, such as pion Form Factor and DVCS on a nucleon, is provided by a single virtual particle, and measured exclusive cross sections are most sensitive to the total momentum transfer from the diffracted hadron, but not to the relative momentum of the quark-antiquark or two gluons from the diffractive hadron.   Consequently, such exclusive processes are only sensitive to the ``moments'' of DAs and/or GPDs, such as  $ \int_0^1 \dd z \, z^{-1} \, \phi_{\pi}(z)$, as discussed in Sec.~\ref{s.numerical}.  The information on such ``moments'' is far from enough to constrain the functional forms of DAs and GPDs, and three-dimensional spatial imaging of quarks and gluons.  That is, we need to seek for more exclusive processes, like the one that we proposed in this paper, to provide better constraints on the hadron tomography.

In this paper, we introduced exclusive production of a pair of high transverse momentum photons in pion-nucleon collisions and demonstrated that this new $2\to 3$ exclusive diffractive process can be systematically factorized into universal pion's distribution amplitude and nucleon's generalized parton distributions, which are convoluted with corresponding infrared safe and perturbatively calculable short-distance hard parts.  The correction to the factorization of this exclusive process is suppressed by powers of $\Lambda_{\rm QCD}/q_T$.  We also showed quantitatively that this new type of exclusive processes is not only complementary to existing processes for extracting GPDs, but also capable of providing an enhanced sensitivity to the parton momentum fraction of both DAs and GPDs from the measured transverse momentum $q_T$ distribution.  This new $2\to 3$ exclusive process could be measured at J-PARC and AMBER.  In addition, our proof of the leading-power factorization for exclusive production of a pair of high transverse momentum photons can be carried through for justifying the factorization of exclusive Drell-Yan production in $\pi N\to \ell^+\ell^-(Q) N'$ when $Q\gg\sqrt{|t|}$, which could be measured at J-PARC and other facilities as well. 

We stress that the sensitivity of observed $q_T$ shape to the functional form of GPDs is because there are {\it two} back-to-back particles coming out of the hard collision and the momentum flow between these two back-to-back particles entangles with the relative momentum of the two active partons from the diffractive hadron.  It is this entanglement of momenta that provides the additional sensitivity to the $x$-dependence of GPDs.  

In addition to the exclusive production of two high transverse momentum photons in the diffractive pion-nucleon collisions, discussed in this paper, we introduce a new class of similar $2\to 3$ exclusive processes for diffractive production of a back-to-back high transverse momentum pair of particles (or jets), $C(p_c)$ and $D(p_d)$, from a hadron $h(p)$ to a hadron $h'(p')$,
\begin{equation}
A(p_1) + h(p)\to C(p_c) + D(p_d) + h'(p')
\label{eq:diff2p}
\end{equation}
\noindent
with $\sqrt{|(p-p')^2|} \ll |p_{c_T}| \sim |p_{d_T}|$. The exclusive process in eq.~\eqref{eq:diff2p} can be viewed effectively as a combination of a diffractive production of the virtual and ``long-lived'' partonic state(s) $B^*$: $h(p) \to B^*(p_2) + h'(p')$ and an exclusive production of two back-to-back high transverse momentum particles (or jets) on such virtual state(s):\ $A(p_1)+B^*(p_2) \to C(p_c)+D(p_d)$. 
The necessary condition for QCD factorization of the exclusive process in eq.~\eqref{eq:diff2p} is that
the virtuality of the intermediate state(s) $B^*$ is much smaller than its energy, 
i.e., the lifetime of $B^*$ is much longer than the timescale of the hard exclusive scattering to produce the two back-to-back high transverse momentum particles (or jets). It is the long lifetime of the intermediate state $B^*$ that effectively suppresses the quantum interference between the diffractive production of the $B^*$ and the hard exclusive scattering between $A(p_1)$ and $B^*(p_2)$.  This necessary condition effectively requires that the transverse momentum $p_{c_T}\sim p_{d_T}$ be much larger than the soft scale, $\sqrt{-t}=\sqrt{-(p-p')^2}$.
We will present the sufficient condition(s) for QCD factorization of various $2\to 3$ exclusive processes of the type in eq.~\eqref{eq:diff2p} in a future publication. 
For example, from the elastic large angle pion-pion scattering, 
$\pi(p_1)+\pi(p_2)\to \pi(p_c) + \pi(p_d)$ we can have a new exclusive $2\to 3$ diffractive production of a back-to-back pair of high transverse momentum pions:
$\pi(p_1)+h(p)\to \pi(p_c) + \pi(p_d) +h'(p')$ with $p_{c_T}\sim p_{d_T}\gg \sqrt{-(p-p')^2}$,
if the $\pi\pi\to\pi\pi$ large-angle elastic scattering is dominated by a single hard scattering.
Similarly, instead of the exclusive pion-baryon scattering to produce two back-to-back high transverse momentum photons, we can switch one of the final-state photon with the initial-state pion to have another $2\to 3$ exclusive process: $\gamma(p_1) + N(p) \to \gamma (q_1)+\pi(p_\pi) + N'(p')$ with the back-to-back high transverse momentum photon-pion pair~\cite{Boussarie:2016qop,Duplancic:2018bum}. Taking the advantage of photon polarization at Jefferson Lab, polarization asymmetries of an exclusive diffractive photo-production of two high transverse momentum and back-to-back particles could provide additional channels of observables to extract various GPDs with better sensitivities on their $x$-dependence, which will be explored in our future publications.

\paragraph*{Acknowledgements.}

We thank Profs. Wen-Chen Chang, Shunzo Kumano and Shin\'ya Sawada for correspondence on measuring GPDs at J-PARC, Profs.~Geoffrey Bodwin, Markus Diehl and C.-P.~Yuan for helpful discussions, and also Prof.~Bernard Pire for correspondence on some important references. This work is supported in part by the US Department of Energy (DOE) Contract No.~DE-AC05-06OR23177, under which Jefferson Science Associates, LLC operates Jefferson Lab, and within the framework of the TMD Collaboration. The work of Z.Y. at MSU is partially supported by the U.S.~National Science Foundation under Grant No.~PHY-2013791, and  the fund from the Wu-Ki Tung endowed chair in particle physics. The Feynman diagrams in this paper were all drawn using JaxoDraw~\cite{Binosi:2003yf, Binosi:2008ig}.

\appendix
\newpage

\def\C{\mathcal{C}}
\def\Co{\widetilde{\mathcal{C}}}
\def\z{\widetilde{z}}
\def\p{\widetilde{p}}
\def\re{{\rm Re~}}
\def\im{{\rm Im~}}

\section{Summary of the hard coefficients}
\label{app:hard coefficients}
\label{a.a}

In this appendix, we present all leading order contributions to the scalar hard coefficients for the exclusive $\pi^++\pi^-\to \gamma+\gamma$ and $p^++\pi^-\to n+\gamma+\gamma$ processes from each diagram.

As defined in eq.~\eqref{eq:gauge-inv expansion}, the scalar coefficients from each diagram can be extracted in the same way as what was done for extracting eq.~\eqref{eq:scalar C A1}. 
For the $p\pi^-$ process, the variable $z_1$ can take values in $[z_m, z_M]$ (see eq.~\eqref{eq:z1limits}), so the $i\varepsilon$ prescription is very important, while for the $\pi^+\pi^-$ case, $z_1$ is only within $[0, 1]$ and the $i\varepsilon$ can be neglected. In the following, we give the hard coefficients for the $p\pi^-$ case, which contains both chiral-even and chiral-odd components. The $\pi^+\pi^-$ process only has chiral-even hard coefficients and can be adapted from those of the $p\pi^-$ case by the replacement $(\Delta, \hat{s}, \hat{\kappa}) \to (p_1, s, \kappa)$ and keeping only the real parts. 

Instead of giving the expression for each individual diagram or scalar coefficient $C_i$ ($\widetilde{C}_i$), we define the following combinations of the hard scalar coefficients based on eq.~\eqref{eq:T p pi},
\begin{align}
\C_0 &= C_0,	\nn\\
\C_1 &= \frac{C_1 + C_2}{4} - \frac{(\Delta\cdot q_1)^2 C_3 + (\Delta\cdot q_2)^2 C_4}{ \hat{s} \, q_T^2} \nn\\
         &= \frac{1}{4} \left[ C_1 + C_2 - \frac{\left(1-\sqrt{1-\hat{\kappa}} \right)^2}{\hat{\kappa}} C_3 
                                      - \frac{\left(1+\sqrt{1-\hat{\kappa}} \right)^2}{\hat{\kappa}} C_4 \right],	\nn\\
\Co_1 &= \frac{(\Delta\cdot q_1) \widetilde{C}_1 - (\Delta\cdot q_2) \widetilde{C}_2}{\hat{s}}
=  \frac{1}{4} \left[ \left(1-\sqrt{1-\hat{\kappa}} \right) \widetilde{C}_1 - \left(1+\sqrt{1-\hat{\kappa}} \right) \widetilde{C}_2 \right],	\nn\\
\Co_2 &= \frac{(\Delta\cdot q_2) \widetilde{C}_3 - (\Delta\cdot q_1) \widetilde{C}_4}{\hat{s}}
=  \frac{1}{4} \left[ \left(1+\sqrt{1-\hat{\kappa}} \right) \widetilde{C}_3 - \left(1-\sqrt{1-\hat{\kappa}} \right) \widetilde{C}_4 \right].
\end{align}
where $\hat{\kappa}=4q_T^2/\hat{s}$, which was first defined in eq.~\eqref{eq:xsec formula0} as an analog of the $\kappa$ for $\pi\pi$ annihilation case in eq.~\eqref{eq:kin q12}.
Then eq.~\eqref{eq:T p pi} and \eqref{eq:T p pi 2} can be written as
 \begin{align} 
\left|\overline{\mathcal{M}}\right|^2
=& \left( \frac{e^2 g^2}{4} \frac{f_{\pi}}{\hat{s}} \frac{C_F}{N_c} \right)^2
\nn \\
&\hspace{1em} \times 
\left[ 
\left| \M[\C_0; \widetilde{\cal F}^{ud}_{pn}, (z_m, z_M); \phi, (0, 1)] \right|^2 + 
\left| \M[\C_1; \widetilde{\cal F}^{ud}_{pn}, (z_m, z_M); \phi, (0, 1)] \right|^2 \right.	\nn\\
& \hspace{2em}
\left. + \left| \M[\Co_1; {\cal F}^{ud}_{pn}, (z_m, z_M); \phi, (0, 1)] \right|^2 + 
\left| \M[\Co_2; {\cal F}^{ud}_{pn}, (z_m, z_M); \phi, (0, 1)] \right|^2
\right]	\\
\simeq &
(1-\xi^2) \left( \frac{e^2 g^2}{4} \frac{f_{\pi}}{\hat{s}} \frac{C_F}{N_c} \right)^2
\nn \\
&\hspace{1em}  \times
\left[ 
\left| \M[\C_0; \widetilde{\cal H}^{ud}_{pn}, (z_m, z_M); \phi, (0, 1)] \right|^2 + 
\left| \M[\C_1; \widetilde{\cal H}^{ud}_{pn}, (z_m, z_M); \phi, (0, 1)] \right|^2 \right.	\nn\\
& \hspace{2em}
\left. + \left| \M[\Co_1; {\cal H}^{ud}_{pn}, (z_m, z_M); \phi, (0, 1)] \right|^2 + 
\left| \M[\Co_2; {\cal H}^{ud}_{pn}, (z_m, z_M); \phi, (0, 1)] \right|^2
\right]
\end{align}

\noindent
The coefficient $\C_0$ is given in terms of its real and imaginary part,
\begin{align}
\re\C_0 =&\, \left(e_u-e_d\right)^2 \left[ \frac{8}{\hat{\kappa}  z_1 z_2 (1-z_1)(1-z_2)} \right]	
   +\left(e_u^2-e_d^2\right) \left[ \frac{4 (z_1 - z_2)}{z_1 z_2 (1-z_1)(1-z_2)} \right]	\nn\\
   &\, + \frac{8 e_u e_d}{z_1 z_2 (1-z_1)(1-z_2)} \frac{(z_1 (1-z_1) + z_2 (1-z_2)) (z_1 z_2 + (1-z_1)(1-z_2))}{ 4 z_1 z_2 (1-z_1)(1-z_2) + \hat{\kappa} (1-z_1-z_2)^2  },	
   \label{eq:ReC0}
   \\
\im\C_0 =&\, - 4 \pi \left(e_u-e_d\right)^2 \left[ \frac{\delta(1 - z_1)}{z_2} + \frac{\delta(z_1)}{1-z_2} \right]	
   + 8 \pi \left(e_u^2-e_d^2\right) \left[ \frac{\delta(z_1) - \delta(1 - z_1)}{\hat{\kappa} z_2 (1-z_2)} \right]	\nn\\
   &\, +\frac{ 8 \pi  e_u e_d }{\hat{\kappa}} 
   \left[ \left(\frac{1}{z_2 - z_{as}} + \frac{\sqrt{1-\hat{\kappa}}}{z_2 (1-z_2)}\right) \delta(z_1 - p(z_2)) \text{sgn}(z_2 - z_{as}) 	\right. \nn \\
   & \hspace{5em}
   +\left(\frac{1}{z_2 - \z_{as}(\hat{\kappa})} - \frac{\sqrt{1-\hat{\kappa}}}{z_2 (1-z_2)}\right)\delta(z_1 - \p(z_2)) \text{sgn}(z_2 - \z_{as}) \nn \\
   & \hspace{5em}
   \left. - \left(\frac{1+\hat{\kappa} }{z_2} + \frac{2}{1-z_2}\right) \delta(1 - z_1)  - \left(\frac{1+\hat{\kappa} }{1-z_2}+\frac{2}{z_2}\right) \delta(z_1)\right]	\, ,
\end{align}
where we defined 
\begin{align}
z_{as} &= \frac{1+\sqrt{1-\hat{\kappa}}}{2}, \quad\quad\quad
\z_{as} = \frac{1-\sqrt{1-\hat{\kappa}}}{2},  \nn \\
p(z_2) &= - z_{as} \frac{1-z_2}{z_2 - z_{as}}, \quad\quad\quad
\p(z_2) = - \z_{as} \frac{1-z_2}{z_2 - \z_{as}}.
\end{align}
The terms without $\delta$-functions contain $z_1$ poles when convoluting with the GPDs, which should be understood in the sense of principal value. 
We have organized the hard coefficient in terms of $(e_u - e_d)^2$, $(e_u^2 - e_d^2)$ and $e_u e_d$, where the $e_u e_d$ part comes exclusively from Type-$A$ diagrams. One immediately notices that the $(e_u - e_d)^2$ and $(e_u^2 - e_d^2)$ terms only give ``moment"-type sensitivity to the functional forms of GPD/DA while the $e_u e_d$ terms contain enhanced sensitivity. 

In order to show how the shape of $\hat{\kappa}$ is sensitive to the $z_1$ or $z_2$ distribution, we show in figure~\ref{fig:kappa z1 z2} the shapes of $\Re \C_0$ distribution in eq.~\eqref{eq:ReC0} as a function of $\hat{\kappa}$ for a few different values of $(z_1, z_2)$, where the $e_u$ and $e_d$ refer to the fractional charges of the two active quark (or antiquark) lines, which are not necessarily the up and down quarks.  For the purpose of illustration, we present two different cases: (a) $(e_u, e_d) = (2/3, -1/3)$ and (2) $(e_u, e_d) = (2/3, 2/3)$.  For the case (a), all three terms in eq.~\eqref{eq:ReC0}, proportional to $(e_u - e_d)^2$, $(e_u^2 - e_d^2)$ and $e_u e_d$, respectively, 
will contribute, while for the case (b), only the term proportional to $e_u e_d$ contributes.  Apart from the evident sensitivity of $\hat{\kappa}$ shape to the $z_1$ and $z_2$ values, which is true for both cases, the sensitivity for the case (b), which does not have the terms proportional to $(e_u - e_d)^2$ and $(e_u^2 - e_d^2)$, is much stronger, as shown in figure~\ref{fig:kappa z1 z2}(b), which indicates that these two terms are larger in the case of $\Re \C_0$.  

\begin{figure}[htbp]
\centering
\includegraphics[scale=0.6]{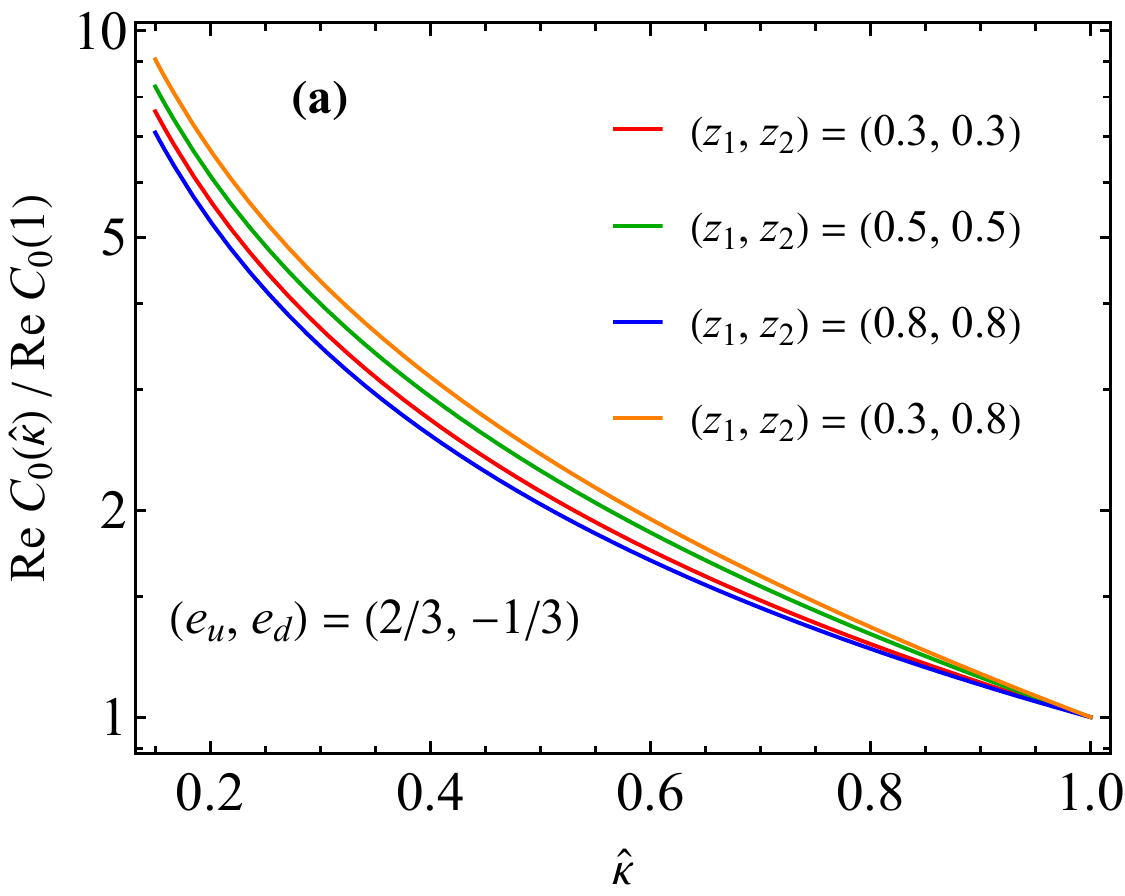} \quad
\includegraphics[scale=0.6]{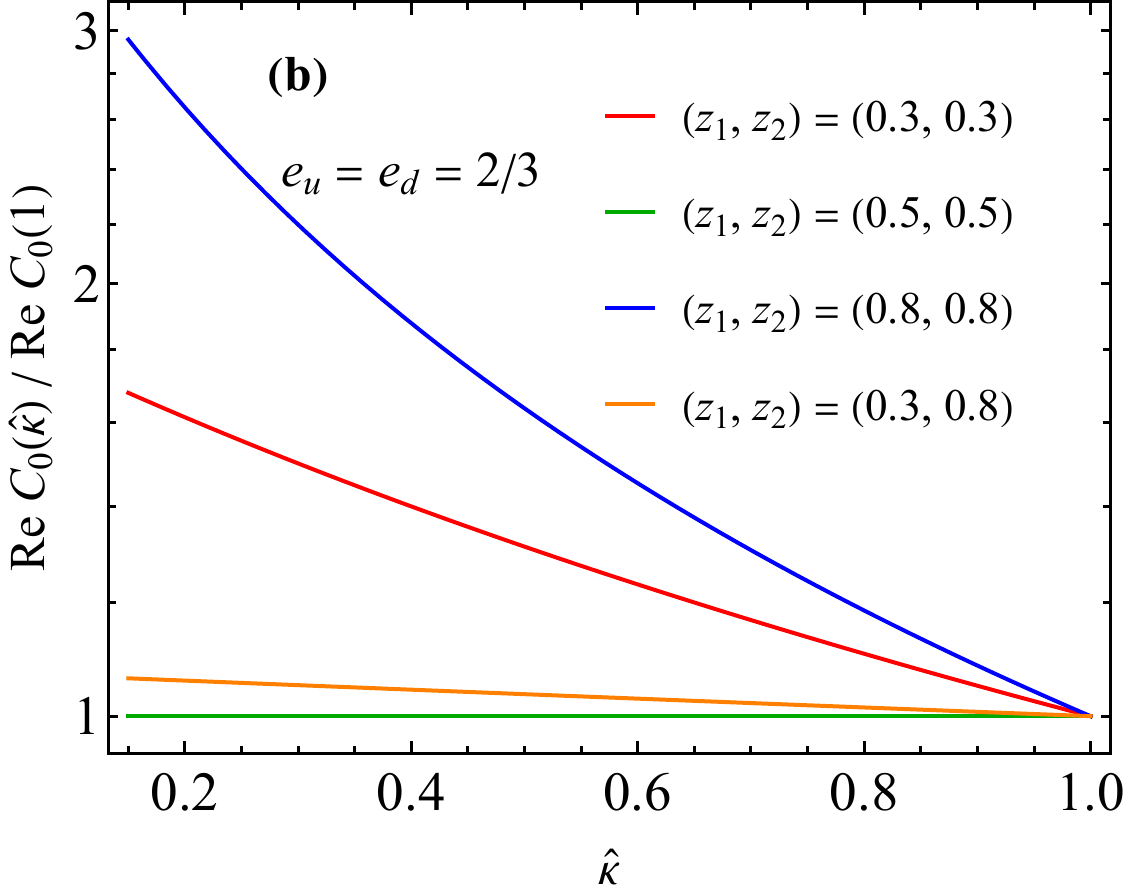}
\caption{The shapes of $\hat{\kappa}$ distribution in $\Re \C_0$ in eq.~\eqref{eq:ReC0}, normalized to its value at $\hat{\kappa} = 1$, for a few different values of $(z_1, z_2)$, for (a) $(e_u, e_d) = (2/3, -1/3)$ and (b) $(e_u, e_d) = (2/3, 2/3)$.}
\label{fig:kappa z1 z2}
\end{figure}

\vskip 0.1in

\noindent
The coefficient $\C_1$ is given by,
\begin{align}
\re\C_1 =&\, \left(e_u-e_d\right)^2 \left[-\frac{8 (1-2 z_1) (1-2 z_2)}{\hat{\kappa}  z_1 z_2 (1-z_1)(1-z_2)}\right] 
   +\left(e_u^2-e_d^2\right) \left[\frac{4 (z_1 - z_2)}{z_1 z_2 (1-z_1)(1-z_2)} \right]\nn\\
	&\, +e_u e_d \left[ \frac{8}{z_1 z_2 (1-z_1)(1-z_2)} \frac{(z_1 (1-z_1) + z_2 (1-z_2)) (z_1 z_2 + (1-z_1)(1-z_2))}{4 z_1 z_2 (1-z_1)(1-z_2)+\hat{\kappa} (1-z_1-z_2)^2}\right]	,\\
\im\C_1 =&\, -4 \pi \left(e_u-e_d\right)^2 \left[ \frac{\delta(z_1)}{1-z_2}+\frac{\delta(1 - z_1)}{z_2} \right]
\nn \\
	&\, - 8 \pi \left(e_u^2-e_d^2\right) \left[ \frac{(1-2 z_2)}{\hat{\kappa}  z_2(1-z_2)} \left( \delta(z_1)+\delta(1 - z_1) \right)\right]
	\nn\\
	&\, + \frac{8 \pi e_u e_d}{\hat{\kappa} } 
	\left[ \left(\frac{\sqrt{1-\hat{\kappa}}}{z_2 (1-z_2)} + \frac{1}{z_2 - z_{as}}\right) \delta(z_1 - p(z_2)) \text{sgn}(z_2 - z_{as}).  \right.
        \nn \\
        &\hspace{4em} 
          -\left(\frac{\sqrt{1-\hat{\kappa}}}{z_2 (1-z_2)} - \frac{1}{z_2 - \z_{as}}\right) \delta(z_1 - \p(z_2)) \text{sgn}(z_2 - \z_{as})
          \nn \\
          &\hspace{4em} \left. 
             -\left(\frac{1 + \hat{\kappa}}{1-z_2}-\frac{2}{z_2}\right) \delta(z_1)    
            -\left(\frac{ 1 + \hat{\kappa}}{z_2} - \frac{2}{1-z_2}\right) \delta(1 - z_1)\right].
\end{align}
We note that for the chiral-even hard coefficients, under the exchange $(z_1, z_2) \leftrightarrow (1-z_1, 1-z_2)$, the $(e_u - e_d)^2$ and $e_u e_d$ terms are invariant, but the $(e_u^2 - e_d^2)$ terms lead to a minus sign, as a direct result of the isospin breaking effect from Type-$B$ diagrams, cf. eqs.~\eqref{eq:C conjugation A}-\eqref{eq:CA symmetry}.
\vskip 0.1in

\noindent
The coefficient $\Co_1$ is given by,
\begin{align}
\re\Co_1 = &\, - \frac{8\left(e_u-e_d\right)^2}{\hat{\kappa}} \frac{ \sqrt{1-\hat{\kappa} }\, (z_1 - z_2) + z_1 z_2 - (1-z_1)(1-z_2) }{z_1 z_2 (1-z_1)(1-z_2)} \nn\\
	&\, - \frac{ 8 e_u e_d }{z_1 z_2 (1-z_1)(1-z_2)} \frac{ \sqrt{1-\hat{\kappa} }\, (z_1 - z_2) (1-z_1-z_2)^2 }{4 z_1 z_2 (1-z_1)(1-z_2) + \hat{\kappa} (1-z_1-z_2)^2},\\
\im\Co_1 = &\, \left(e_u^2-e_d^2\right) \frac{8 \pi}{\hat{\kappa}} 
	\left[ \left(\frac{1}{1-z_2}+\frac{\sqrt{1-\hat{\kappa} }\,}{z_2}\right) \delta(1 - z_1)+\left(\frac{1}{z_2}+\frac{\sqrt{1-\hat{\kappa} }\,}{1-z_2}\right) \delta(z_1)\right] 	\nn\\
	&\,
	+\frac{ 8 \pi  e_u e_d }{\hat{\kappa}  z_2 (1-z_2)} 
	\bigg[ \left(|z_2 - z_{as}| +\frac{\hat{\kappa} }{4 |z_2 - z_{as}| }\right) \delta(z_1 - p(z_2))  \nn \\
	&\hspace{1em} 
	-\left(|z_2 - \z_{as}| +\frac{\hat{\kappa} }{4 |z_2 - \z_{as}| }\right) \delta(z_1 - \p(z_2)) \nn \\
	&\hspace{1em} 
	+\left(\sqrt{1-\hat{\kappa} }\, (1-z_2) + 2 z_2\right)\delta(1 - z_1)    
	-\left(\sqrt{1-\hat{\kappa} }\, z_2 + 2 (1-z_2)\right) \delta(z_1)\bigg]
\end{align}

\noindent
The coefficient $\Co_2$ is given by,
\begin{align}
\re\Co_2 = &\, -\frac{8\left(e_u-e_d\right)^2 }{\hat{\kappa}  z_1 z_2 (1-z_1)(1-z_2)} \left[ \sqrt{1-\hat{\kappa} }\, (z_1 - z_2)-z_1 z_2 + (1-z_1)(1-z_2)\right]		\nn\\
	&\, - \frac{8 e_u e_d}{z_1 z_2 (1-z_1)(1-z_2)} \frac{ \sqrt{1-\hat{\kappa} }\, (z_1 - z_2) (1-z_1-z_2)^2 }{4 z_1 z_2 (1-z_1)(1-z_2)+\hat{\kappa}  (1-z_1-z_2)^2 },\nn\\
\im\Co_2 = &\, \frac{8 \pi \left(e_u^2-e_d^2\right)}{\hat{\kappa}} \left[ \left(\frac{\sqrt{1-\hat{\kappa} }}{z_2} - \frac{1}{1-z_2} \right) \delta(1 - z_1)+\left(\frac{\sqrt{1-\hat{\kappa} }}{1-z_2}-\frac{1}{z_2}\right) \delta(z_1) \right]
	\nn\\
	&\, +\frac{8 \pi  e_u e_d }{\hat{\kappa}  z_2 (1-z_2)}
	\bigg[ \left(|z_2 - z_{as}| +\frac{\hat{\kappa} }{4 |z_2 - z_{as}| }\right) \delta(z_1 - p(z_2)) \nn \\
	&\hspace{1em} 
	-\left(|z_2 - \z_{as}| +\frac{\hat{\kappa} }{4 |z_2 - \z_{as}| }\right) \delta(z_1 - \p(z_2))  \nn \\
	&\hspace{1em}
		+\left(\sqrt{1-\hat{\kappa} }\, (1-z_2)-2 z_2\right) \delta(1 - z_1)	
		-\left(\sqrt{1-\hat{\kappa} }\, z_2 - 2 (1-z_2)\right) \delta(z_1)\bigg].
\end{align} 
The chiral-odd hard coefficients have opposite symmetry behavior to the chiral-even ones under the exchange $(z_1, z_2) \leftrightarrow (1-z_1, 1-z_2)$. Namely, the $(e_u - e_d)^2$ and $e_u e_d$ terms lead to a minus sign, but the $(e_u^2 - e_d^2)$ terms are invariant, cf. eqs.~\eqref{eq:C conjugation A tilde}-\eqref{eq:CA O symmetry}.

\bibliographystyle{JHEP}
\bibliography{bibliography}

\end{document}